\documentclass[structabstract]{aa}  
\usepackage{graphicx,natbib}
\usepackage{lscape}
\usepackage{longtable}
\usepackage{color}
\usepackage{txfonts}

\newcommand{\mi}{\relax \ifmmode {\mu{\mbox m}}\else $\mu$m\fi}
\newcommand{\sii}{\relax \ifmmode {\mbox S\,{\scshape ii}}\else S\,{\scshape ii}\fi}
\newcommand{\siii}{\relax \ifmmode {\mbox S\,{\textsc {iii}}}\else S\,{\scshape iii}\fi}
\newcommand{\siv}{\relax \ifmmode {\mbox S\,{\textsc {iv}}}\else S\,{\scshape iv}\fi}
\newcommand{\nii}{\relax \ifmmode {\mbox N\,{\scshape ii}}\else N\,{\scshape ii}\fi}
\newcommand{\neii}{\relax \ifmmode {\mbox Ne\,{\textsc {ii}}}\else Ne\,{\scshape ii}\fi}
\newcommand{\neiii}{\relax \ifmmode {\mbox Ne\,{\textsc {iii}}}\else Ne\,{\scshape iii}\fi}
\newcommand{\oiii}{\relax \ifmmode {\mbox O\,{\scshape iii}}\else O\,{\scshape iii}\fi}
\newcommand{\oii}{\relax \ifmmode {\mbox O\,{\scshape ii}}\else O\,{\scshape ii}\fi}
\newcommand{\oi}{\relax \ifmmode {\mbox O\,{\scshape i}}\else O\,{\scshape i}\fi}
\newcommand{\ha}{\relax \ifmmode {\mbox H}\alpha\else H$\alpha$\fi}
\newcommand{\hep}{\relax \ifmmode {\mbox H}\epsilon\else H$\epsilon$\fi}
\newcommand{\hdel}{\relax \ifmmode {\mbox H}\delta\else H$\delta$\fi}
\newcommand{\hgam}{\relax \ifmmode {\mbox H}\gamma\else H$\gamma$\fi}

\newcommand{\pa}{\relax \ifmmode {\mbox Pa}\alpha\else Pa$\alpha$\fi}
\newcommand{\hb}{\relax \ifmmode {\mbox H}\beta\else H$\beta$\fi}
\newcommand{\rdostres}{\relax \ifmmode {\,\mbox{R}}_{\rm 23}\else \,\mbox{R}$_{\rm 23}$\fi}
\newcommand{\ergs}{\relax \ifmmode {\,\mbox{erg\,s}}^{-1}\else \,\mbox{erg\,s}$^{-1}$\fi}
\newcommand{\me}{\relax \ifmmode {\,}^{-1}\else \,$^{-1}$\fi}

\newcommand{\msun}{\relax \ifmmode {\,\mbox{M}}_{\odot}\else \,\mbox{M}$_{\odot}$\fi}

\newcommand{\cmtres}{\relax \ifmmode {\,\mbox{cm}}^{-3}\else \,\mbox{cm}$^{-3}$\fi}
\newcommand{\cmdos}{\relax \ifmmode {\,\mbox{cm}}^{-2}\else \,\mbox{cm}$^{-2}$\fi}
\newcommand{\cmseis}{\relax \ifmmode {\,\mbox{cm}}^{-6}\else \,\mbox{cm}$^{-6}$\fi}
\newcommand{\hi}{\relax \ifmmode {\mbox H\,{\scshape i}}\else H\,{\scshape i}\fi}

\newcommand{\arcsecond}{\hbox{$^{\prime\prime}$}}
\newcommand{\Spi}{{\it Spitzer}}
\newcommand{\Her}{{\it Herschel}}
\newcommand{\DE}{{\tt DustEM}}
\newcommand{\ytot}{$Y_{\rm TOTAL}$}
\newcommand{\ypah}{$Y_{\rm PAH}$}
\newcommand{\yvsg}{$Y_{\rm VSG}$}
\newcommand{\ybg}{$Y_{\rm BG}$}
\newcommand{\chidos}{$\chi^{2}_{red}$}
\newcommand{\hii}{\relax \ifmmode {\mbox H\,{\textsc ii}}\else H\,{\scshape ii}\fi}

\usepackage[colorlinks]{hyperref}
\begin{document}

   \title{ Spatially resolving the dust properties and submillimetre excess in M\,33}
   \author{
     M.\,Rela\~{n}o\inst{1,2,5}  \and
     I. \,De Looze\inst{3,4} \and
     R. C. Kennicutt\inst{5} \and
     U.\,Lisenfeld\inst{1,2} \and
     A.\,Dariush\inst{5} \and
     S.\,Verley\inst{1,2} \and
     J.\,Braine\inst{6} \and 
     F.\,Tabatabaei\inst{7,8,9} \and 
     C.\,Kramer\inst{10}\and 
     M.\,Boquien\inst{11} \and
     M.\,Xilouris\inst{12}\and 
     P.\,Gratier\inst{6}}
   \institute{
     Dept. F\'{i}sica Te\'orica y del Cosmos, Universidad de Granada, Spain -- \email{mrelano@ugr.es}
    \and
      Instituto Universitario Carlos I de F\'isica Te\'orica y Computacional, Universidad de Granada, 18071, Granada, Spain
     \and
     Department of Physics and Astronomy, University College London, Gower Street, London WC1E 6BT, UK
     \and 
     Sterrenkundig Observatorium, Universiteit Gent, Krijgslaan 281 S9, B-9000 Gent, Belgium
     \and
     Institute of Astronomy, University of Cambridge, Madingley Road, Cambridge, CB3 0HA, UK
     \and 
     Laboratoire d'Astrophysique de Bordeaux, Univ. Bordeaux, CNRS, B18N, All\'ee Geoffroy Saint-Hilaire, 33615 Pessac, France
     \and
    Instituto de Astrof\'isica de Canarias, V\'ia L\'actea S/N, E-38205 La Laguna, Spain 
    \and 
    Departamento de Astrof\'isica, Universidad de La Laguna, E-38206 La Laguna, Spain 
    \and
    Max-Planck-Institut fur Astronomie, Konigstuhl 17, D-69117 Heidelberg, Germany  
    \and
    Instituto Radioastronom\'{i}a Milim\'{e}trica, Av. Divina Pastora 7, N\'ucleo Central, E-18012 Granada, Spain
     \and
     Unidad de Astronom\'{i}a, Fac. Cs. B\'asicas, Universidad de Antofagasta, Avda. U. de Antofagasta 02800, Antofagasta, Chile
     \and
     Institute of Astronomy and Astrophysics, National Observatory of Athens, P. Penteli, 15236 Athens, Greece
   }

   \date{Received ; accepted }
 
  \abstract
   {The relative abundance of the dust grain types in the interstellar medium is directly linked to physical quantities that trace the evolution of galaxies. Because of the poor spatial resolution of the infrared/submillimetre data, we are able to study the dependence of the resolved infrared Spectral Energy Distribution (SED) across regions of the interstellar medium (ISM) with different physical properties in just a few objects.}  
  {We study the dust properties of the whole disc of M\,33 at spatial scales of $\sim$170\,pc. This analysis allows us to infer how the relative dust grain abundance changes with the conditions of the ISM, study the existence of a submillimetre excess and look for trends of the gas-to-dust mass ratio (GDR) with other physical properties of the galaxy.}
 {For each pixel in the disc of M\,33 we fit the infrared SED using a physically motivated dust model that assumes an emissivity index $\beta$ close to 2.  We apply a Bayesian statistical method to fit the individual SEDs and derive the best output values from the study of the probability density function of each parameter. We derive the relative amount of the different dust grains in the model, the total dust mass, and the strength of the interstellar radiation field (ISRF) heating the dust at each spatial location.}
  {The relative abundance of very small grains (VSGs) tends to increase, and for big grains (BGs) to decrease, at high values of \ha\ luminosity. This shows that the dust grains are modified inside the star-forming regions, in agreement with a theoretical framework of dust evolution under different physical conditions.  The radial dependence of the GDR is consistent with the shallow metallicity gradient observed in this galaxy. The strength of the ISRF derived in our model correlates with the star formation rate (SFR) in the galaxy in a pixel by pixel basis. Although this is expected it is the first time that a correlation between both quantities is reported. We produce a map of submillimetre excess in the 500\,\mi\ SPIRE band for the disc of M\,33. The excess can be as high as 50\% and increases at large galactocentric distances.  We further study the relation of the excess with other physical properties of the galaxy and find that the excess is prominent in zones of diffuse ISM outside the main star-forming regions, where the molecular gas and dust surface density are low.}
   {}

   \maketitle
%

\section{Introduction}\label{sec:intr}
The infrared/submillimetre spectral energy distribution (SED) of galaxies provides a unique set of data to study how the dust is processed in galaxies and in the interstellar medium (ISM) in general. The emission from dust has been proposed as a probe of the amount of star formation within a galaxy, therefore understanding better how the dust physical properties change under different conditions of the ISM will help us to get a better picture on how the dust traces  star formation rate (SFR) and how galaxies evolve with time.

 The physical properties of the dust are directly linked to those of the ISM where it is located. The dust is not only heated by the {\it interstellar radiation field}ISRF but it is also affected by other mechanisms, such as destruction or coagulation of dust grain species \citep[see][for a review]{2004ASPC..309..347J} that are at play in the ISM and can lead to a change in its physical properties.  \citet{1994ApJ...433..797J,1996ApJ...469..740J} set up a theoretical framework for dust destruction mechanisms in regions of the ISM affected by strong shocks. These studies have been recently updated by \citet{2014A&A...570A..32B} \citep[see also][]{2015ApJ...803....7S}. There is some observational evidence that dust is evolving under different environments: dust grains can fragment in the ISM  \citep[e.g.][]{2011ApJ...735....6P,2014ApJ...784..147S} and very small grains (VSGs) can coagulate on the surfaces of big grains (BGs) \citep{2003A&A...398..551S,2009A&A...506..745P,2012A&A...548A..61K}. In an observational study \citet[][]{2016A&A...595A..43R}\defcitealias{2016A&A...595A..43R}{Paper I} \citepalias[hereafter][]{2016A&A...595A..43R} found that the  abundance of VSGs is higher in intense star-forming regions, where shocks might be present, than in more quiescent diffuse-type \hii\ regions. Besides, it is also well known that the amount of polycyclic aromatic hydrocarbons (PAHs) are reduced in regions of intense ISRF \citep[e.g.][]{2006A&A...446..877M} and that the dust abundance changes with the metallicity of the environment \citep{2003A&A...407..159G,2005A&A...434..867G,2008ApJ...672..214G}. 

The analysis of integrated SEDs in galaxies with different metallicity has shown that some low metallicity objects exhibit an excess emission above the modelled dust continuum in the submillimetre wavelength range, which is referred as {\it submillimetre excess}: in the Large Magellanic Cloud (LMC) and Small Magellanic Clouds (SMC) \citep{2010A&A...519A..67I,2010A&A...523A..20B} and in nearby galaxies \citep{2003A&A...407..159G,2005A&A...434..867G, 2011A&A...532A..56G,2012ApJ...745...95D,2013ApJ...778...51K,2013A&A...557A..95R}. The origin of this excess is so far unknown. Some studies have tried to locally isolate this excess to shed light on to the origin of this phenomenon: \citet{2011A&A...536A..88G} found that the excess is located in regions of low dust mass surface density in the LMC, and \citet{2014MNRAS.439.2542G} studied a sample of KINGFISH (Key Insights on Nearby Galaxies: a Far-Infrared Survey with Herschel) galaxies \citep{2011PASP..123.1347K} and found that the excess was located in the outer parts of the discs. Also \citet{2014ApJ...797...85G} found that the excess in the 500\,\mi\ SPIRE band was located in the outer parts of the most intense star-forming regions in the LMC and SMC. \citet{2014A&A...561A..95T} showed that the emissivity coefficient $\beta$ tends to take lower values in the outer parts of the disc of M\,33. A similar result was found by \citet{2012ApJ...756...40S} for M\,31. A lower $\beta$ would be the required coefficient to fit the excess of the SED in the submillimetre wavelength range. Finally, \citet{2016A&A...590A..56H} presented observational evidence of excess in M\,33 located mainly in the diffuse part of the disc of this galaxy. These authors made a careful analysis to fit the excess with the emission produced by different proposed mechanisms: they found that a change in the physical properties of the dust would be the best explanation for the excess in this galaxy. In this present study, the submillimetre excess is defined as it was done by \citet{2016A&A...590A..56H}: the fraction of emission in the submillimetre range that is above a dust model having an emissivity coefficient $\beta$\,=\,2. SEDs fitted with dust models with variable $\beta$ would not, in principle, present any excess, as lower values of $\beta\leq$\,2 would be able to fit the observed SED.

The ratio between the gas and dust mass, GDR, gives an estimation of the amount of metals locked up in the dust compared to the ones still present in the gas phase. It is directly linked to the metallicity of the ISM \citep{1998ApJ...501..643D,1998ApJ...496..145L,2007ApJ...663..866D,2014A&A...563A..31R}, thus providing information about the evolutionary stage of the galaxy. Quantifying the GDR requires an estimation of the gas and dust masses of galaxies. Total gas masses are derived using neutral hydrogen and CO observations. The CO observations would lead, assuming a CO-to-$\rm H_{2}$ conversion factor $X(\rm CO)$, to an estimate of the amount of molecular mass traced by CO. Some difficulties in obtaining accurate molecular gas masses reside in the existence of a 'CO-dark gas component', a fraction of molecular gas that under certain physical conditions is not traced by CO observations \citep[][]{2014A&A...570A.121P,2013ARA&A..51..207B,2012IAUS..284..141M,2008AJ....136..919B}, and in the variation of $X(\rm CO)$ with the metallicity of the galaxy \citep{2002Ap&SS.281..127B,2011ApJ...737...12L,2012AJ....143..138S,2013ApJ...777....5S,2017MNRAS.470.4750A}. The dust mass estimates are also affected by certain uncertainties: they are subject to the dust model assumed to explain the far infrared peak of the SED, e.g. the assumed emissivity coefficient, and affected by the evolution of dust under different physical conditions \citep{2011A&A...536A..88G}.  

In this paper we present a study of the SED of the whole disc of M\,33 in a pixel-by-pixel basis. We want to better understand better which mechanisms might produce the excess in the submillimetre wavelength range and how the GDR depends on other properties of the galaxy. M\,33 is an ideal target to perform such as a study because the distance to the galaxy   \citep[840\,kpc,][]{1991ApJ...372..455F} and the high resolution of the available infrared (IR) observations allow us to perform a SED analysis at small galactic scales of $\sim$170\,pc. Besides, the disc covers a wide range of physical interstellar conditions over a large amount of pixels to give statistically significant conclusions. The submillimetre excess for the total SED of M\,33 was observed for the first time by \citet{2016A&A...590A..56H}, who found that the diffuse part of the galaxy exhibits a higher fraction of excess than the SED corresponding to the integrated emission coming from the star-forming regions. GDR measurements for M\,33 were obtained by  \citet{2017A&A...600A..27G} who, following the method derived by \citet{2011ApJ...737...12L} and \citet{2013ApJ...777....5S}, and assuming a single modified black body (MBB) with variable $\beta$ for the dust emission derived in \citet{2014A&A...561A..95T}, were able to estimate the GDR over the disc of M\,33 taking into account the fraction of dark gas not traced by CO observations. 

Our analysis goes a step further than these studies as we apply a dust model that assumes different dust grain species. Besides, we vary the fraction of the different dust grains to fit the SED at local $\sim($170\,pc) scales in the disc of M\,33. With this approach we are able to isolate the regions where the submillimetre excess is located, and thus see if these locations share similar physical conditions. Moreover, we are able to better estimate the dust mass at local scales as the variation of the different grain types in the model allows us to mimic the grain evolution across the disc of M\,33.

The paper is outlined as followed. In Section~\ref{sec:dataset} we present the data used in this study and the preliminary analysis to create the data cube that will be used to perform the SED fitting. In Section~\ref{sec:sed_fit} we present the dust model and describe the fitting technique applied to each pixel in the disc of M\,33. In Section~\ref{sec:results} we show the direct results from the fitting procedure: dust masses and grain abundances. The GDR and its relation to other physical properties is described in Section~\ref{sec:GtD}. We analyse how well the strength of the ISRF is linked to the amount of star formation at local scales in Section~\ref{sec:G0_SFR}. In Section~\ref{sec:subm} we study the observed submillimetre excess and its trend with the spatial location within the disc of the galaxy. Finally, the conclusions are presented in Section~\ref{sec:conc}. 

\section{The data sets}\label{sec:dataset}
For the purpose of this study we make use of the available data of M\,33 covering the mid-IR to submillimetre wavelength range. We use \Spi\ IRAC (3.6\,\mi, 4.5\,\mi, 5.8\,\mi, and 8.0\,\mi) and \Spi\ MIPS (24\,\mi, 70\,\mi) data already presented in \citet{2013A&A...552A.140R}. We refer the reader to this paper for further description of these particular datasets. We cover the gap between 8.0\,\mi\ and 24\,\mi\ with 12\,\mi\ ({\it W}3), and 22\,\mi\ ({\it W}4) {\it Wide-field Infrared Survey Explorer} ({\it WISE}) \citep{2010AJ....140.1868W} data. The data reduction has already been reported in \citetalias{2016A&A...595A..43R}.  \Her\ data (PACS: 70\,\mi, 100\,\mi, and 160\,\mi, and SPIRE: 250\,\mi, 350\,\mi, and 500\,\mi) were presented within the HerM33es  (Herschel M33 extended survey HerM33es) collaboration \citep{Kramer:2010p688,2012A&A...543A..74X,Boquien:2011p764,2015A&A...578A...8B}. 
We use the latest version of 100\,\mi\ PACS data reprocessed with Scanamorphos v16 \citep{2013PASP..125.1126R} as described in \citet{Boquien:2011p764}, and the new observed 70\,\mi\ and 160\,\mi\ PACS data, obtained in a follow-up open time cycle 2 programme of \Her\ \citep{2015A&A...578A...8B}. The SPIRE images were obtained using version 10.3.0 of  the \Her\ Data Processing System \citep[HIPE,][]{2010ASPC..434..139O,2011ASPC..442..347O} with the updated calibration scheme (spire$_{-}$cal$_{-}$10$_{-}$1).  The beam areas for the three SPIRE bands (250\,\mi, 350\,\mi\ and 500\,\mi) are 431.7, 766.0, 1572.7\,$\rm arcsec^{2}$. For a description of the data reduction the reader is referred to \citet{2012A&A...543A..74X}. Finally, $^{12}$CO(J=2--1) and HI (21\,cm) intensity maps were obtained from \citet{2014A&A...567A.118D} and \citet{2010A&A...522A...3G}, respectively.  Each image in our data set was registered first and then convolved to the spatial resolution of the SPIRE 500\,\mi\ image using the convolution kernels of G. Aniano \citep{2011PASP..123.1218A}. The final data set has a spatial resolution of $\sim$38.1\arcsecond$\times$35.1\arcsecond\ and a pixel size of 14\arcsecond\ ($\sim$56\,pc) and will produce SEDs with the fluxes of 14 different filters spanning between 3.6\,\mi\ and 500\,\mi. 

\subsection{Uncertainties}
We assumed the following uncertainties for our data set. For IRAC data we assume a conservative value of 10\% uncertainty for all IRAC bands, as reported in the {\it IRAC Instrument Handbook}\footnote{\url {http://irsa.ipac.caltech.edu/data/SPITZER/docs/irac/iracinstrumenthandbook/17}}. For WISE 12\,\mi\ ({\it W}3) and 22\,\mi\ ({\it W}4) we adopt a flux uncertainty of 10\% to account for discrepancy between red and blue calibrators as it is reported in {\it Explanatory Supplement to the WISE All-Sky Data Release Products}. MIPS 24\,\mi\ and 70\,\mi\ uncertainties are 4\% \citep{2007PASP..119..994E} and 10\% \citep{2007PASP..119.1019G}, respectively. We assume an uncertainty of 10\% in the three PACS bands, which is a combination of the absolute uncertainty in the calibration plus an additional uncertainty to account for extended emission \citep{2014ApJ...797...85G}. Finally, the uncertainty for SPIRE data is 15\% \citep{2010A&A...518L...4S,Kramer:2010p688}.

\subsection{Masking}\label{sec:mask}
\begin{figure}[h]
  \includegraphics[width=0.49\textwidth]{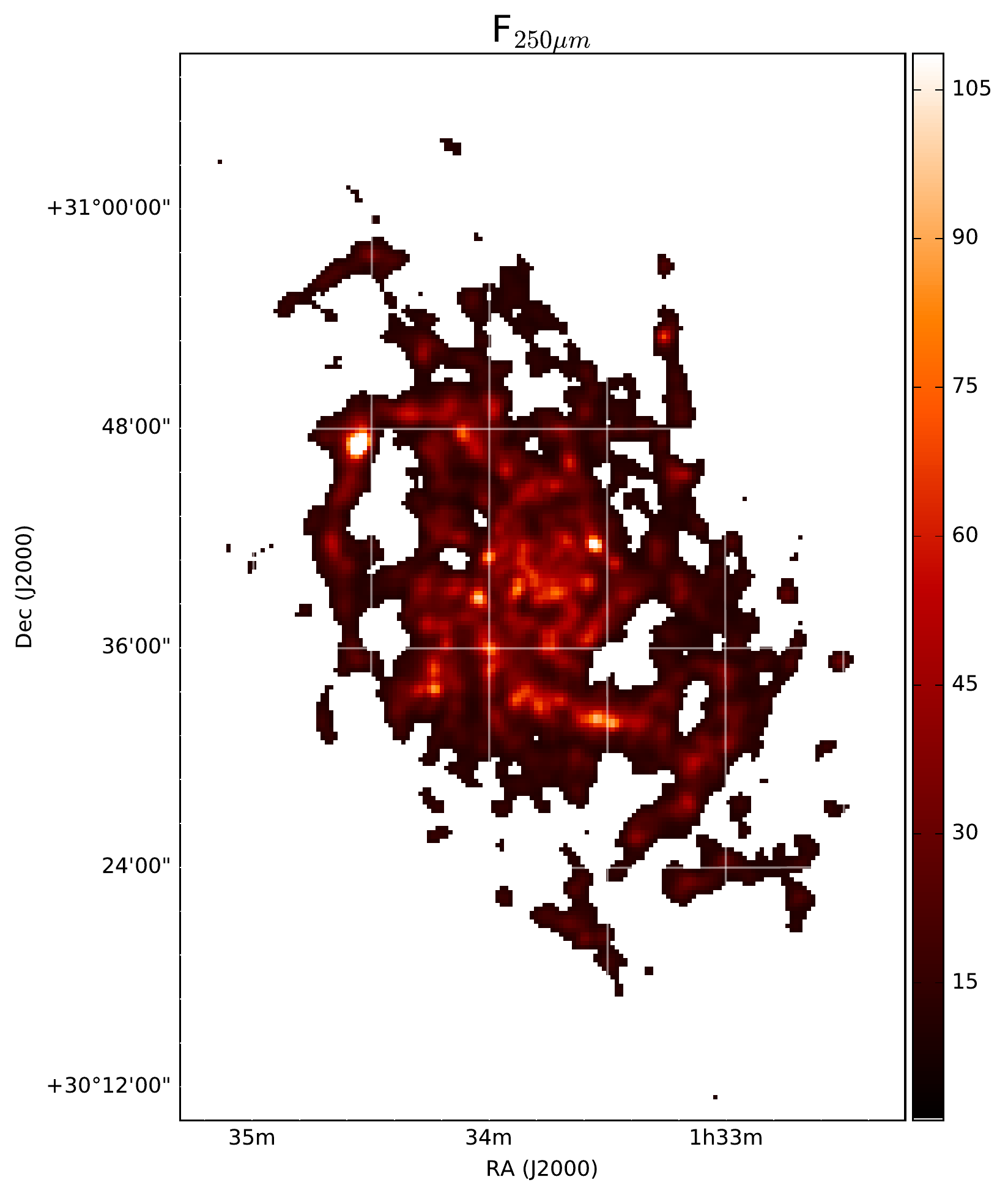}   
   \caption{250\,\mi\ intensity image in MJy/sr masked using the constraints presented in Section~\ref{sec:mask}.}
   \label{fig:250ima}
\end{figure}

We are interested in fitting the SEDs of each pixel in the disc of M\,33. However, in order to avoid bad fits which produce misleading conclusions we fit only the SED with reliable flux measurements in all bands. We thus first masked in each image all the pixels with fluxes below 3 times the standard deviation of the background value. Then we produced another mask using the 250\,\mi\ image: all the pixels with fluxes below F(250\,\mi)=10 MJy/sr were eliminated. The threshold is chosen to cover a wide area of the disk with diffuse and star-forming regions while avoiding low surface brightness areas that would lead to unreliable fits. The spatial coverage of the M\,33 disc is shown in Fig\,\ref{fig:250ima}. We apply the mask to all the images in our data set and create a masked data-cube which will be used in our SED fitting procedure.  

\section{SED fitting technique}\label{sec:sed_fit}

We make use of the tool \DE\ \citep{2011A&A...525A.103C} to obtain the emission of the different dust grain types in our dust model. The code allows the use of different dust models and ISRFs as inputs and gives the emissivity per hydrogen atom ($\rm N_{H} = N_{HI}+2N_{H_{2}}$) for each grain type based on the incident radiation field strength and the grain physics in the optically thin limit. We will adopt here the classical dust model from \citet[][]{1990A&A...237..215D} and the ISRF from \citet{Mathis:1983p593}. Although more sophisticated dust models can be used with \DE\ we decide not to adopt them here. The reason for this choice is that we are interested in studying the submillimetre excess, which is directly related to the emissivity index $\beta$ (see section~\ref{sec:subm}) and most of the submillimetre excess reported in the literature is based on MBB fitting with $\beta$=2. Making use of the average extinction of the dust distribution provided by \DE\ we obtain a grain absorption cross section of  $\rm \kappa_{350\,\mu m}=4.8\,cm^{2}g^{-1}$ and $\beta$\,=\,2. Therefore, the submillimetre excess detected using the dust model from \citet[][]{1990A&A...237..215D} can be directly compared with the results obtained from fitting the SED with a MBB with  $\beta$\,=\,2. This model consists of three dust grain types: PAHs, VSGs of carbonaceous material, and BGs of astronomical silicates. For the purpose of this study we will change the mass relative to hydrogen of each dust grain population ($\rm Y_{i}=M_{i}/M_{H}$, for i = PAH, VSG, and BG) and the ISRF parameter $\rm G_{0}$, which is the scaling factor relative to the solar neighbourhood ISRF given in \citet{Mathis:1983p593}. The ISRF is measured as energy density per unit frequency as: $\rm G_{nu}=G_{0}\times G_{nu}^{\rm M83}$, where $\rm G_{nu}^{\rm M83}$ is the ISRF for the solar neighbourhood. Finally, in order to fit the contribution of the stellar emission in the near-IR, we add a near-IR continuum modelled using a black body with a temperature of 1000\,K following \citet{2016A&A...595A..43R}.

The fit of the observed SED in each pixel of the data cube generated in Section~\ref{sec:mask} was performed using a Bayesian approach similar to the one followed in \citet{2008MNRAS.388.1595D}. We outline here the steps to perform the fit. We first create a library of models with different values of $Y_{i}$, $\rm G_{0}$, and $G_{\rm NIR}$ (the scale factor of the near-IR 1000\,K black body continuum). Making use of the previous analysis of \hii\ regions in M\,33  \citep{2016A&A...595A..43R}, the parameter values are generated within a wide range that covers possible solutions for this galaxy. For each combination of the input parameters, we obtained the dust emissivity per hydrogen atom with the use of \DE\ and created a library of 7\,695\,000 dust models. In Table~\ref{tab:param} we show the range and the sampling of the input parameters. We convolve the SEDs of each dust model in the library with the corresponding filter bandpass of our observations to obtain the fluxes in each band for each library model.

\begin{table}
\begin{center}
\begin{tabular}{ccc}
\hline
Parameter & Range  \\
\hline
$\rm G_{0}$ & 0.1-60, Unevenly spaced grid$^a$  \\
\ypah\ & \ypah(DBP)$\times$10$^{n}$, with $n$=[-0.5,1], $\Delta n=0.1$ \\
\yvsg\ & \yvsg(DBP)$\times$10$^{n}$, with $n$=[-0.5,1], $\Delta n=0.1$ \\
\ybg\ & \ybg(DBP)$\times$10$^{n}$, with $n$=[-0.5,1], $\Delta n=0.1$ \\
$G_{\rm NIR}$ & 1$\times$10$^{n}$, with $n$=[-2.0], $\Delta n=0.1$ \\
\hline
\end{tabular}
\caption{$a$: $\rm G_{0}$ range of values is: [0.1,5] in steps of 0.1, [5, 30] in steps of 0.5 and [30, 60] in steps of 2. \ypah(DBP), \yvsg(DPB), and \ybg(DBP) are the mass abundances relative to hydrogen in the dust model from \citet[][]{1990A&A...237..215D},  $4.3\times 10^{-4}$,  $4.7\times 10^{-4}$, and $6.4\times 10^{-3}$, respectively. } 
\label{tab:param}
\end{center}
\end{table}

\begin{figure} 
\includegraphics[width=0.49\textwidth]{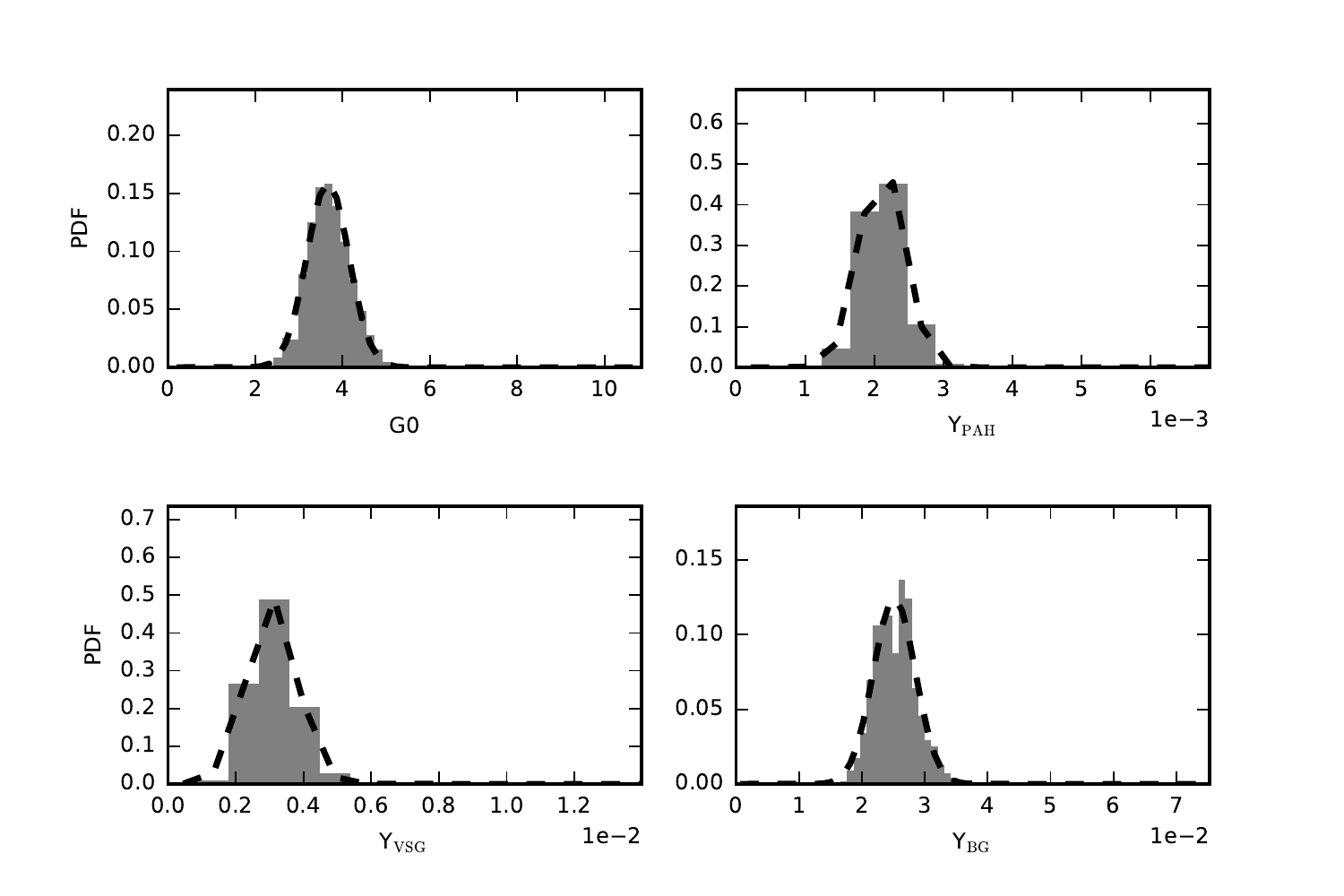}
\includegraphics[width=0.49\textwidth]{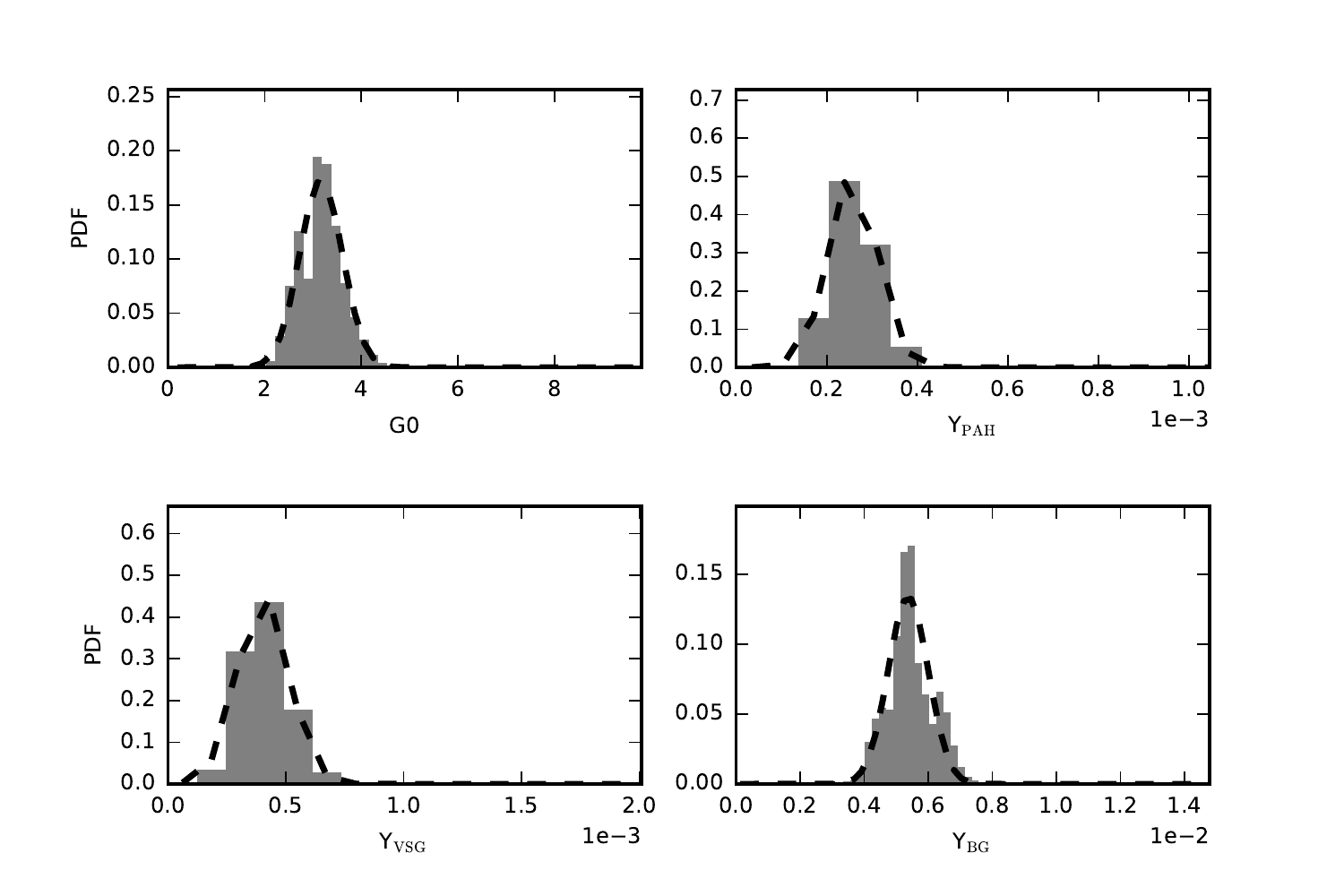}

   \caption{Histograms derived in our fitting procedure for two different pixels (70, 123) (top) and (88, 144) (bottom) corresponding to star-forming and diffuse regions, respectively. The values of the best fit parameters obtained as described in the text are: $\rm G_{0}$ = 3.69$\pm$ 0.97, \ypah = $(2.1\pm0.4)\times 10^{-3}$, \yvsg =  $(3.1\pm1.8)\times 10^{-3}$, and \ybg = $(2.5\pm0.6)\times 10^{-2}$ for the top panel,  and $\rm G_{0}$ = 3.17$\pm$ 0.97, \ypah = $(2.6\pm0.7)\times 10^{-4}$, \yvsg =  $(4.1\pm2.5)\times 10^{-4}$, and \ybg =  $(5.5\pm1.5)\times 10^{-3}$  for the bottom panel. The dashed line corresponds to the Gaussian fitted to each PDF. We compare the results from the best fit parameters obtained with our procedure and those derived from the Gaussian fitting in Section~\ref{App:robust} of the Appendix.}
   \label{fig:histoparam}
\end{figure}

We obtained the parameters that best reproduce the observed data for each pixel in the following way. For each model in the library we compare the theoretical fluxes with the observed ones using the $\chi^{2}$ parameter and build up the probability density function (PDF) by weighting the input parameter distribution in each model $i$ by the likelihood $e^{-\chi_{i}^{2}/2}$. We take the mean of the PDF as the best parameter value and the 16th-84th percentile range as an estimate of its uncertainty. In Fig.~\ref{fig:histoparam} we show two typical examples of PDFs derived with our fitting procedure.

\begin{figure} 
 \includegraphics[width=0.49\textwidth]{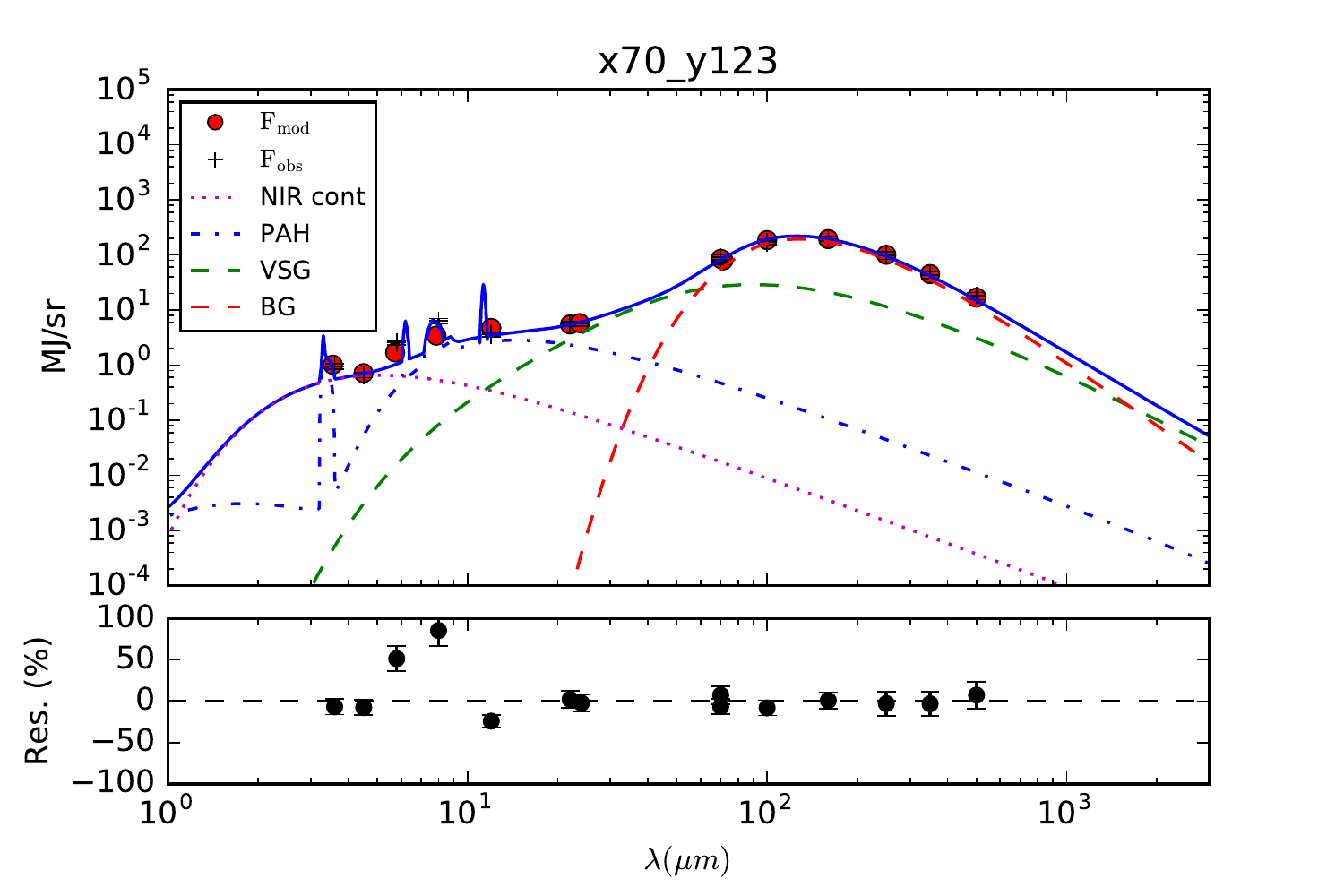}   
 \includegraphics[width=0.49\textwidth]{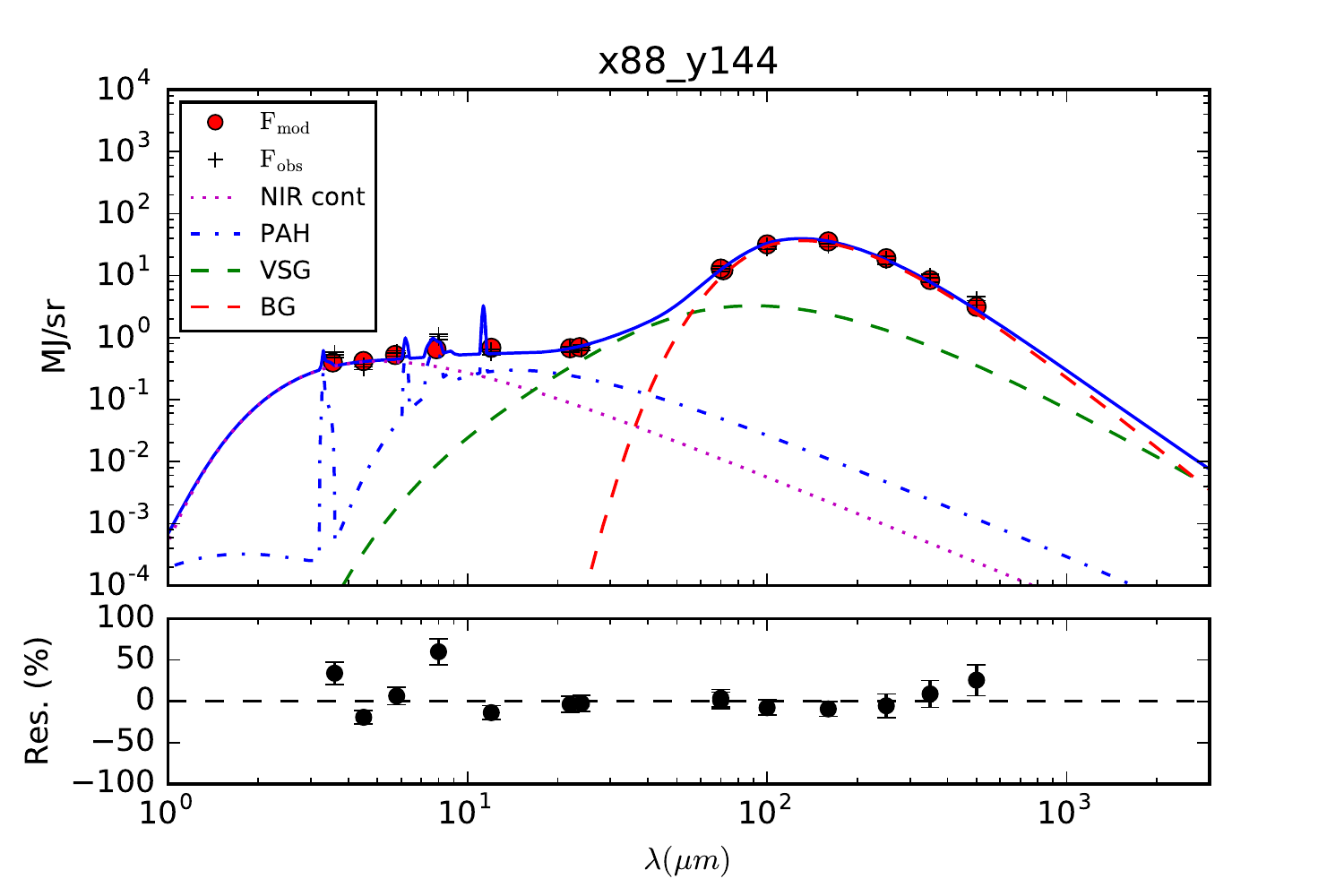}   
   \caption{Examples of SED fitting for pixels (70,123) (top) and (88, 144) (bottom) corresponding to star-forming and diffuse regions, respectively. \chidos\ values are 5.1 and 3.7 for the top and bottom panels, respectively. The bottom panel shows submilimetre excess in the 500\,\mi\ band. We will study this excess in Section\,\ref{sec:subm}.}
   \label{fig:sed}
\end{figure}

In Fig.~\ref{fig:sed} we show two examples of the best fit SEDs for two different pixels within the disc of M\,33. The far-IR peak is nicely fitted with this routine. The PAH features are not always well fitted with the dust model from \citet{1990A&A...237..215D}, as was already reported in \citetalias{2016A&A...595A..43R}. However, for the purpose of this study we are interested in deriving good estimates of the dust mass and the submilimeter excess, thus the lack of good fits in the PAH domain does not affect the conclusions of this paper. We analyse the robustness of the fit and the existence of double peaks in the PDF in Section~\ref{App:robust} and ~\ref{App:doub} of the Appendix. In  Fig.~\ref{fig:chi2_dist} we show the \chidos\ distribution for all the fits. In general we are able to get good fits (\chidos$\lesssim$8.0) for most of the pixels in our sample. For the rest of the study we take into account only those pixels with fits having \chidos$\lesssim$8.0.
The relations between the best fit parameters are shown in Fig.~\ref{fig:allparamplot}.

\section{Results from the fitting}\label{sec:results}
In this section we analyse the direct results from our fitting procedure. Although we show here maps of the parameters derived from the fit for the whole disc of M\,33 (e.g. Fig.~\ref{fig:sd_dust}), in order to make a meaningful statistical analysis of the correlations presented in this paper we take into account only independent pixels. This is done by choosing pixels spaced roughly the spatial resolution of the data set (see Section\,\ref{sec:dataset}), following \citet{2013A&A...552A..19T}. In Section\,\ref{sec:dustmass_comp} we study the dust mass derived in a pixel-by-pixel basis for the disc of M\,33, in Section\,\ref{sec:grain_filters} we analyse the contribution of the different grain types to the emission of each filter. Finally, in Section\,\ref{sec:grain_abund} we compare the relative abundances for each grain type to other physical properties of the ISM.

\subsection{Dust Mass}\label{sec:dustmass_comp}

\begin{figure} 
  \includegraphics[width=0.49\textwidth]{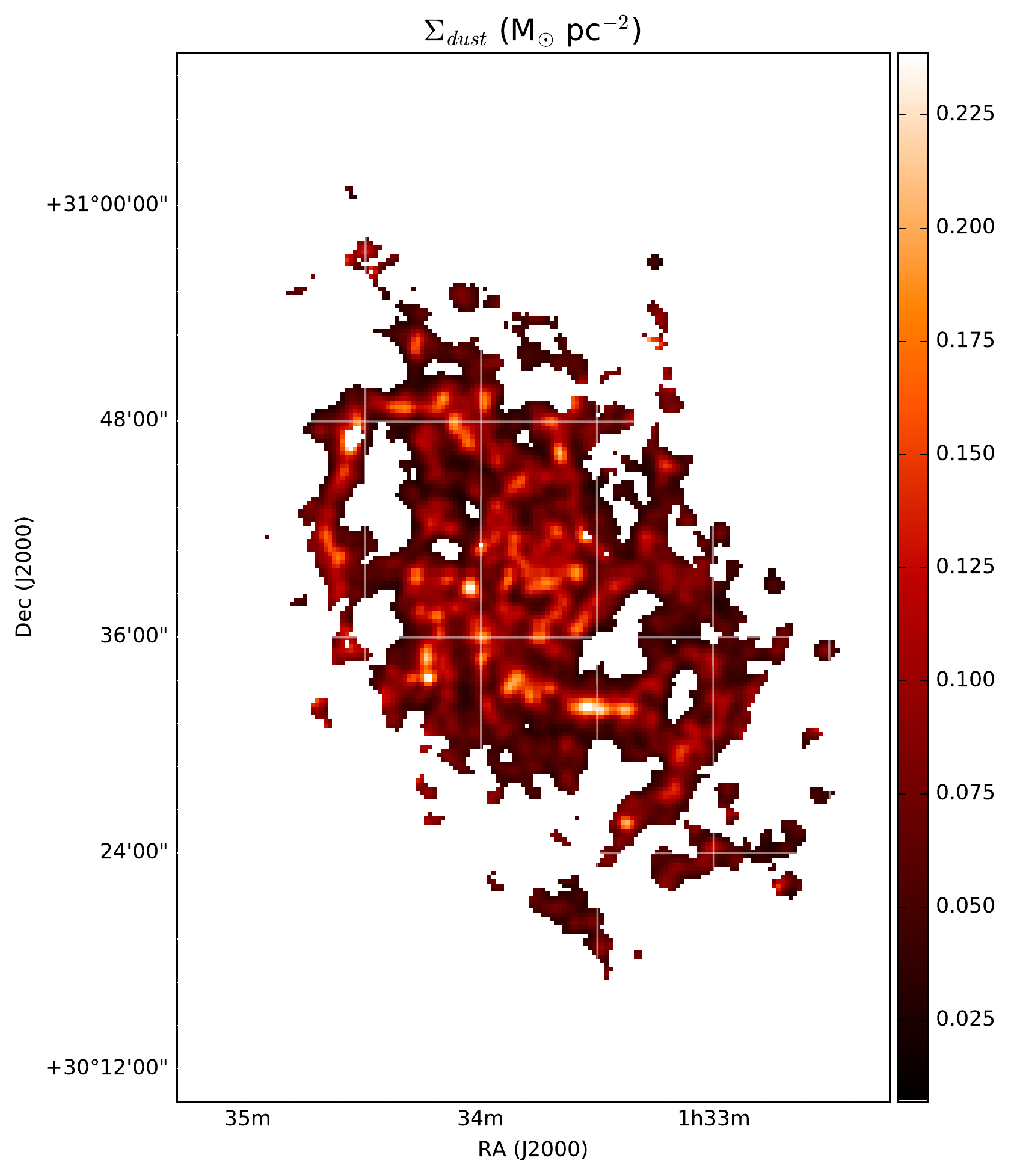}   
   \caption{Map of the dust surface density in units of \msun\,pc$^{-2}$ for M\,33 derived from our fitting technique.}
   \label{fig:sd_dust}
\end{figure}

From the fitting we can estimate the dust mass in every pixel in the disc of M\,33. The result is shown in Fig.~\ref{fig:sd_dust}. We can see the spiral arms delineated in the map, showing that in general the dust mass surface density is higher at the location of the star-forming regions. This agrees with the correlation between the SFR and dust mass surface density found by \cite{2012ApJ...756...40S} for M31 using single-MBB fitting (top-left panel of Fig.\,11 in that paper). A similar trend was found by \citet{2014A&A...561A..95T} for the cold dust mass derived using two-MBBs corresponding to the warm and cold dust  (see their Fig.~5). These authors kept fixed $\beta_{w}$ ($\beta$ coefficient for the warm dust emission) to 1.5, while the one corresponding to the cold dust, $\beta_{c}$, was left free. 
We regrid our images to a pixel size of 10\arcsec, which is the one used by \citet{2014A&A...561A..95T}, and perform the fitting routine for each pixel in the new dataset. We then produce a new dust mass map of M\,33, which is directly comparable in a pixel by pixel basis with the results from \citet{2014A&A...561A..95T}. 

\begin{figure}
\centering 
\includegraphics[width=0.49\textwidth]{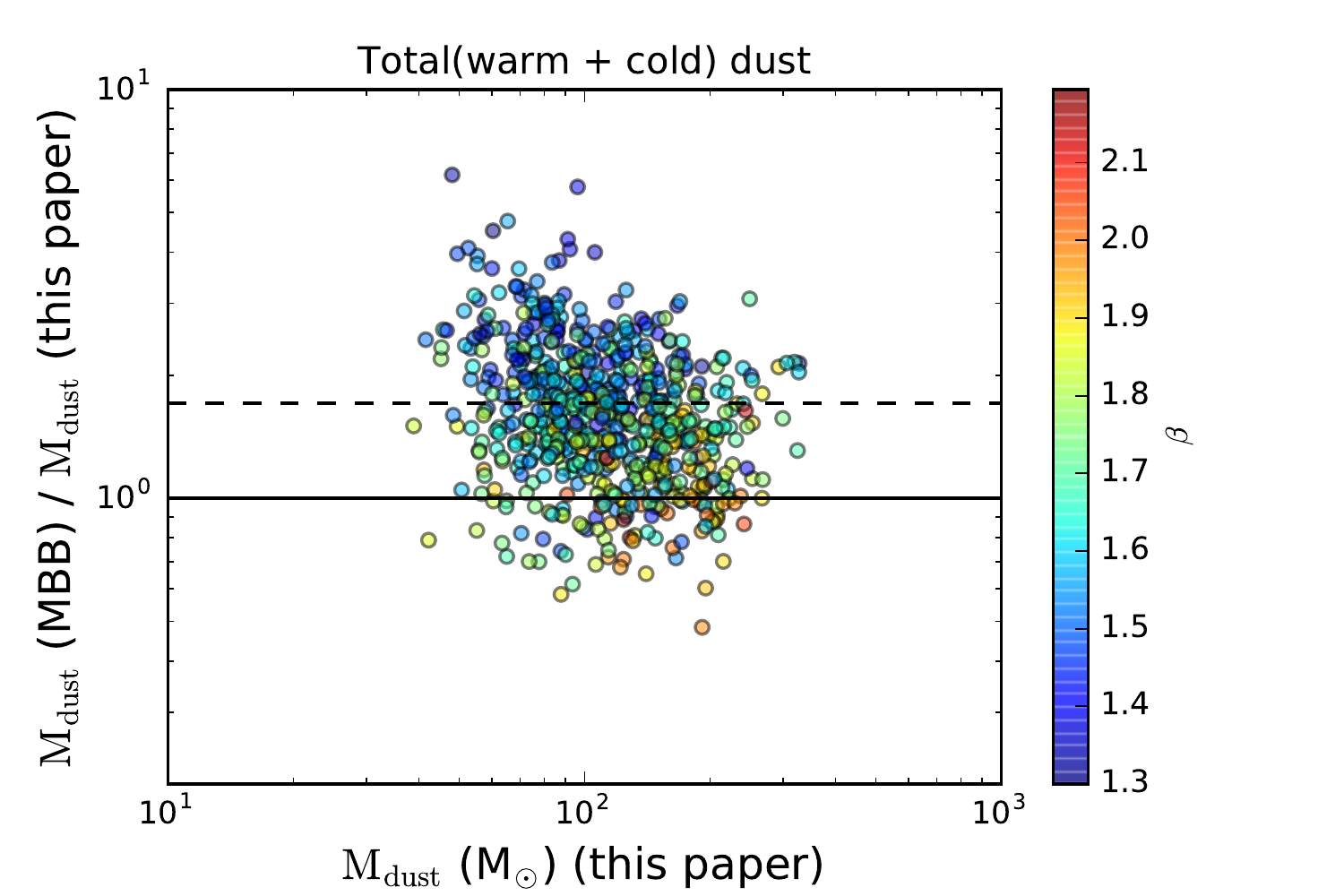} 
\caption{Comparison of the dust masses obtained in this paper and those derived by \citet{2014A&A...561A..95T} with two MBBs. Each point corresponds to one pixel on the sky. Only independent data points have been considered to compare both masses (see text for details). The mean value and standard deviation of the ratio between the dust mass derived from \citet{2014A&A...561A..95T} and from this study is 1.7 (dashed line) and 0.7, respectively. The relative uncertainties for the dust masses derived in this paper are $\sim$25\%.}
\label{fig:dustcomp}
\end{figure}

In Fig.~\ref{fig:dustcomp} we show the comparison of the sum of the warm and cold dust mass derived by  these authors and the dust masses derived with our fitting technique. In general the MBB dust masses are above those derived with the dust model from \citet[][]{1990A&A...237..215D}. The mean value of the ratio between the dust mass derived from \citet{2014A&A...561A..95T} and the one derived in this study is 1.7, although there are ratios up to 4-5. The grain absorption cross section per unit mass that is used in \citet{2014A&A...561A..95T} is $\rm \kappa=0.04\,(\nu/250\,GHz)^{2}\,m^{2}kg^{-1} $, which corresponds to $\rm \kappa_{350\,\mu m}=4.7\,cm^{2}g^{-1}$, similar to the one used in \citet[][]{1990A&A...237..215D},  $\rm \kappa_{350\,\mu m}=4.8\,cm^{2}g^{-1}$. Therefore, the discrepancies cannot be attributed to different grain absorption cross sections.  

The differences shown in Fig.~\ref{fig:dustcomp} could be due to the use of a free $\beta$ coefficient and a $\kappa_{350\,\mu m}$ from a dust model that assumes $\beta$ close to 2. As it has been explained in \citet{2013A&A...552A..89B}, this procedure gives dust masses that depend on the wavelength chosen for normalisation, yielding differences of 50\% between normalisations at 250\,\mi\ and 500\,\mi. Finally, differences between MBB fitting and more sophisticated dust models have been reported previously: \citet{2012MNRAS.425..763G} performed a two-MBB fitting to the integrated SEDs of the KINGFISH galaxies with photometric data from 24\,\mi\  to 500\,\mi\ and found discrepancies up to 35\% between the dust masses from the MBB fitting and \citet{2007ApJ...657..810D} dust models. 

\subsection{Grain emission in different filters}\label{sec:grain_filters}

Before looking for trends between the grain abundances and other physical properties of the ISM it is interesting to estimate how much emission of the different grain types is contributing to the fluxes in the \Spi\ and \Her\ bands. In Table~\ref{tab:frac} we show the mean values of the fraction of the emission of each grain type to the total emission in the 8\,\mi, 24\,\mi\ filters from \Spi, and 70\,\mi, 100\,\mi\ and 160\,\mi\ filters from \Her. We can see, as expected, that the main contributors to the emission in the 8\,\mi\ filter are the PAHs, while the BGs contribute $\sim$90\% of the emission in the 100\,\mi\ and 160\,\mi\ filters. The emission in the 24\,\mi\ band is dominated by the VSGs ($\sim$60\%), but there is a significant contribution ($\sim$40\%) of the PAHs to the flux in this band. In the 70\,\mi\  filter,  $\sim$70\% of the emission is coming from the BGs, whereas the remaining 30\% are from VSGs. Thus, the emission of VSGs needs to be taken into account when including the 70\,\mi\ flux in the SED fit with MBBs, as not all the emission in 70\,\mi\ band comes from grains in thermodynamic equilibrium.

\begin{table*}
\begin{center}
\begin{tabular}{cccccc}
\hline
& 8\mi\ & 24\mi\ & 70\mi\ & 100\mi\ & 160\mi\  \\
\hline
PAH &  0.97$\pm$0.02 &   0.4$\pm$0.1  & $<$0.01 & $<$0.01 & $<$0.01 \\
VSG & 0.03$\pm$0.02 &  0.6$\pm$0.1 & 0.3$\pm$ 0.1 &  0.12$\pm$0.07  & 0.07$\pm$ 0.05  \\
BG & $<$0.01 & $<$0.01 & 0.7$\pm$0.1 &  0.88$\pm$0.07 & 0.93$\pm$0.05 \\  
\hline
\end{tabular}
\caption{Average relative fraction of emission corresponding to PAHs, VSGs, and BGs in the 8\,\mi\ and 24\,\mi\ bands from \Spi, and in the 70\,\mi, 100\,\mi, and 160\,\mi\  filters from \Her.} 
\label{tab:frac}
\end{center}
\end{table*}

\subsection{Grain Abundances}\label{sec:grain_abund}
\begin{figure*} 
\includegraphics[width=0.32\textwidth]{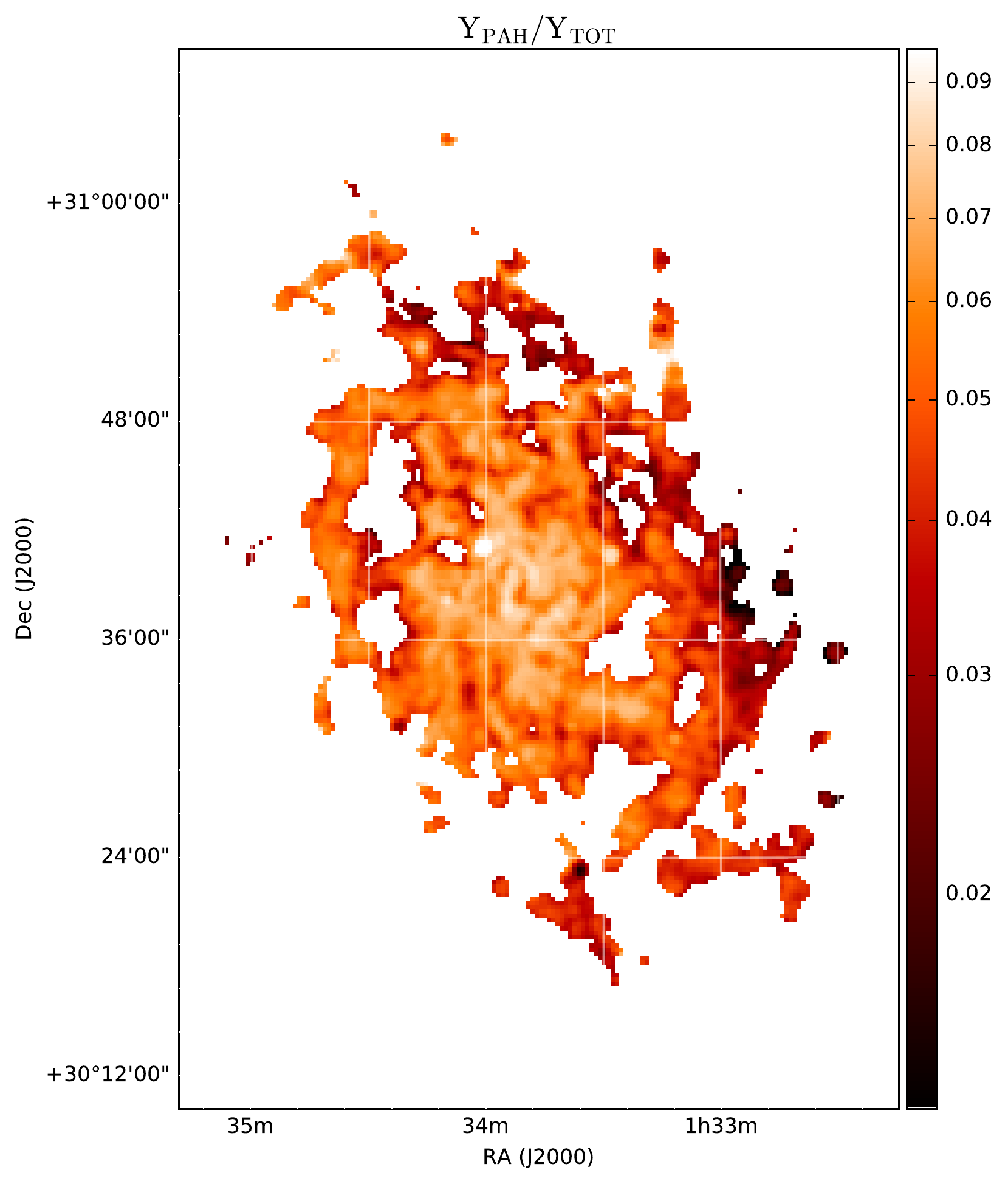}
\includegraphics[width=0.32\textwidth]{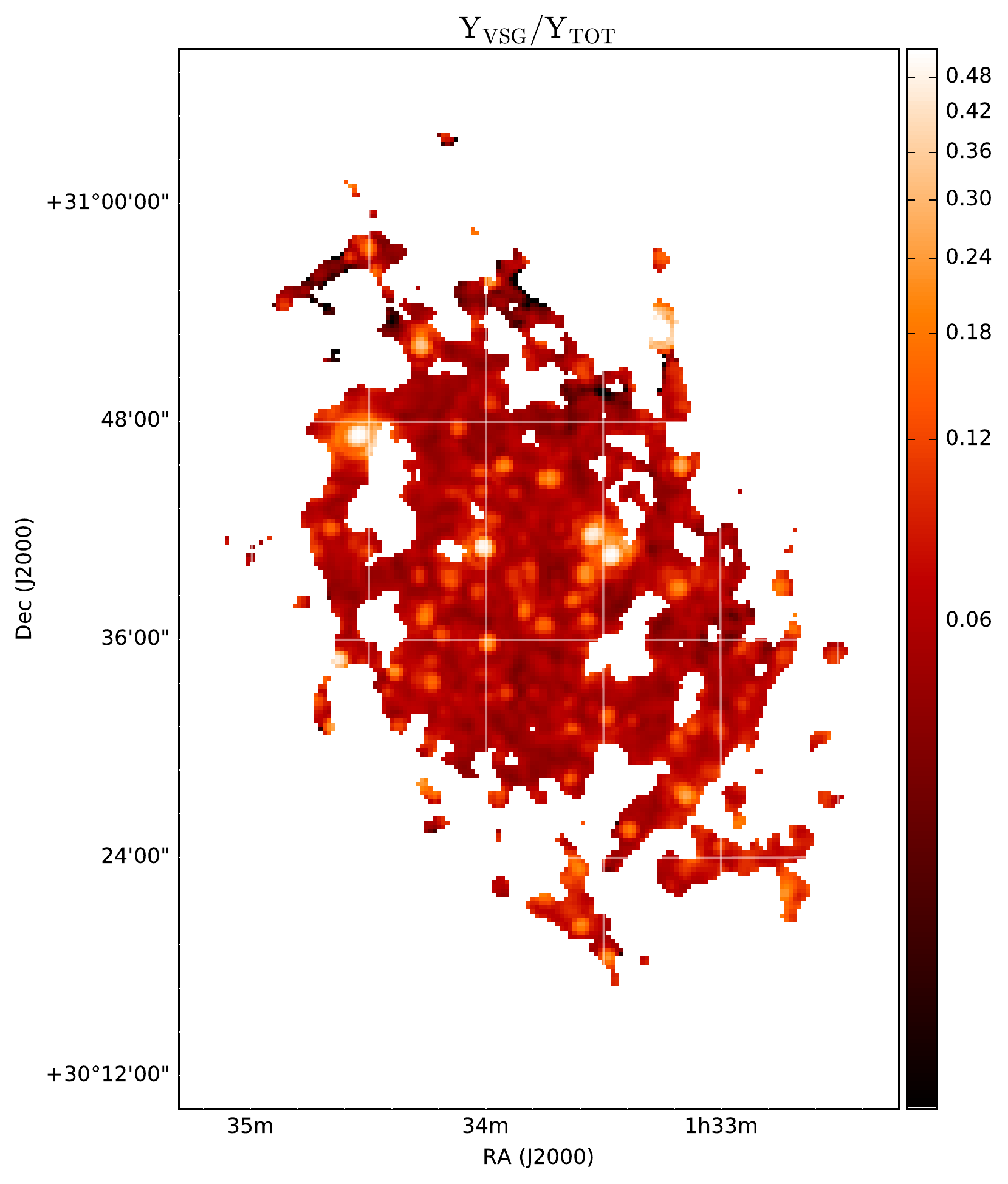}
\includegraphics[width=0.32\textwidth]{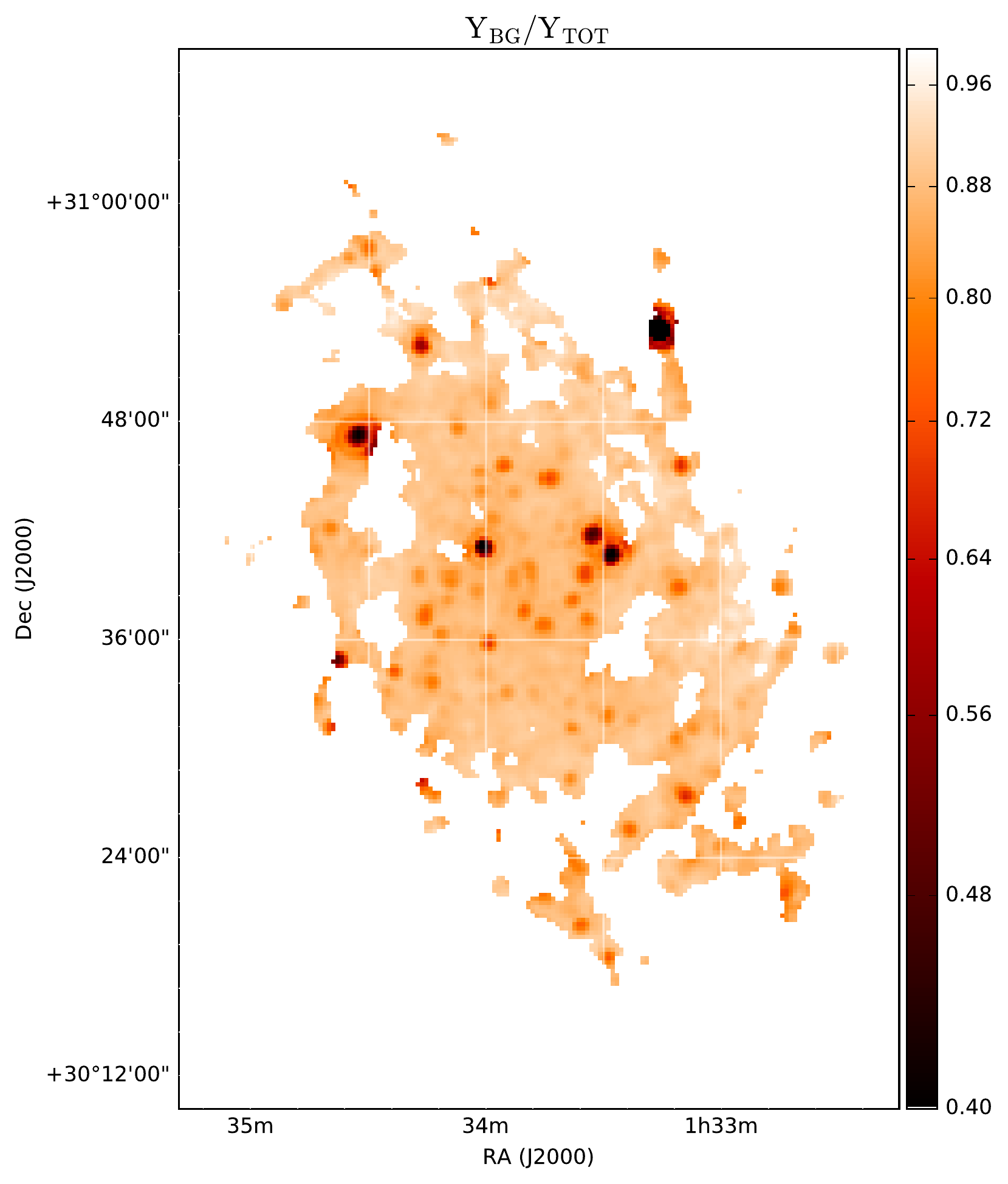}
   \caption{\ypah/\ytot\ (left), \yvsg/\ytot\ (middle), and \ybg/\ytot\ (right) maps for the disc of M\,33 derived using our fitting procedure.}
   \label{fig:grainfrac_map}
\end{figure*}

Our fitting technique gives the dust mass fraction relative to the hydrogen mass for each grain type in our model. With this information we can create maps of the dust abundance for each grain type for the whole surface in the disc of M\,33. In Fig.~\ref{fig:grainfrac_map} we show the maps for each grain type. In general, while the \yvsg/\ytot\  and  \ybg/\ytot\ maps present distributions similar to the one of the star-forming regions, the  \ypah/\ytot\ map shows a more diffuse distribution across the disc of the galaxy. 

\begin{figure*} 
\includegraphics[width=0.5\textwidth]{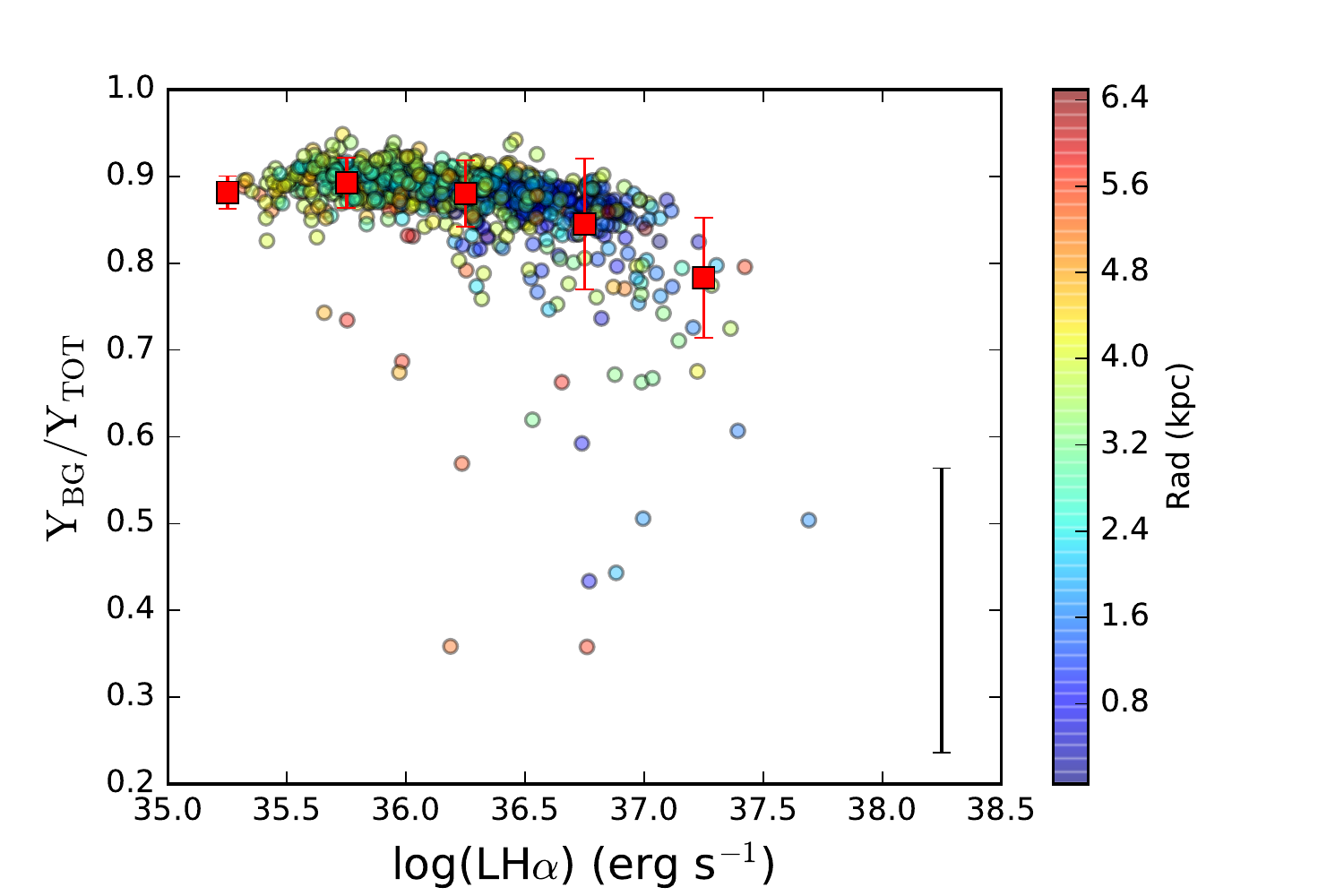}
\includegraphics[width=0.5\textwidth]{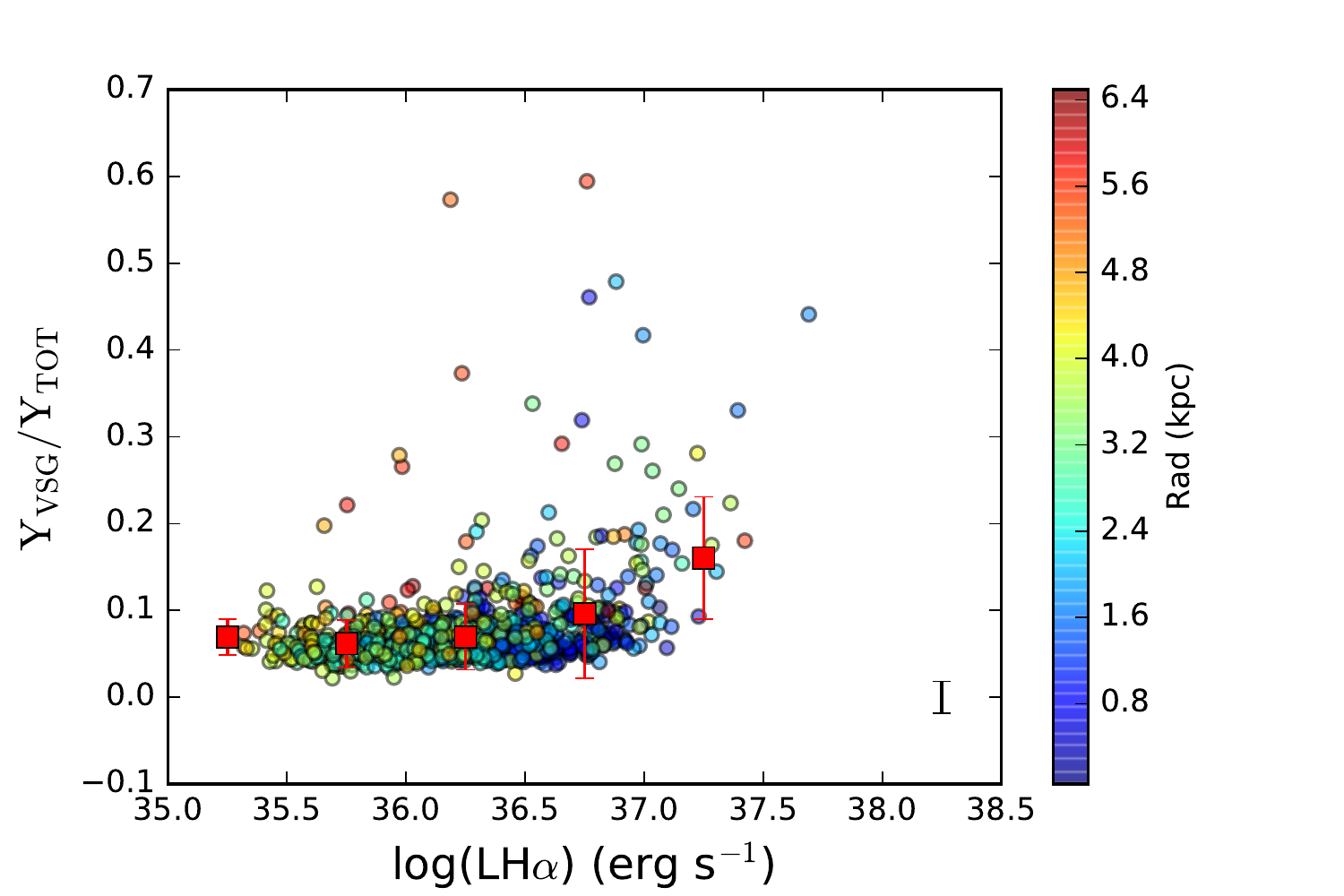}
\includegraphics[width=0.5\textwidth]{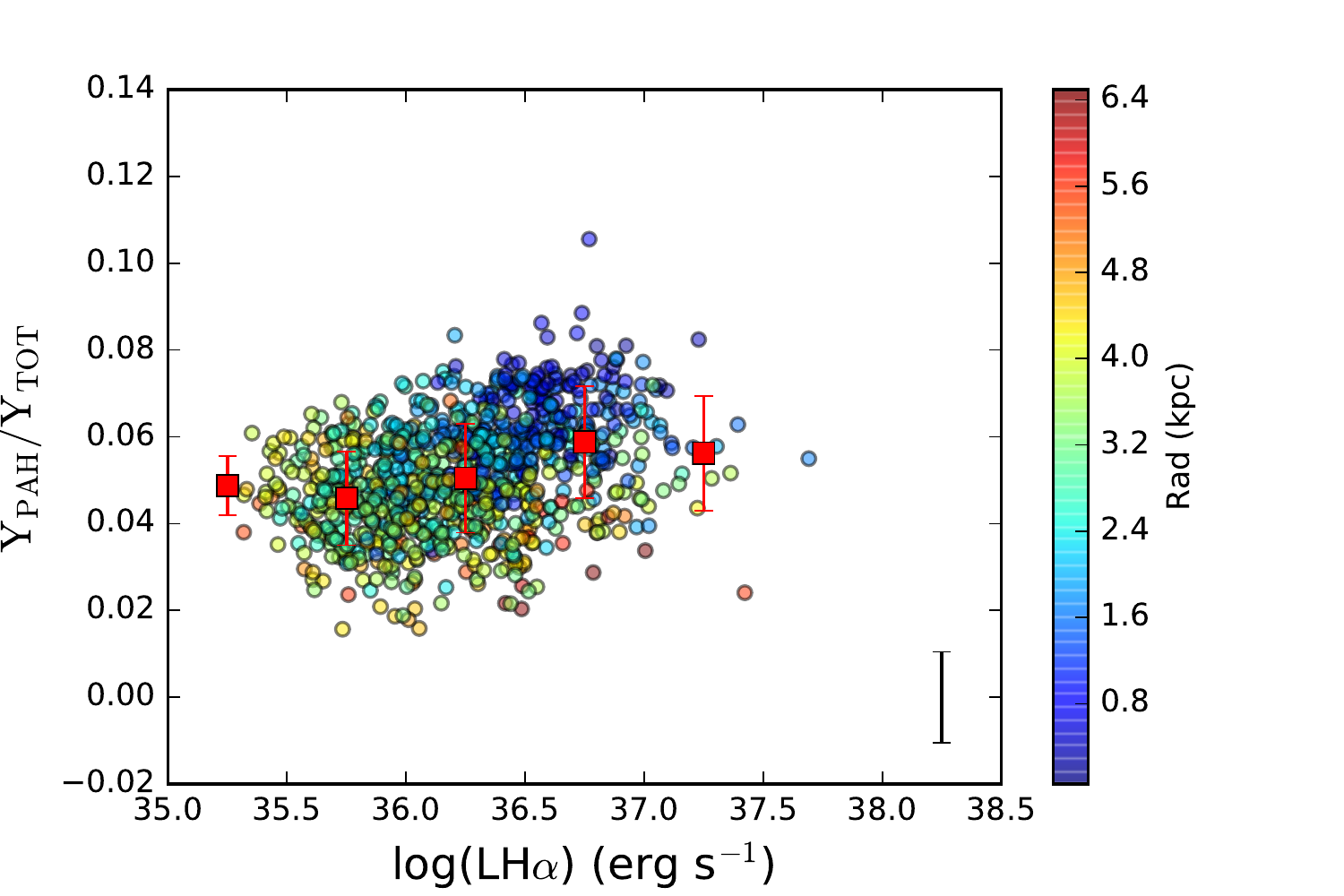}
\includegraphics[width=0.5\textwidth]{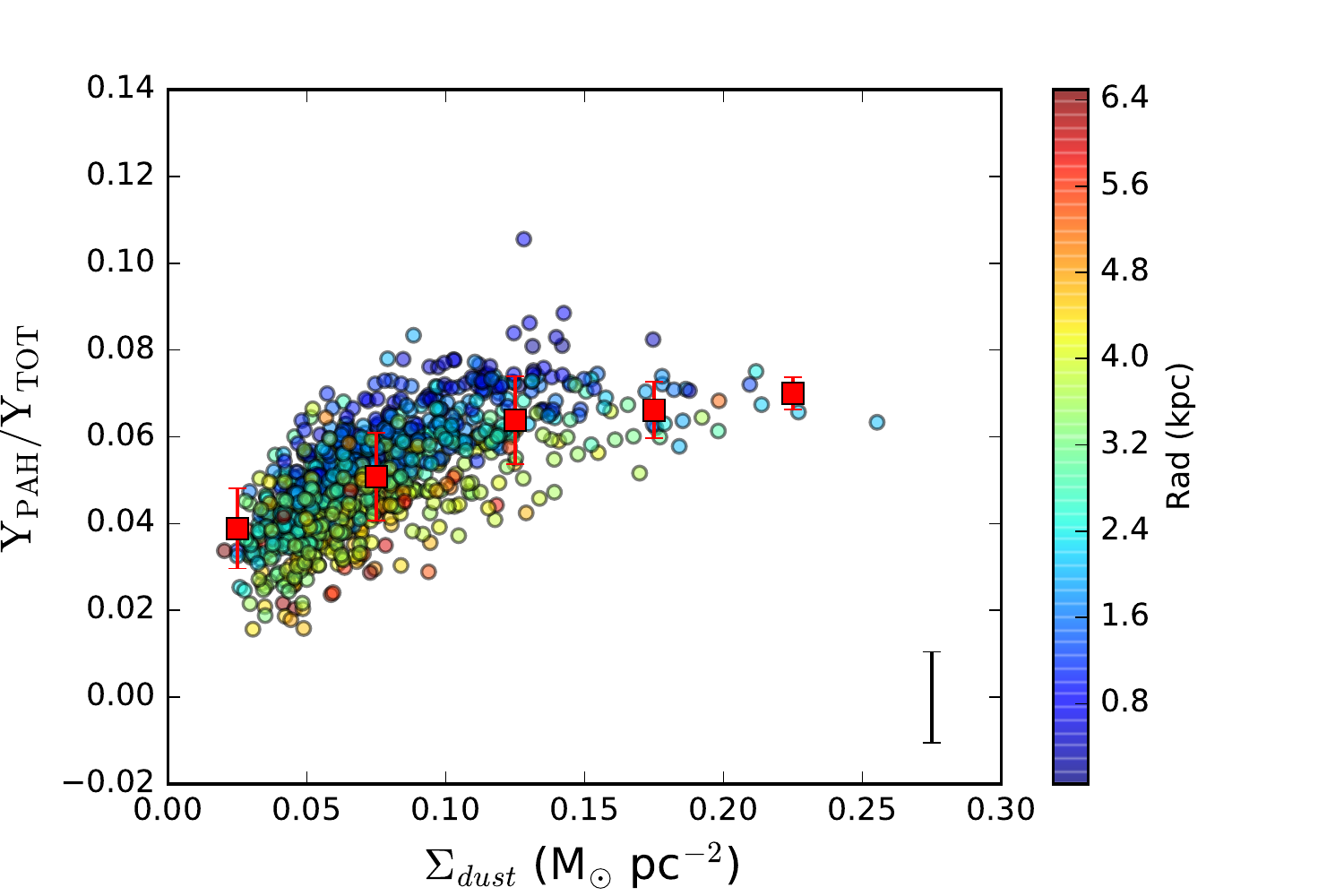}
   \caption{Top: \ybg/\ytot\ (left), \yvsg/\ytot\ (right) versus the logarithmic \ha\ luminosity for the individual pixels in the disc of M\,33 fitted with our code. Bottom: \ypah/\ytot\ versus logarithmic \ha\ luminosity (left) and dust surface density (right). The black dots and error bars correspond to the mean and standard deviation of the grain fractions for different bins in logarithmic \ha\ luminosity, which are overplotted to better show the trend between the quantities in both axis. The error bar corresponds to the median error: 0.04 for \yvsg/\ytot, 0.02 for \ypah/\ytot, and 0.32 \ybg/\ytot.}
   \label{fig:grainfrac_Ha}
\end{figure*}

In Fig.~\ref{fig:grainfrac_Ha} we show the fraction of BGs (top-left panel) and VSGs (top-right panel) versus the logarithmic \ha\ luminosity for the individual pixels of the disc of M\,33. Although both panels present a high dispersion we can see that \yvsg/\ytot\  tends to increase at locations of intense star formation, while  \ybg/\ytot\ follows the opposite correlation. There are some data points outside of the main trend showing higher values of \yvsg/\ytot, these correspond to the same pixels with low \ybg/\ytot. For these pixels, a single ISRF might not be enough to describe the radiation field heating the dust, and a combination of several ISRFs might describe better the dust temperature distribution at these locations \citep[see][]{2017A&A...601A..55C}. 

The high values of  \yvsg/\ytot\, and the corresponding low values of \ybg/\ytot\ at location where the \ha\ luminosity is high, could be explained by the reprocessing of the BGs fragmenting into VSGs. \citet{2016A&A...595A..43R} shows that for a sample of \hii\ regions in M\,33 catalogued in terms of morphology, the most intense regions tend to have a higher fraction of VSGs than the more diffuse shell-type \hii\ regions. The high velocity shocks occurring inside the most luminous \hii\ regions can destroy the BGs giving as consequence an increase of the VSG fraction. The results shown here agree with those presented in \citet{2016A&A...595A..43R} and are consistent with the framework of the dust evolution models in \citet{1994ApJ...433..797J,1996ApJ...469..740J}, suggesting dust grain destruction/fragmentation by interstellar shocks in the warm medium. Finally, the data points in the top-right panel with values of \yvsg/\ytot\ higher than 0.5 correspond to pixels in the outer parts of IC133,  an \hii\ complex with a very high emission at 24\,\mi. In these pixels the 22\,\mi\ and 24\,\mi\ emission from MIPS and WISE, respectively, is so high that the fluxes are comparable to the IR peak of the SED. 

Interestingly, we do not find any relation between the fraction of PAHs and the \ha\ luminosity, (bottom-left panel of Fig.~\ref{fig:grainfrac_Ha}), nor the UV luminosity and $\rm G_{0}$ (plots not shown here). The reason for the lack of relation is that the linear scales of the individual SEDs, $\sim$170\,pc, include the \hii\ region, Photodissociation Region, and molecular cloud, and thus, do not allow us to isolate the central parts of the \hii\ regions where the strong ISRF would destroy the PAHs. There is a slight relation between \ypah/\ytot\  and galactocentric radius (albeit with high dispersion), as it can be seen in the bottom-left panel of Fig.~\ref{fig:grainfrac_Ha}: \ypah/\ytot\ in the centre of the disc is around 0.08 and decreases to values of $\sim$0.03 at R$\geq$\,3\,kpc. This trend could be related to the metallicity \citep[e.g.][]{2008ApJ...678..804E}: the metallicity gradient of M\,33 is quite shallow \citep[0.045\,dex/$\rm R_{kpc}$,][]{2011ApJ...730..129B}, which is consistent with the small differences of  \ypah/\ytot\  between the centre and the outer parts of the galactic disc. \ypah/\ytot\ correlates with surface density of the dust (bottom-right panel of Fig.~\ref{fig:grainfrac_Ha}), but the correlation seems to be modulated by the radial distribution of the surface density of the dust. 

\begin{figure*} 
\includegraphics[width=0.5\textwidth]{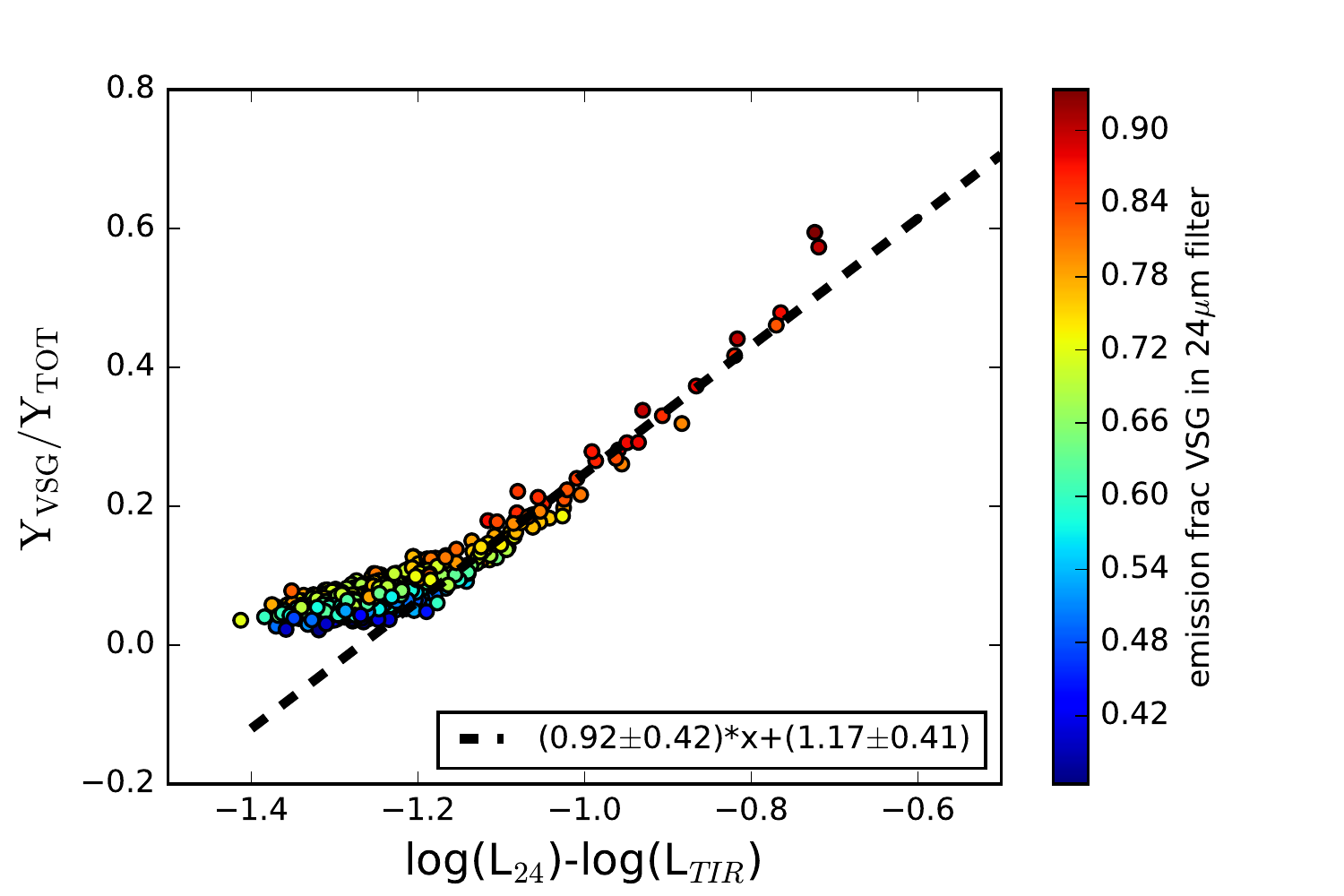}
\includegraphics[width=0.5\textwidth]{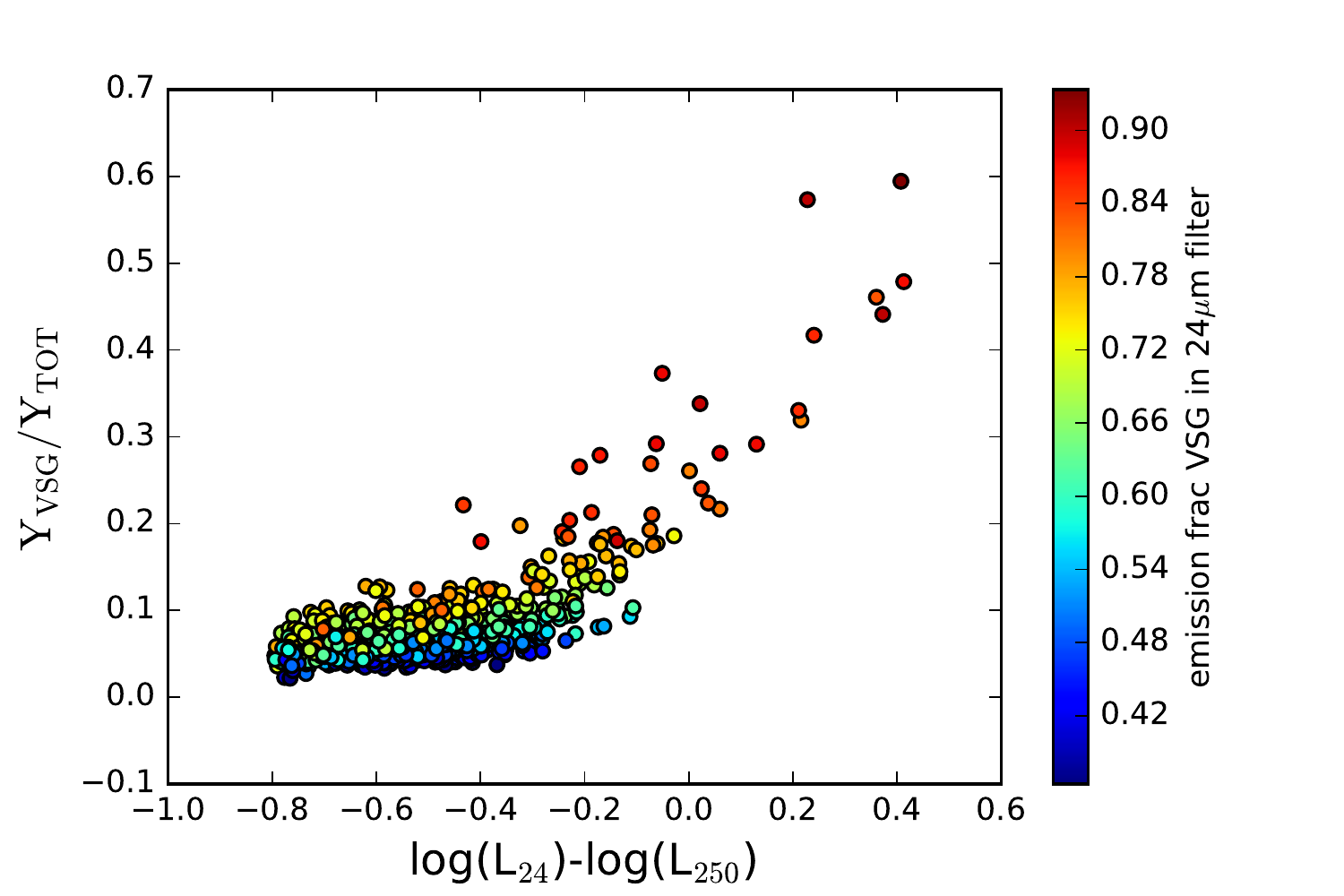}
   \caption{\yvsg/\ytot\ versus the ratio of 24\,\mi\ and the TIR luminosity (left) and the ratios of the 24\,\mi\ and 250\,\mi\ luminosities (right). The linear fit is performed for points with \yvsg/\ytot$>$0.2, or alternatively for log(L($_{24\,\mu m}$)/L($_{TIR}$))$>$-1.08. The color code corresponds to the emission fraction of the VSGs contributing to the 24\,\mi\ filter.}
   \label{fig:grainfrac_cal}
\end{figure*}

Finally, we have investigated how well the ratio between the 24\,\mi\ and the total-IR (TIR) luminosity (and between  24\,\mi\ and 250\,\mi\ luminosities) traces the abundance of VSGs. In the left-hand panel of Fig.~\ref{fig:grainfrac_cal}, we show \yvsg/\ytot\  versus log(L($_{24}$)/L($_{TIR}$)). For \yvsg/\ytot$>$0.2 (log(L($_{24}$)/L($_{TIR}$))$>$-1.08) there is a tight correlation between the fraction of VSGs and log(L($_{24}$)/L($_{TIR}$)). The linear fit to the correlation is \yvsg/\ytot$\rm =(1.036\pm0.003)\times log(L(_{24})/L(_{TIR})) + (1.282\pm0.003)$. The correlation between \yvsg/\ytot\ and log(L($_{24}$)/L($_{250}$)) (right-hand panel of  Fig.~\ref{fig:grainfrac_cal})  is not so tight. Finally, it is interesting that we found no correlation between   \ypah/\ytot\ and the ratio of L($_{8}$) and L($_{TIR}$).

\section{Gas-to-dust mass ratio}\label{sec:GtD}
\begin{figure} 
\includegraphics[width=0.49\textwidth]{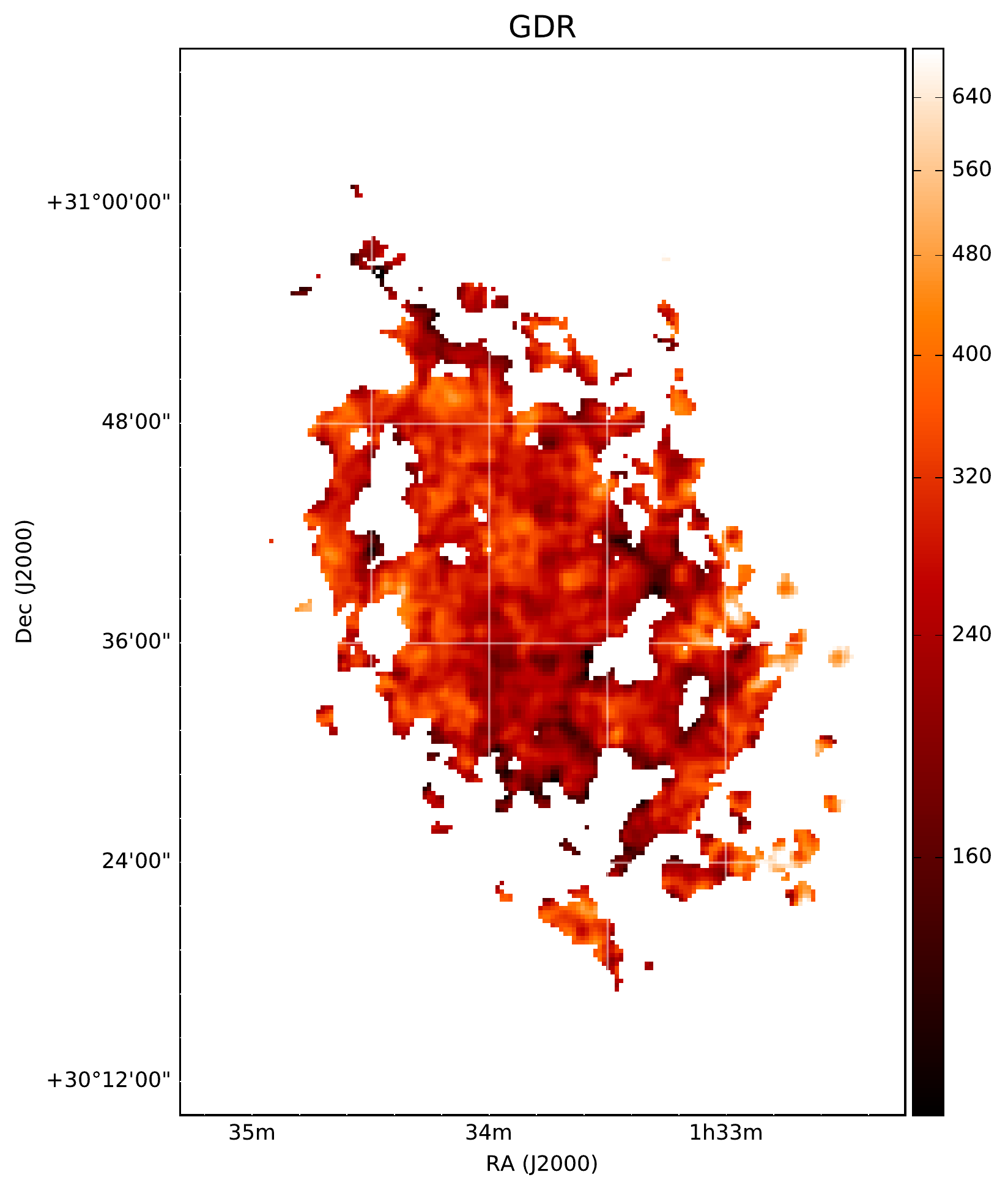}
   \caption{GDR map of the whole disc of M\,33 obtained from the dust masses derived with our fitting procedure and total gas masses from the HI and CO observations of \citet{2010A&A...522A...3G} and \citet{2014A&A...567A.118D}, respectively. }
   \label{fig:GtD_map}
\end{figure}

The dust mass derived in the previous section can be combined with the gas mass obtained from observations to estimate the GDR over the whole face of M\,33. The gas mass was obtained using the $^{12}$CO(J=2--1) and \hi\ intensity maps presented in \citet{2014A&A...567A.118D} and \citet{2010A&A...522A...3G}, respectively. For the neutral hydrogen gas mass we used a conversion factor of $\rm X(H\,\textsc{i})=1.8\times10^{18}\,cm^{-2}/(K\,km\,s^{-1})$, as in \citet{2010A&A...522A...3G}. The molecular gas mass  was obtained using a  CO-to-$\rm H_{2}$ conversion factor of $X(\rm CO)=\frac{N_{H_{2}}}{I_{CO(1-0)}}=4.0\times10^{20}\,cm^{-2}/(K\, km\,s^{-1}$), consistent with the metallicity of M\,33. A constant $\rm I_{CO(2-1)}/I_{CO(1-0)}=0.8$ was assumed following \citet{2014A&A...567A.118D}.  Both masses include a factor of 1.37 to account for the helium contribution. The total gas mass is the sum of the neutral \hi\ and the molecular $\rm H_{2}$ mass ($\rm M_{tot}=M_{HI}+M_{H_{2}}$). In Fig.~\ref{fig:GtD_map} we show the GDR map for the whole disc of M\,33. The map seems quite smooth with no significant structures across the disc. 

\subsection{Radial trend}

\begin{figure} 
  \includegraphics[width=0.49\textwidth]{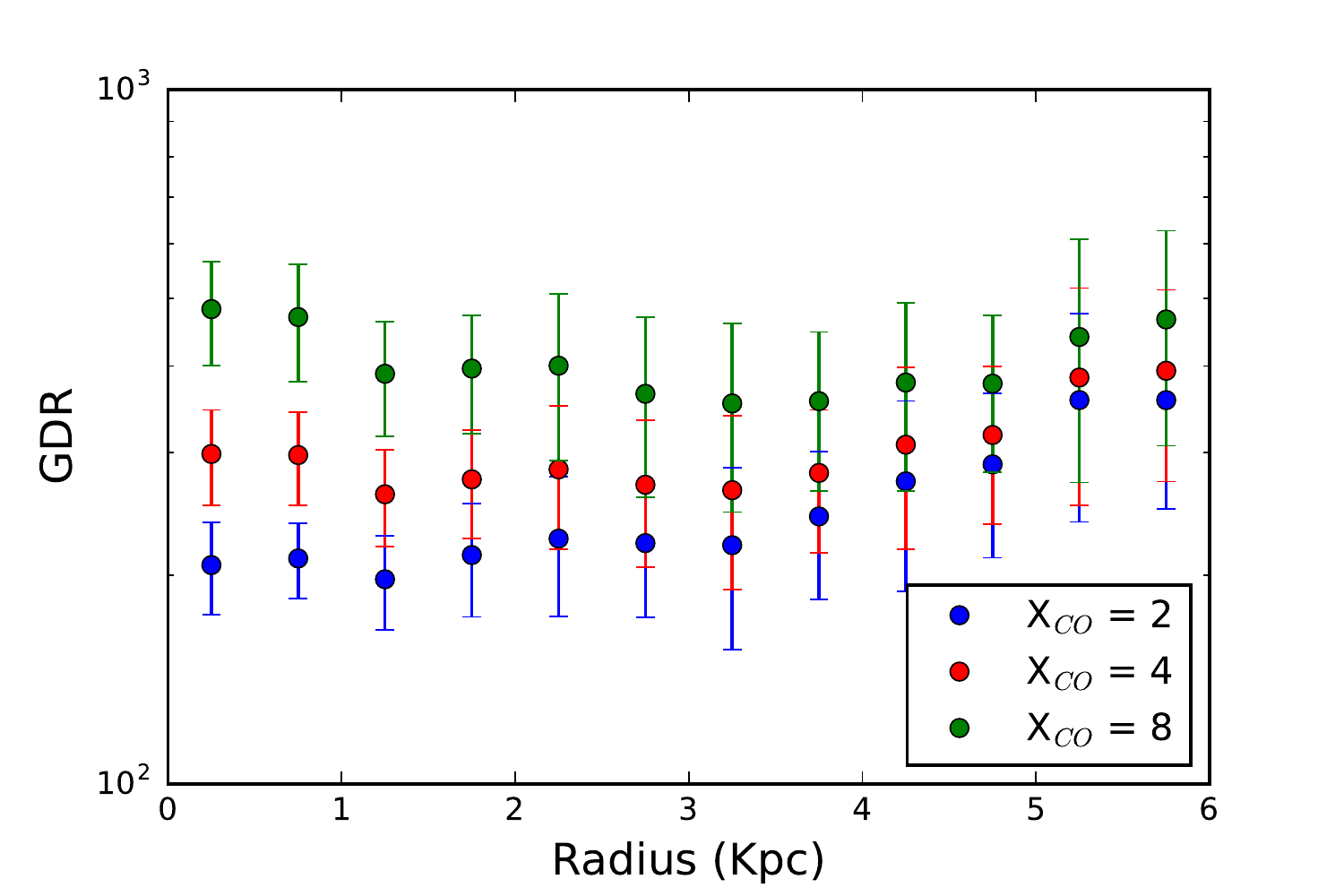}   
   \caption{Elliptical profiles of the GDR for M\,33 for different values of $X(\rm CO)$. The differences in the profiles are more notable at R$<$\,4\,kpc, where the molecular gas mass fraction is high. At  R$>$\,4\,kpc, the molecular hydrogen surface density, $\Sigma_{\rm H_{2}}$, is considerable smaller than the atomic hydrogen surface density, $\Sigma_{\rm HI}$ \citep{2014A&A...567A.118D,2007A&A...473...91G}.}
   \label{fig:GtDprof}
\end{figure}

In Fig.~\ref{fig:GtDprof} we show the radial profile of the GDR for different values of $X(\rm CO)$. In general the profiles are quite constant up to R$\sim$4\,kpc, where they start to slightly increase. The value of $X(\rm CO)=4.0\times10^{20}\,cm^{-2}/(K\, km\,s^{-1})$ assumed in this paper to estimate the molecular gas mass gives a GDR of $\sim$300 for the inner 4~kpc. This value is slightly higher than previous studies. \citet{2010A&A...518L..67K} derived a GDR of 200 for the entire galaxy with variations between 120 and 200 at different annuli. Using radiation transfer models to fit the whole SED of M\,33, \citet{2016A&A...590A..56H} derived a GDR of $\sim$100, lower than what would be expected from the metallicity of M\,33. These authors argue that dust properties different from those assumed in their radiation transfer model are the main explanation for the low values of the GDR. In a novel approach \citet{2011ApJ...737...12L} solved simultaneously the conversion factor $\alpha_{\rm CO}$\footnote{$\alpha_{\rm CO}$ is the conversion factor from integrated CO luminosity to molecular gas mass. It scales linearly with $X(\rm CO)$, which is the conversion factor from integrated CO luminosity to column density of $\rm H_{2}$.} and the GDR, assuming a constant value of GDR at local scales larger than typical individual molecular clouds. These authors found values for the GDR in the range of 100-200, lower than expected for the metallicity of M\,33. The dust mass in  \citet{2011ApJ...737...12L} was derived using the \citet{2007ApJ...657..810D} dust model. The dust masses derived from this model are on average a factor of 2 too high \citep{2016A&A...586A.132P}, thus accounting for a dust mass lower by a factor of 2 would bring the GDR to 200-400, closer to the expected value for the metallicity of M\,33. A similar method as applied in \citet{2011ApJ...737...12L}, but accounting for a CO-dark gas component, was presented in \citet{2017A&A...600A..27G}. These authors find values of 250 to 400 for the GDR in the whole disc of M\,33: in the inner R$<$3~kpc the GDR has a constant value of $\sim$250, at R$>$3~kpc the GDR starts to increase up to 400 at R$\sim$5~kpc.  Assuming $\beta$\,=\,2 they found constant values for the GDR ratio within the disc that are lower and do not increase with radius as much as when a variable $\beta$ is adopted. The radial trend observed in Fig.~\ref{fig:GtDprof} is in agreeement with the trend found in \citet{2017A&A...600A..27G}. 

\begin{figure} 
  \includegraphics[width=0.49\textwidth]{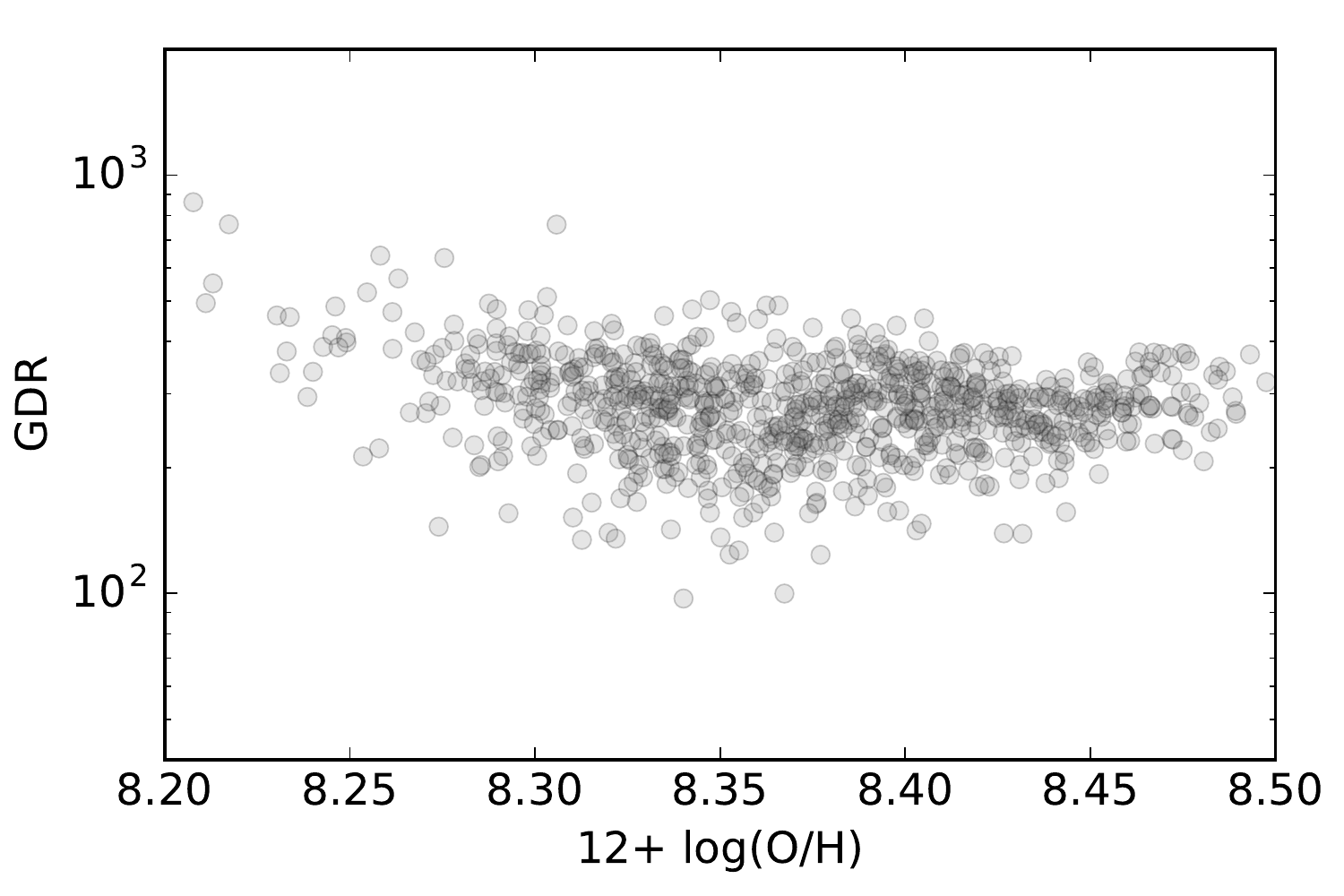}   
   \caption{GDR as a function of oxygen abundance for the disc of M\,33. The abundances have been derived using the radial gradient obtained by \citet{2011ApJ...730..129B}.}
   \label{fig:GtDmeta}
\end{figure}

The GDR radial profile increases slightly in the outer parts (R\,$\gtrsim$\,4\,kpc) of the galaxy. This mild increase is consistent with the shallow oxygen abundance gradient derived in \citet{2011ApJ...730..129B}:  oxygen abundances vary from 12+log(O/H)=8.50 in the inner part of the galaxy to 12+log(O/H)=8.14 at 8\,kpc. Based on the radial gradient given in \citet{2011ApJ...730..129B}, we assign a metallicity to each pixel in the M\,33 disc and present the relation between GDR and oxygen abundance for the disc of M\,33 in Fig.~\ref{fig:GtDmeta}. The GDR tends to increase for 12+log(O/H)\,$\lesssim$\,8.32, which is the oxygen abundance at  R\,$\sim$\,4\,kpc. As a comparison with M31, \cite{2012ApJ...756...40S} obtained a well defined radial increase of GDR that is consistent with the radial oxygen abundance gradient of M31 derived by \citet{1999AJ....118.2775G}, but steeper than the abundance gradient derived more recently using electron temperature determinations for this galaxy \citep[][]{2012MNRAS.427.1463Z}. For M33, we find that the relation between the GDR and 12+ log(O/H) at local scales agrees with the general relation found for nearby galaxies by \citet{2014A&A...563A..31R}.

\subsection{Dependence with other parameters}\label{sec:GtD:param}

There are recent studies which claim a variation of the GDR with properties other than the metallicity of the ISM \citep{2011A&A...536A..88G,2014ApJ...797...86R,2017ApJ...841...72R}. In particular, \citet{2011A&A...536A..88G} found a correlation between the GDR and the mean value of the radiation field intensity ($<U>$) at local scales of 54\,pc in the LMC (see Fig.~13 in that paper). These authors proposed two explanations for the correlation: at locations with a weak radiation field, the GDR can be underestimated by the presence of undetected gas, while at the positions with an intense radiation field, shocks might destroy the grains and thus the dust mass derived by their models can be underestimated \citep[see Fig.~B.2 in][]{2011A&A...536A..88G}, producing an artificial correlation between the GDR and $<U>$. 

\begin{figure} 
\includegraphics[width=0.45\textwidth]{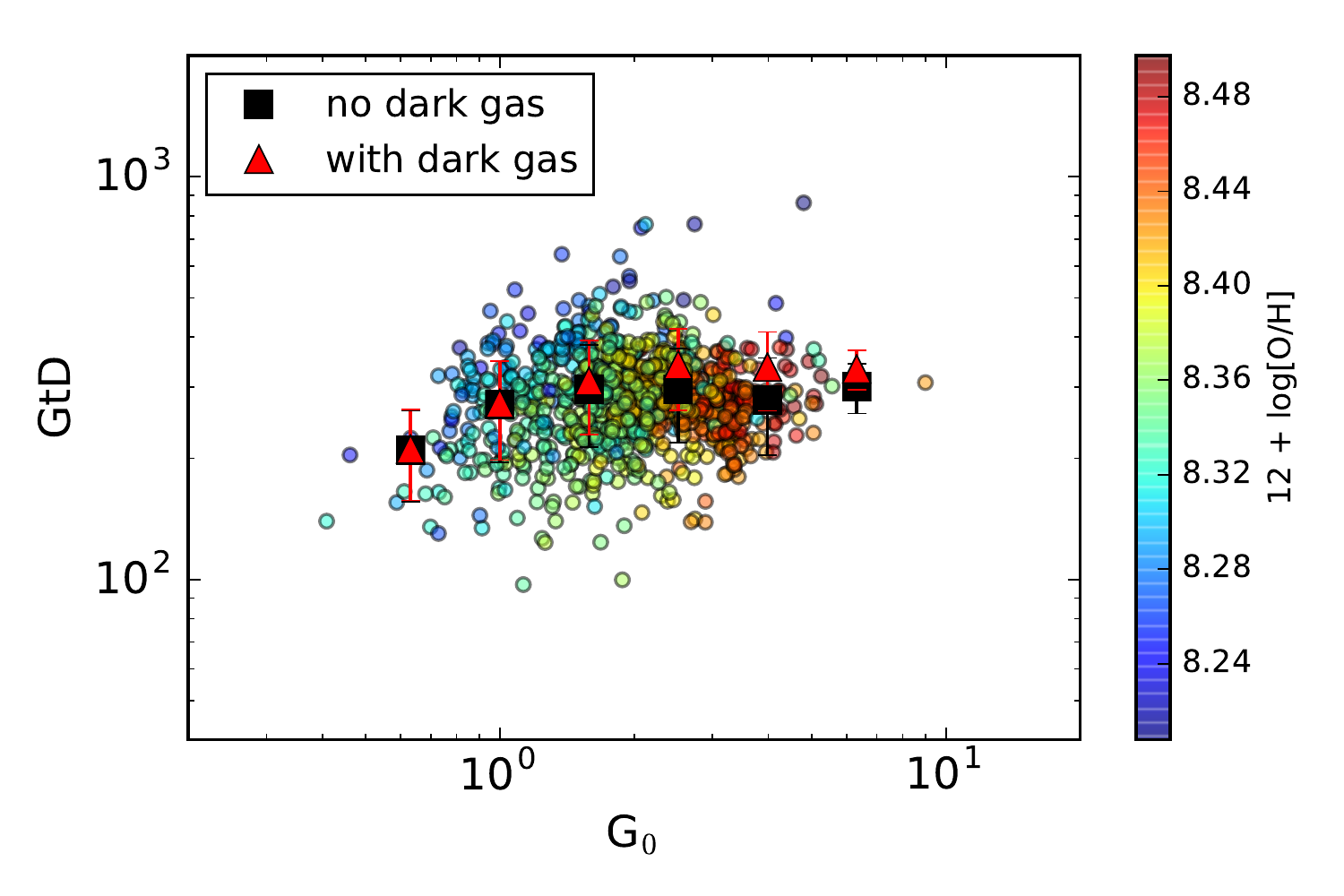}
\includegraphics[width=0.45\textwidth]{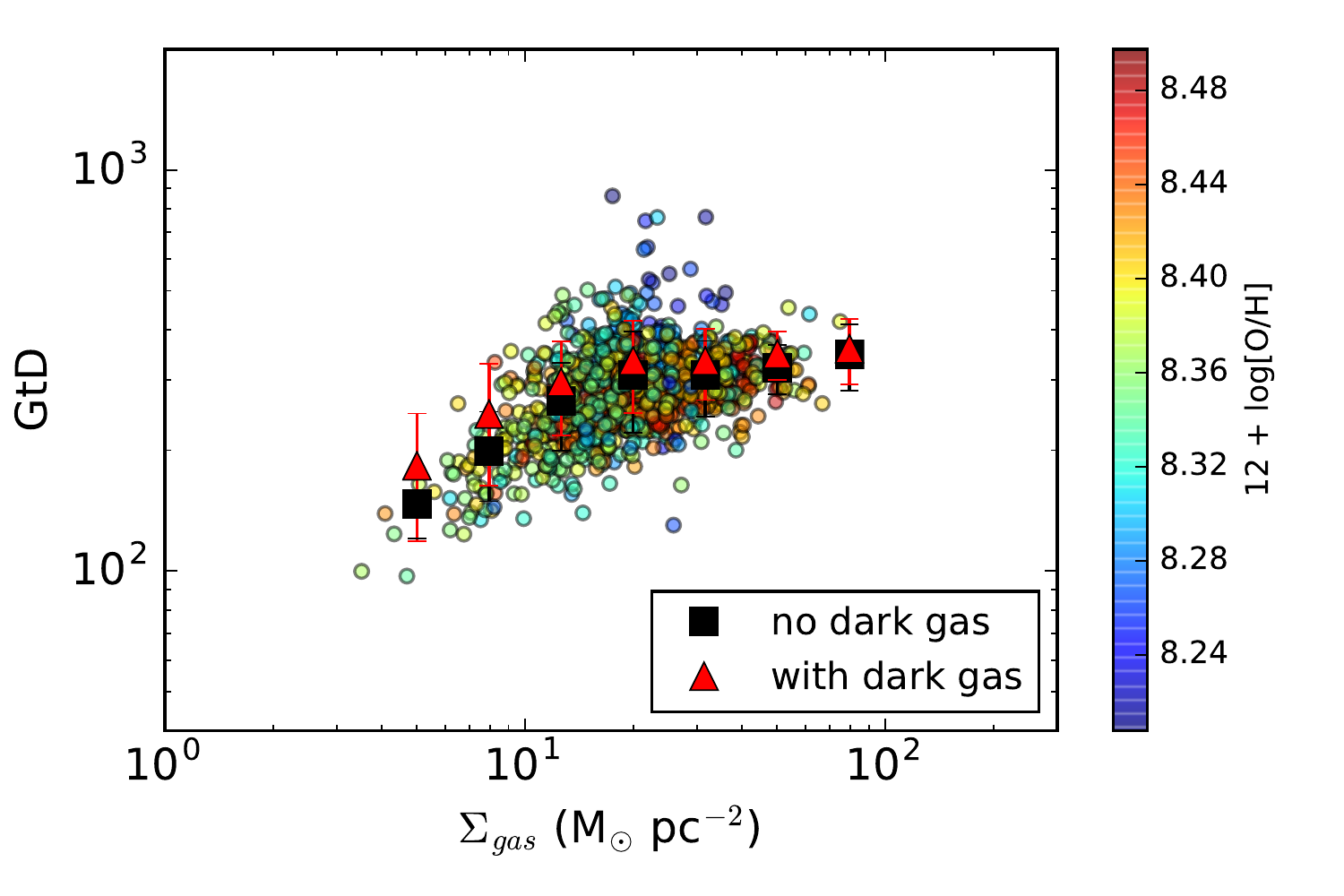}
\includegraphics[width=0.45\textwidth]{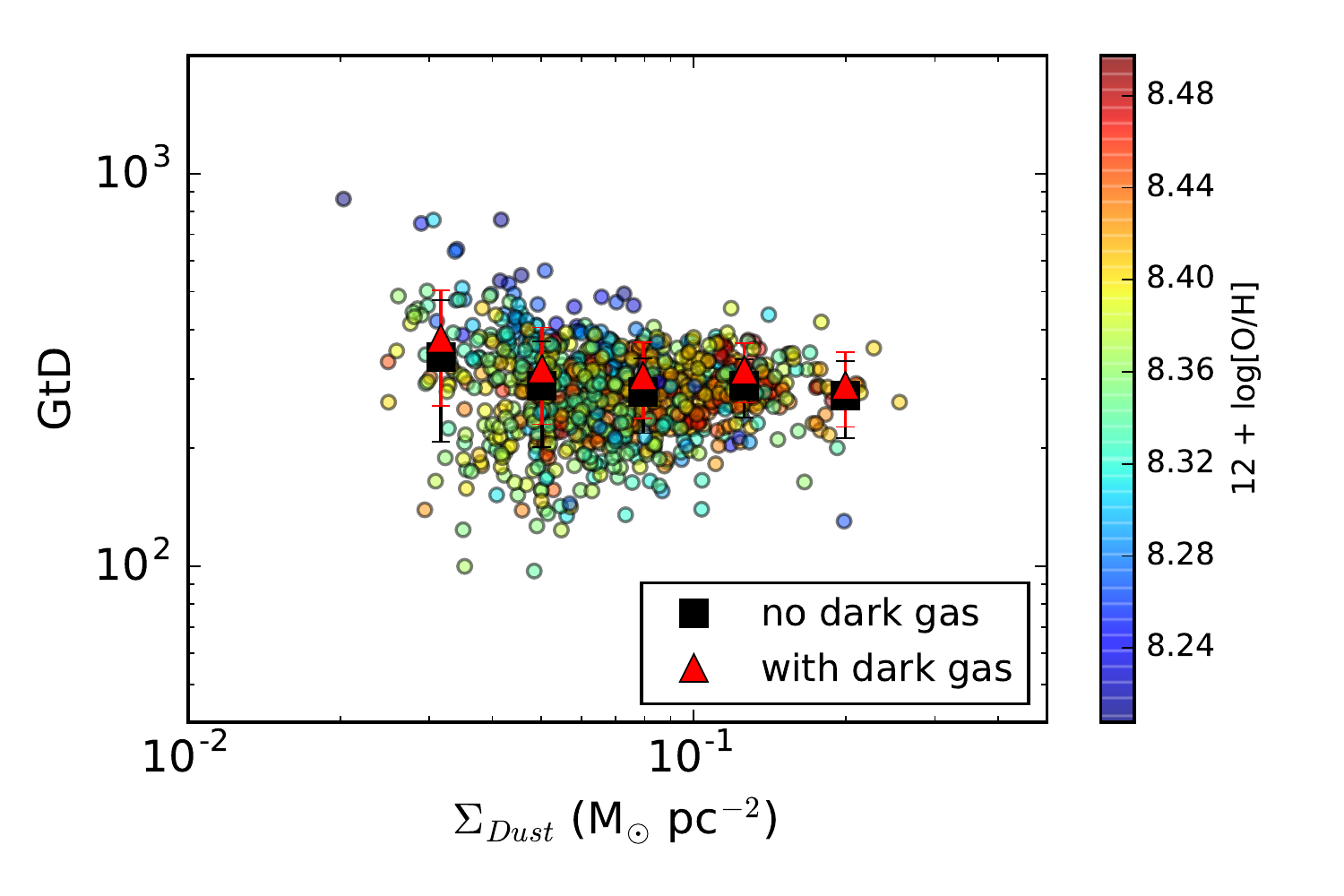}
   \caption{GDR versus $\rm G_{0}$ (top), $\Sigma_{gas}$ (middle), and $\Sigma_{dust}$ (bottom). We assumed a constant value of $X(\rm CO)=4.0\times10^{20}\,cm^{-2}/(K\, km\,s^{-1})$ to derive the gas mass in the disc of M\,33. We colour-code the data with the oxygen abundance of each point obtained using the radial metallicity gradient from \citet{2011ApJ...730..129B}.}
   \label{fig:GtD_G0}
\end{figure}

In Fig.~\ref{fig:GtDprof} we see that the GDR for M\,33 is relatively constant within a radius of $\sim$4\,kpc, which agrees with the results from \citet{2017A&A...600A..27G}. In this section we investigate if the GDR varies with other properties of the ISM. In Fig.~\ref{fig:GtD_G0} we investigate the relation with $\rm G_{0}$, the dust and gas surface densities. We show in top-panel of Fig.~\ref{fig:GtD_G0} that at the spatial scales of this study ($\sim$170\,pc) the GDR is relatively constant with $\rm G_{0}$, with some hints of a decrease for low values of  $\rm G_{0}$.  This decrease of the GDR has also been seen by \citet{2011A&A...536A..88G} for the LMC, but these authors found a stronger variation than the one presented in the top-panel of Fig.~\ref{fig:GtD_G0}. 
We investigate if at low $\rm G_{0}$, the GDR can be underestimated by the presence of CO-dark gas as suggested by \citet{2011A&A...536A..88G}. We make use of the radial dependence of the dark gas fraction obtained by \citet{2017A&A...600A..27G} and created a new GDR map where the gas surface density is the combination of the gas surface density derived from CO observations and the fraction related to dark gas. The red triangles in the top panel of Fig.~\ref{fig:GtD_G0} represent the mean values of the GDR in bins of $\rm G_{0}$ when the dark gas fraction has been considered. We can see that the dark gas fraction does not change the values of GDR at low  $\rm G_{0}$, in disagreement with the idea proposed by \citet{2011A&A...536A..88G} to explain the GDR and  $\rm G_{0}$ relation. 

In the bottom panel of Fig.~\ref{fig:GtD_G0} we present the relation between the GDR and $\Sigma_{\rm dust}$. There is slight trend of increasing GDR at low $\Sigma_{\rm dust}$, while for  $\Sigma_{\rm dust}\gtrsim$0.05 \msun/pc$^{2}$ the GDR is basically constant. Higher GDR values at low  $\Sigma_{\rm dust}$ was also found by \citet{2011A&A...536A..88G} and \citet{2014ApJ...797...86R} for the LMC with a stronger trend than the one presented here. The relation between the GDR and $\Sigma_{\rm gas}$ is presented in the middle panel of Fig.~\ref{fig:GtD_G0}. Lower values of GDR are found at low $\Sigma_{\rm gas}$, while at $\Sigma_{\rm gas}\gtrsim$20 \msun/pc$^{2}$, the GDR is constant. \citet{2011A&A...536A..88G} also found a constant GDR for high values of $\Sigma_{\rm gas}$ and a hint of a decrease of GDR at low $\Sigma_{\rm gas}$. However, their data show a too high dispersion at low $\Sigma_{\rm gas}$ to establish a clear trend. We find agreement with the trends for the molecular data points of \citet{2017ApJ...841...72R}, which seems plausible as the gas in M\,33 is molecular dominated.

The trends presented here show that the GDR might vary with the physical conditions of the ISM. We also note here that we do not find any significant correlation in the three panels of Fig.~\ref{fig:GtD_G0} with the metallicity of each pixel (colour-code in the plots) and therefore we can conclude that the trends observed in Fig.~\ref{fig:GtD_G0} cannot be explained by the relation between the GDR and the metallicity in each point of the disc of M\,33. Dark gas does not seem to explain the trends shown here. Therefore, reprocessing of dust, as \citet{2011A&A...536A..88G} suggested, might play a role in the variations. Further investigation of the parameters that drive the variations in the GDR requires a larger dynamic range in GDR than the one M\,33 has.

\section{$\rm G_{0}$ as a tracer of the star formation rate}\label{sec:G0_SFR}

$\rm G_{0}$ is the scale factor of the integral between 6 to 13.6 eV of the ISRF of the solar neighbourhood given by \citet{Mathis:1983p593}, which is the one used in our models. Therefore, $\rm G_{0}$ represents the energy density heating the dust in our models and the strength of the ISRF. Since the dust is heated by stars in the ISM,  $\rm G_{0}$ should be related in principle to the amount of star formation. Despite the fact that some radiation from the stars could leak out of the regions where the dust is heated, it is interesting to investigate the relation between $\rm G_{0}$ and the SFR.

It is true that the concept of SFR at small spatial scales may not be valid for the following main reasons \citep{2012ARA&A..50..531K}: incomplete sampling of the Initial Mass Function for SFR below $\sim$0.01\,\msun\,yr$^{-1}$, the breakdown of the assumption of continuous star formation implicit in the SFR calibrations, and larger spatial extension than the considered spatial scale of the emission that is used to calibrate the SFR.  Keeping these warnings in mind, we look for a trend between the strength of the ISRF, $\rm G_{0}$, and the SFR estimations at the local scales of $\sim$170\,pc.  Although at these small spatial scales, \citet{2015A&A...578A...8B} warns about the concept of SFR not being very suitable for the reasons we have presented above, they show that at local scales hybrid SFR estimators (those based on more than a single luminosity) reproduce better the SFR than the monochromatic SFR estimators. Therefore, we make use of the combination of \ha\ and 24\,\mi\ luminosities as well as the far-UV (FUV) and TIR luminosities to trace the SFR at these scales. 

\begin{figure*} 
\includegraphics[width=0.5\textwidth]{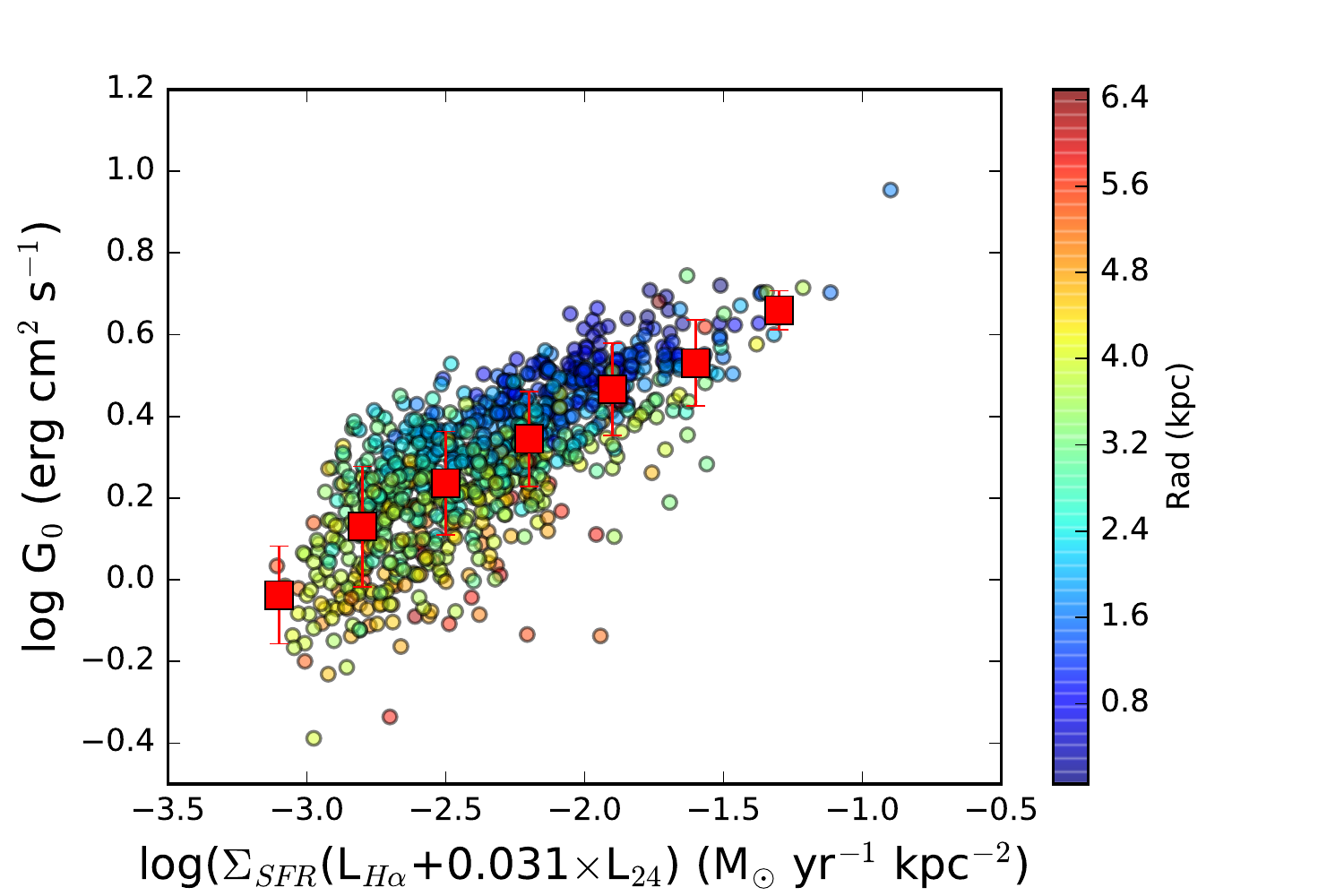}
\includegraphics[width=0.5\textwidth]{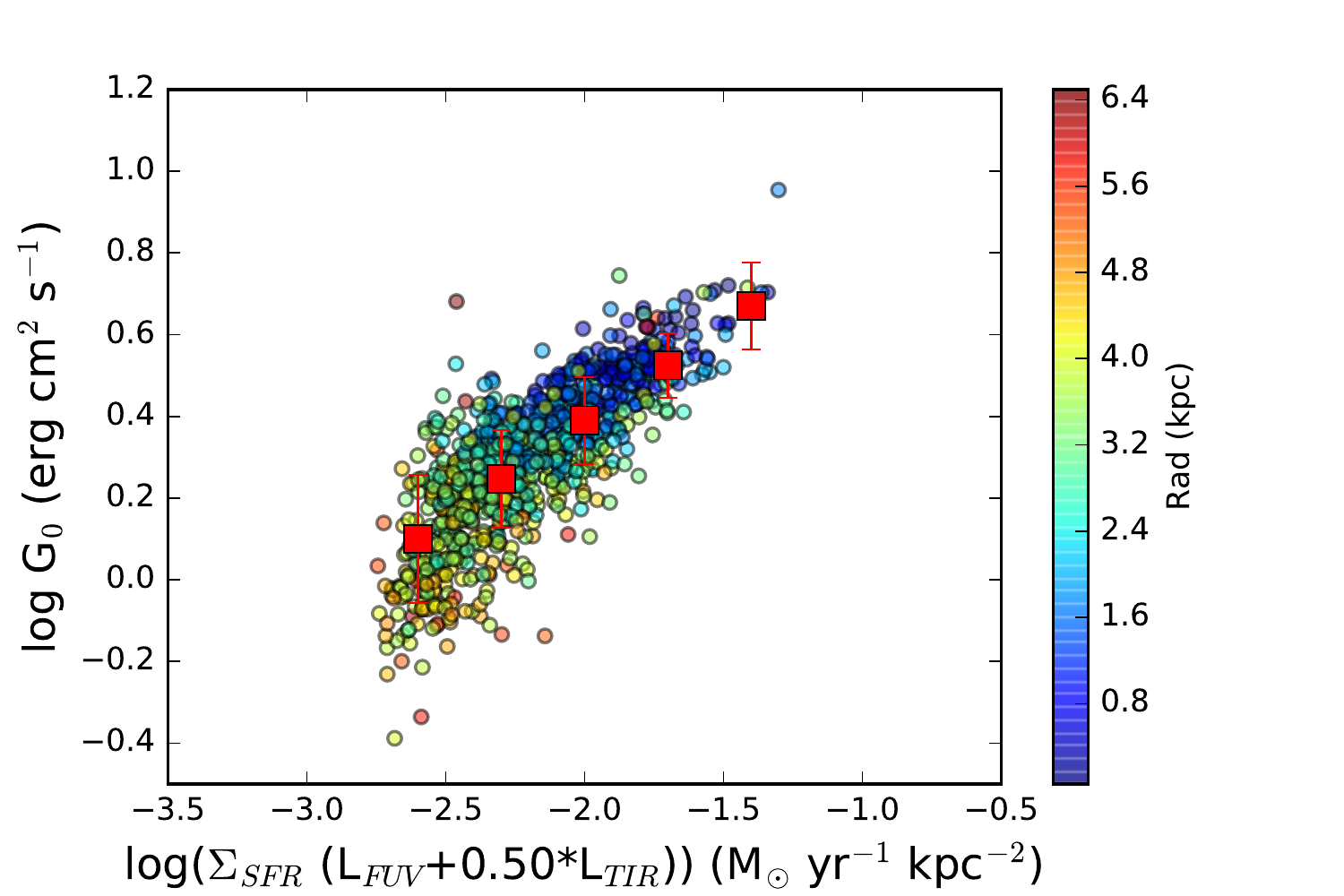}
   \caption{$\rm G_{0}$ versus $\Sigma_{SFR}$ derived from two different combinations of luminosities: \ha\  and 24\mi\ luminosities (left panel), and FUV and TIR luminosities (right panel). The color bar corresponds to galactocentric distances in kpc. The coefficients of the linear combinations are taken from  \citet{2007ApJ...666..870C}  and \citet{2012ApJ...761...97M} for \ha\ plus 24\mi\ luminosities and FUV plus TIR luminosities, respectively.}
   \label{fig:G0_SFR}
\end{figure*}

In the left panel Fig.~\ref{fig:G0_SFR} we show the correlation between $\rm G_{0}$ and $\Sigma_{SFR}$ derived from \ha\ and 24\,\mi\ luminosities following the calibration of 
\citet{2007ApJ...666..870C} which is obtained for scales between $\sim$\,30\,pc to $\sim$\,1.2\,kpc, and is therefore valid for the spatial scales considered here ($\sim$\,170\,pc). In the right panel we show the correlation when the SFR is derived using the combination of FUV and TIR luminosities from \citet{2012ApJ...761...97M}, derived for a star-forming complexes of $\sim$1\,kpc size. We have checked that this calibration is consistent with the one presented in  \citet{2016A&A...591A...6B}, who parameterised the linear coefficient for the combination of FUV and TIR luminosities with FUV-3.6 colours. It is interesting that the correlation holds for all values of  $\Sigma_{SFR}$, not only at the location of the star forming regions corresponding to pixels with high SFR but also for pixels in the more diffuse ISM. 

The SFR distribution derived from the \ha\ and  24\,\mi\ luminosities seems to extend to lower values than the SFR distribution derived from the FUV and TIR luminosities. The reason for this could be that at low SFRs the combination of \ha\ and  24\,\mi\ luminosity might underestimate the SFR due to stochasticity of the Initial Mass Function \citep{2009ApJ...706..599L}, or that at local scales of $\sim$170\,pc we cannot account for some leakage of ionising photons which would underestimate the SFR when \ha\ luminosity is used \citep{2012MNRAS.423.2933R}. Another possible explanation is that the radiation field of old stars contributing to the TIR emission is more extended than the radiation field responsible for the dust emission at 24\,\mi, which is more related to star forming regions. In this situation, at local scales the TIR emission can be contaminated by radiation of stars located at further distances, giving higher values of SFR than the ones estimated using the emission at \ha\ and 24\,\mi. For both hybrid SFR estimators, we find a correlation between the strength of the ISRF and the SFR. We note that although a relation between the amount of energy heating the dust and the amount of star formed in the same area is expected, this is the first time such a relation is empirically demonstrated.

\section{Submillimetre excess}\label{sec:subm}

We define the submillimetre excess as the fraction of observed luminosity in the SPIRE 500\,\mi\ band above the luminosity derived from the model in the same band. We note that the dust model we apply in this study has an emissivity coefficient $\beta$\,=\,2 and therefore our results can be directly compared to what has been referred to in the literature as a {\it submillimetre excess} using MBBs (where a constant $\beta$\,=\,2 is normally assumed). We parameterise the excess as: 
\begin{equation}
e_{500}=\frac{(F^{500}_{obs}-F^{500}_{mod})}{F^{500}_{obs}}
\label{eq:ex}
\end{equation}
where $F^{500}_{obs}$ and $F^{500}_{mod}$ are the observed and modelled fluxes in the SPIRE 500\,\mi\ band, respectively. 

\begin{figure}[h]
  \includegraphics[width=0.49\textwidth]{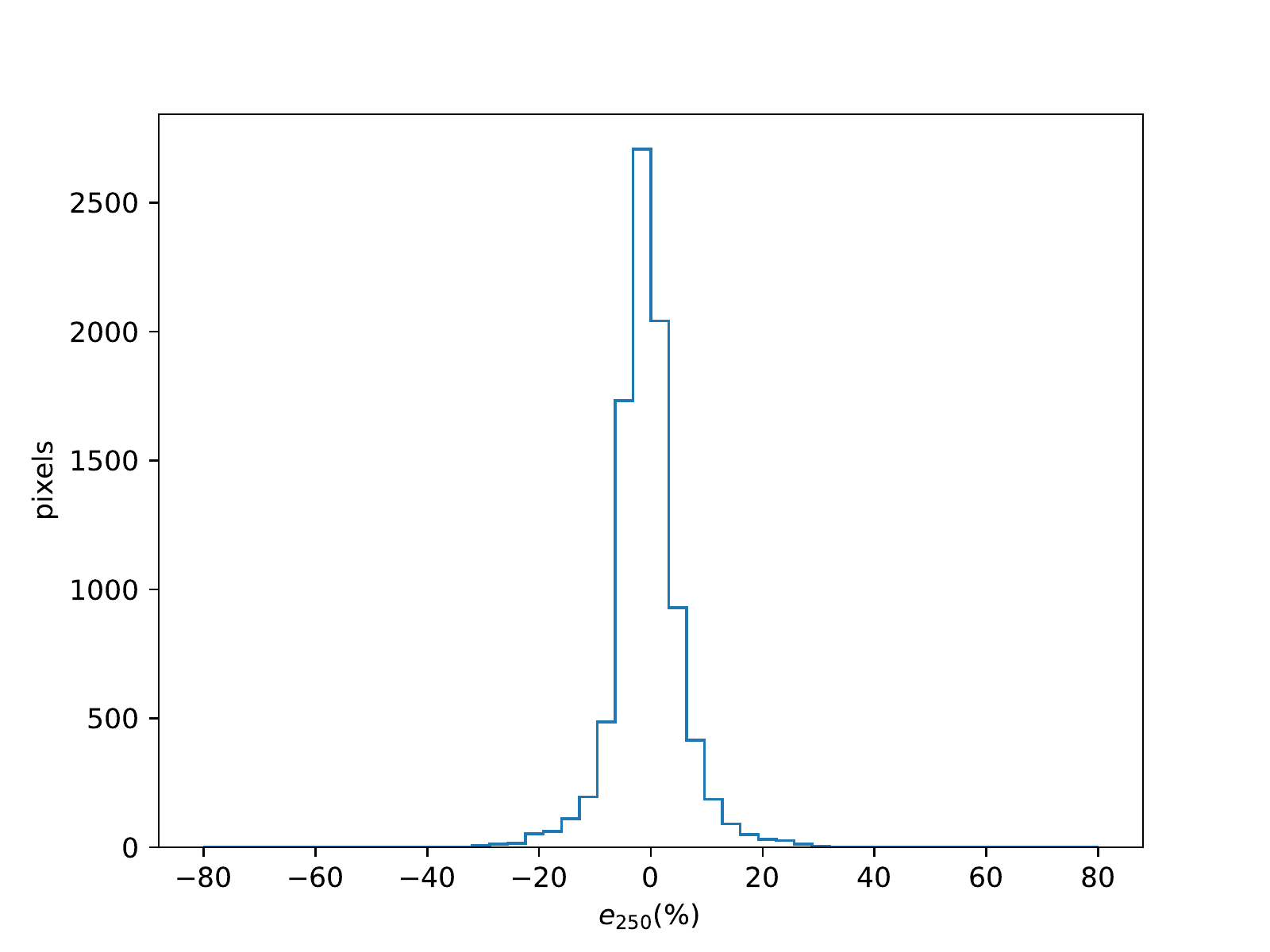}   
  \includegraphics[width=0.49\textwidth]{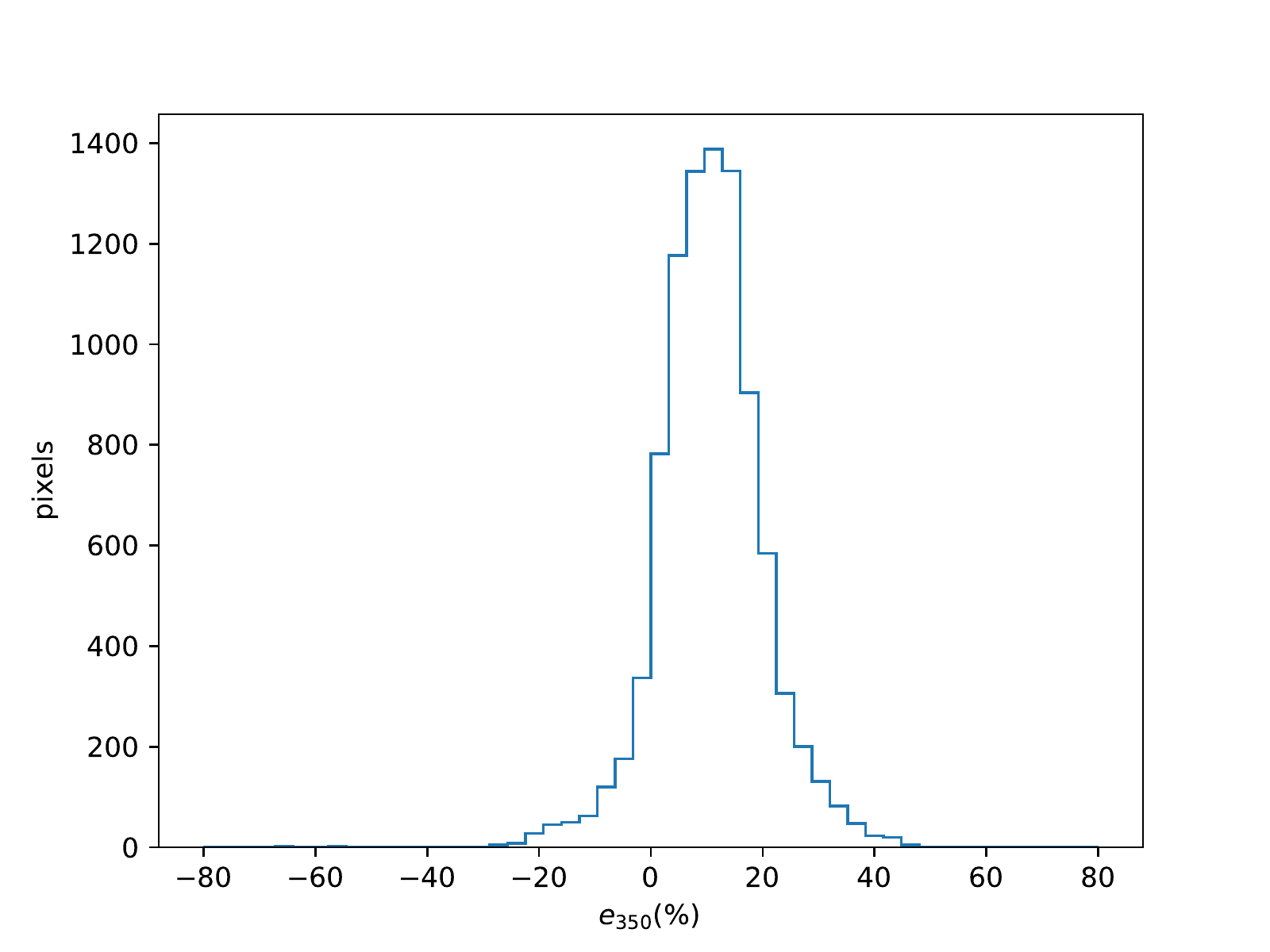}   
  \includegraphics[width=0.49\textwidth]{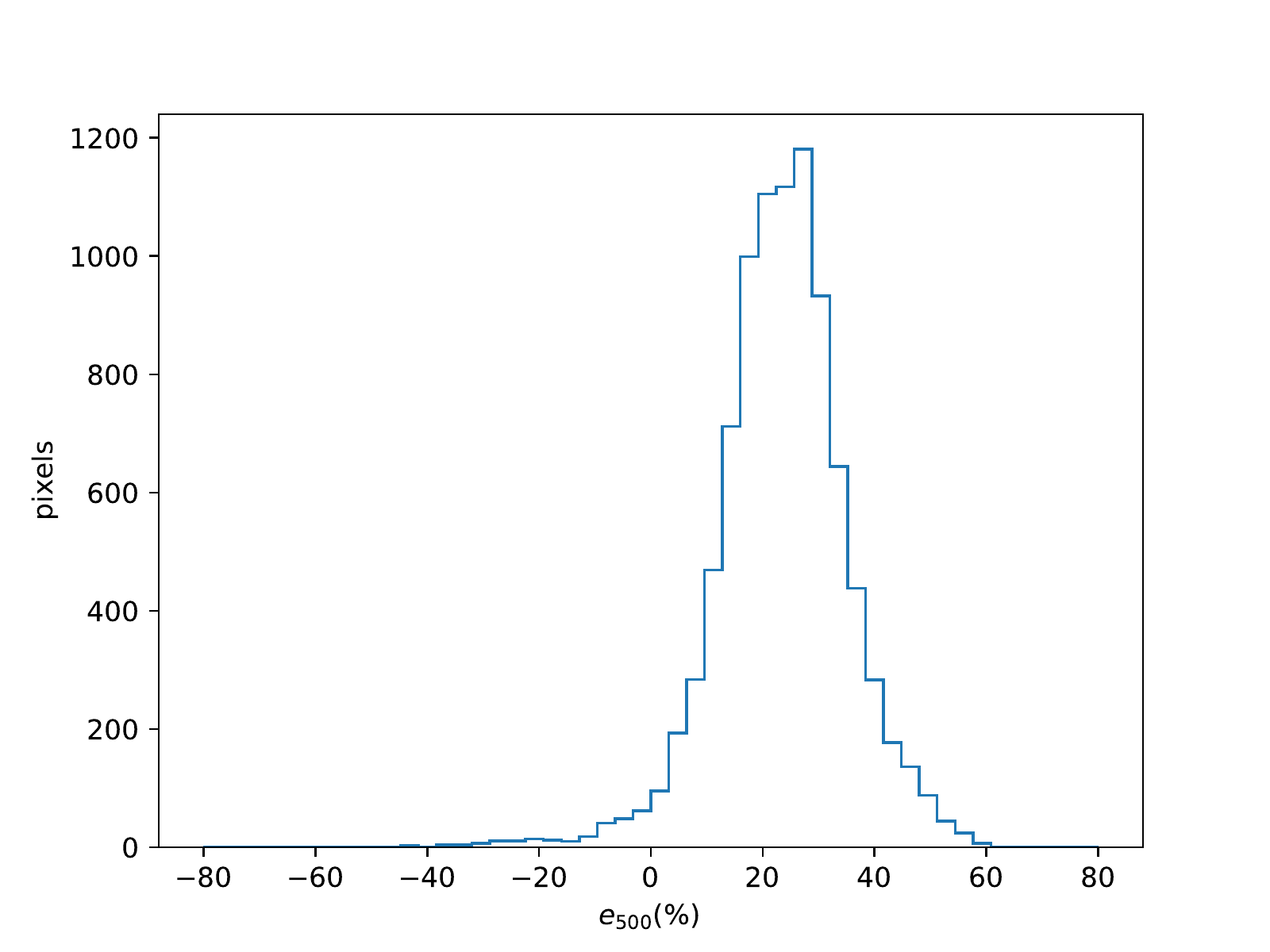}   
   \caption{Histograms of the excess  in \% obtained using Eq.~\ref{eq:ex} for the three SPIRE bands: 250\,\mi, 300\,\mi, and 500\,\mi. There is no sign of excess at 250\,\mi, the excess starts to appear at 350\,\mi\ and clearly visible at 500\,\mi.}
   \label{fig:excesshist}
\end{figure}

In Fig.~\ref{fig:excesshist} we show histograms of $e_{500}$ for all the individual pixels fitted with our model in the three bands of SPIRE. The fluxes at 250\,\mi\  are very well fitted with our dust models: the histogram is centred at zero with a width of 15-20\%, close to the uncertainty of the SPIRE fluxes (15\%, see Section\,\ref{sec:dataset}). However, in the 350\,\mi\ SPIRE band the centre of the histogram is shifted to positive values and at 500\,\mi\ the excess is as high as $\sim$50\%, clearly above the uncertainty of the 500\,\mi\ flux. 

\begin{figure}[h] 
\includegraphics[width=0.49\textwidth]{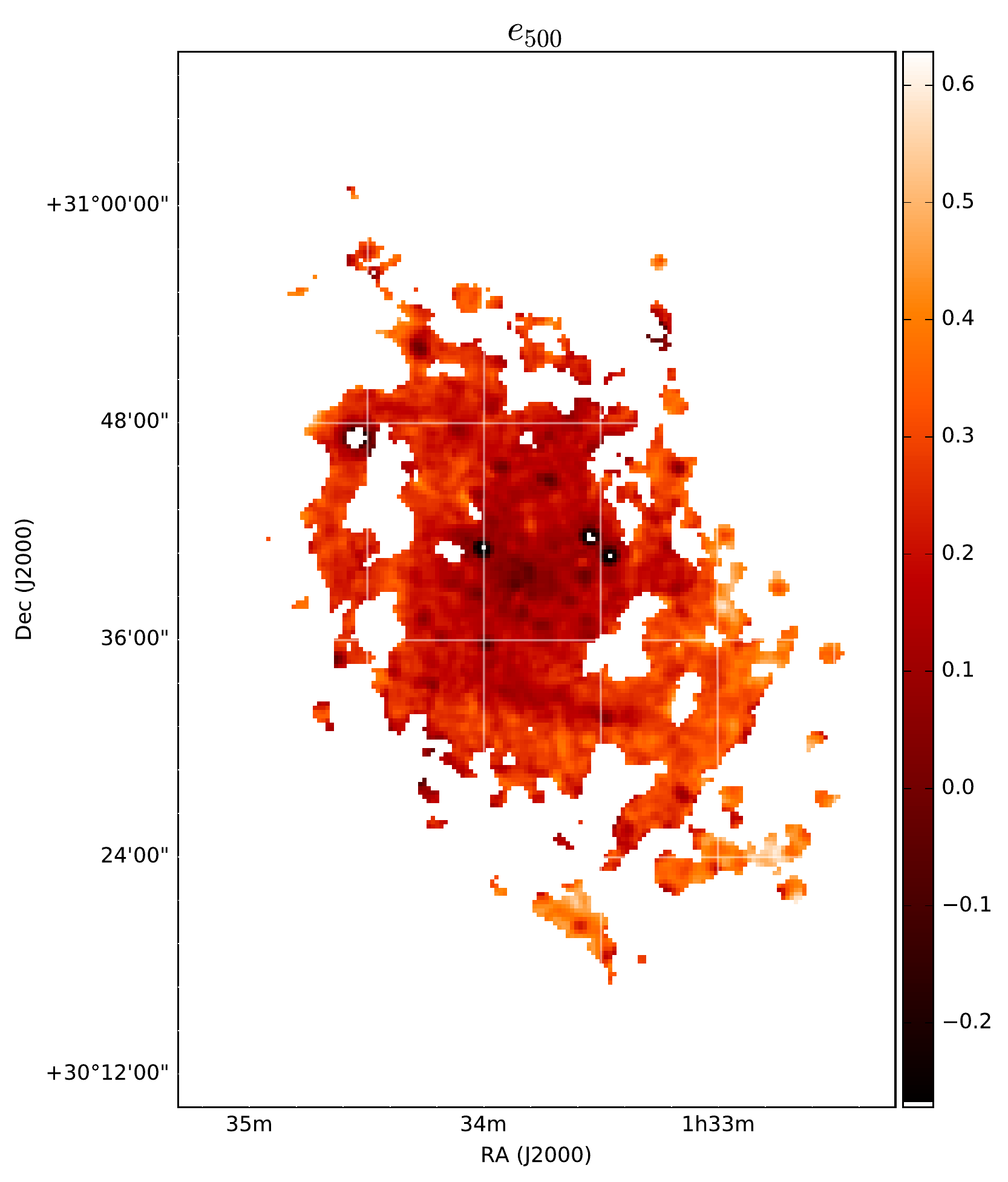}
   \caption{Excess at at 500\,\mi\ derived using Eq.~\ref{eq:ex} for the whole disc of M\,33.}
   \label{fig:e500_map}
\end{figure}

In Fig.~\ref{fig:e500_map} we show the map of excess at 500\,\mi\ for M\,33. We confirm with these plots that M\,33 exhibits an excess in the submillimetre wavelength range, which has already been reported before by \citet{2016A&A...590A..56H}. Since we fit here the SED of the whole disc of M\,33 in a pixel-by-pixel basis we are able to localise the regions of the disc where the excess is present. This will help us to better understand the main mechanisms which might produce this excess.

\subsection{Radial trend}

\begin{figure}[t] 
  \includegraphics[width=0.49\textwidth]{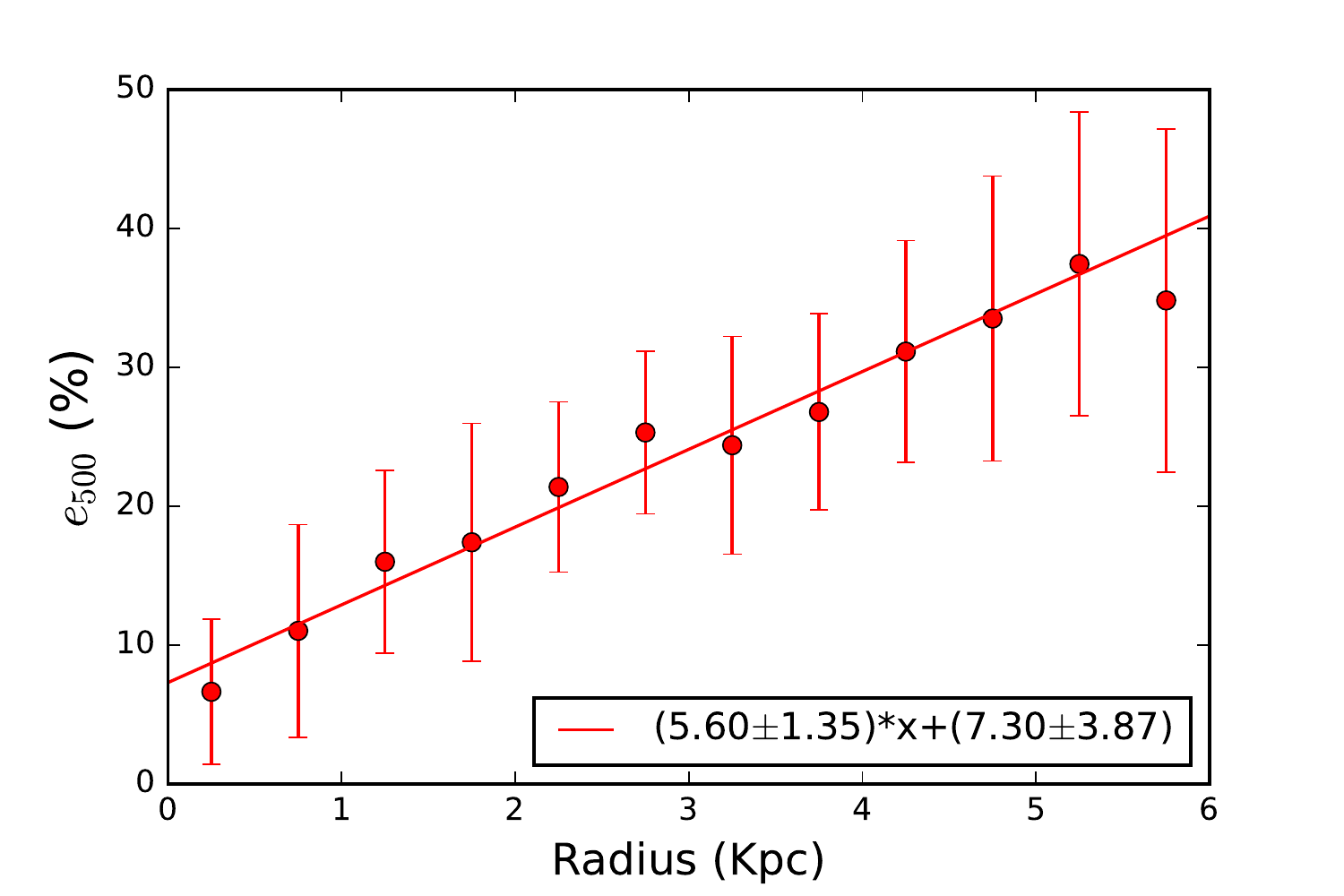}   
   \caption{Radial profile of $e_{500}$. Each dot and its error bars correspond to the mean and the standard deviation of the  $e_{500}$ in each concentric 500\,pc wide elliptical ring.}
   \label{fig:excess}
\end{figure}

In Fig.~\ref{fig:excess} we show the variation of  the mean $e_{500}$ with the galactocentric distance. We see that the excess increases towards the outer parts of the galaxy and reaches values up to 38\% in agreement with the analysis of the histograms shown in the previous section. The radial trend is in agreement with previous results shown in the literature. Using MBB fitting for M\,33, \citet{2014A&A...561A..95T} report a decline of the emissivity coefficient $\beta$ with the radial distance of the galaxy (Figure~3 in their paper): lower values of $\beta$ correspond to a shallower slope in the SED (compared to $\beta$\,=\,2), and therefore to an expected excess in the submillimetre wavelength range. An excess in 500\,\mi\ of 38\% corresponds to a change of $\beta$ from 2 to 1.1, which fits nicely with the $\beta$ radial variation shown by  \citet{2014A&A...561A..95T}. The increase of  $e_{500}$ with the galactocentric radius is also seen in other galaxies: \citet{2012A&A...537A.113P} studied the 500\,\mi\ excess in the Galactic plane with \Her\ observations and found that the excess is located in the peripheral regions of the Galactic plane; \citet{2012MNRAS.425..763G,2014MNRAS.439.2542G} found that when the dust temperature and $\beta$ are left free in the MBB fitting, $\beta$ tends to decrease radially towards low surface brightness regions. Using MBB fitting, \cite{2012ApJ...756...40S} found a radial decrease of $\beta$ for R\,$>$\,3.1\,kpc in M31. The range of $\beta$ values for M31 are however higher than those found in M33 by  \citet{2014A&A...561A..95T}:  $\beta$ is $\sim$\,1.8 at the centre of M33 and decreases radially towards the outer parts of the galaxy, while for M31 $\beta\sim$\,2.5 at R$\sim$3.1\,kpc.

\subsection{Relation to physical parameters}
\begin{figure*}[h]
\includegraphics[width=0.49\textwidth]{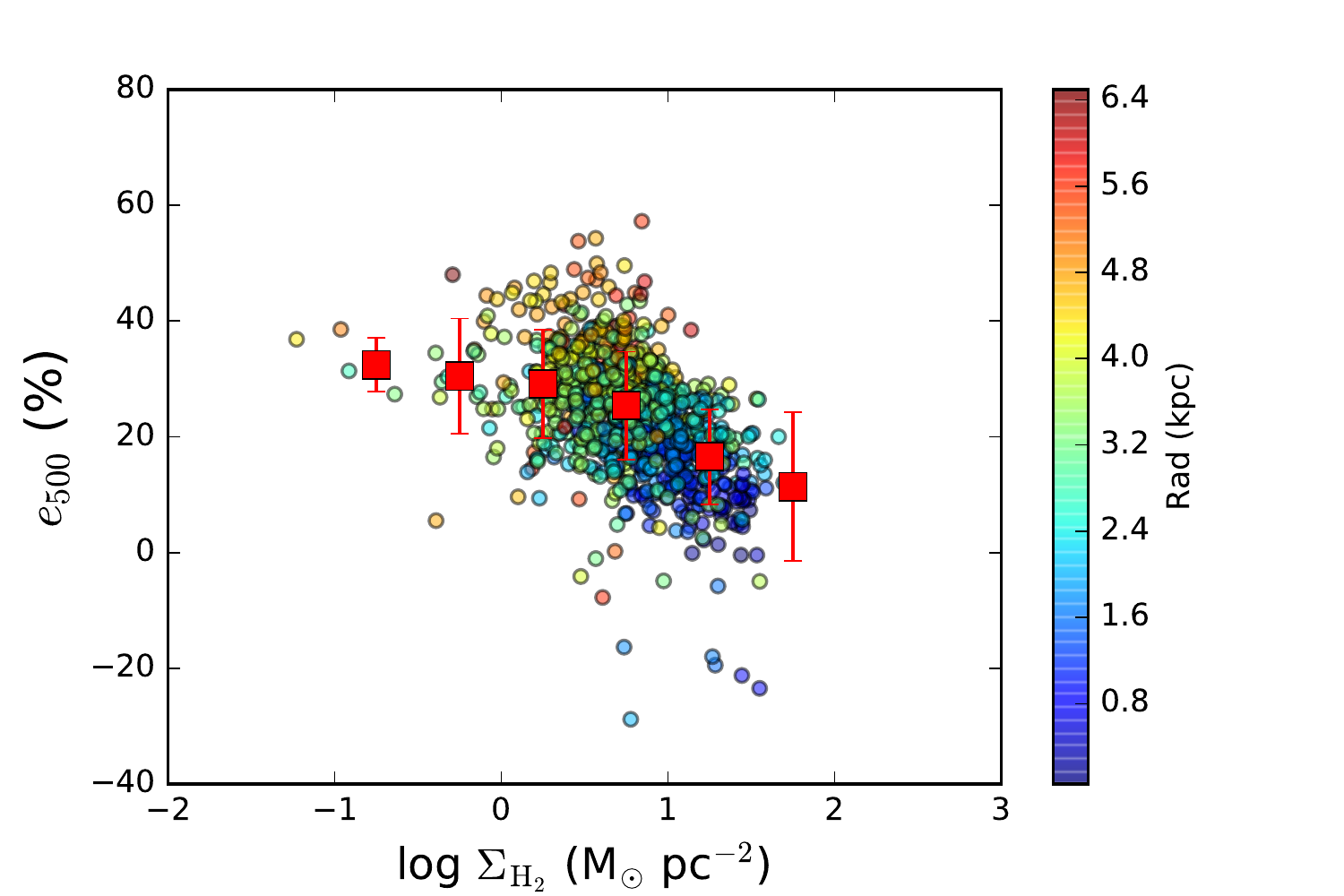}
\includegraphics[width=0.49\textwidth]{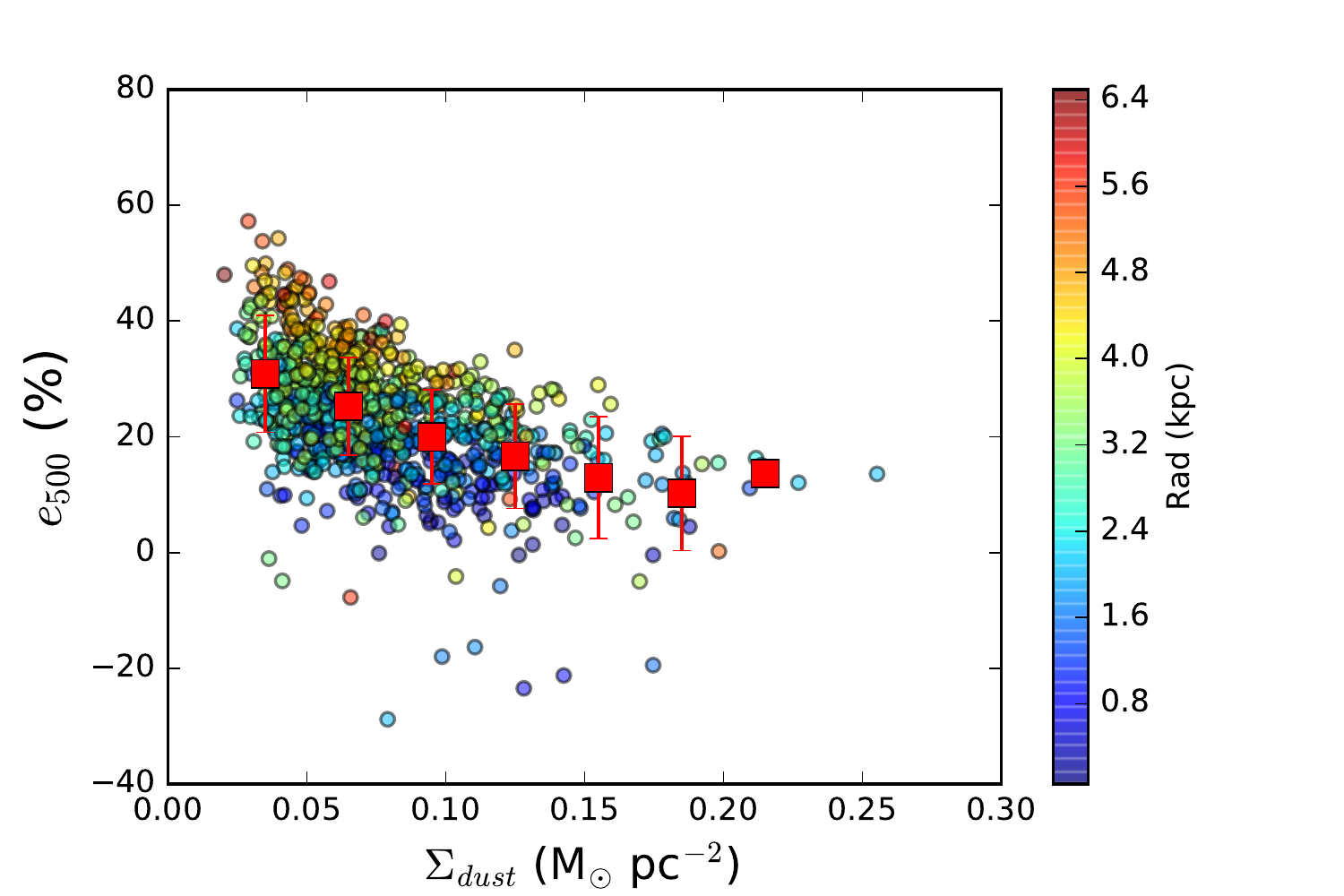}
\includegraphics[width=0.49\textwidth]{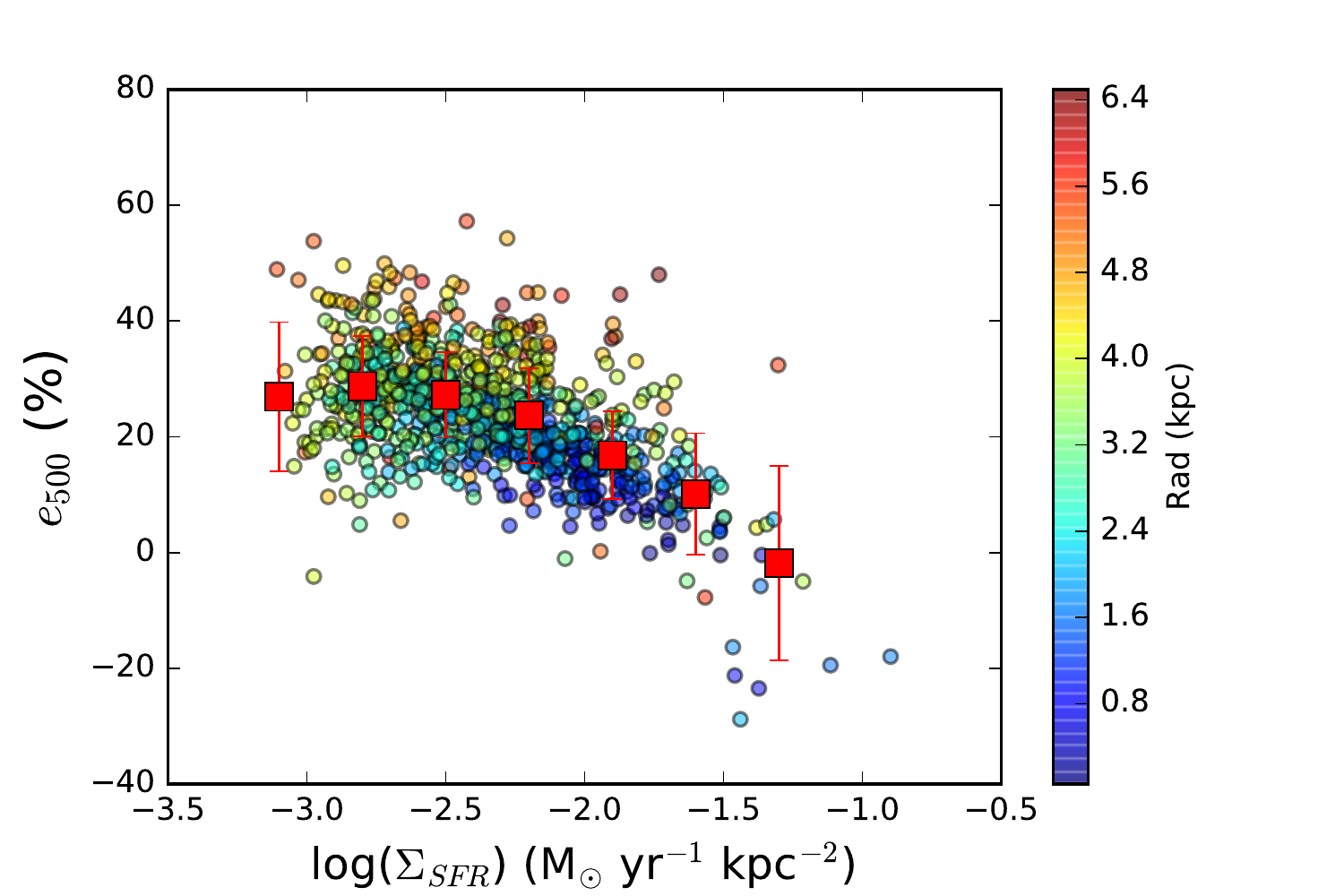}
\includegraphics[width=0.49\textwidth]{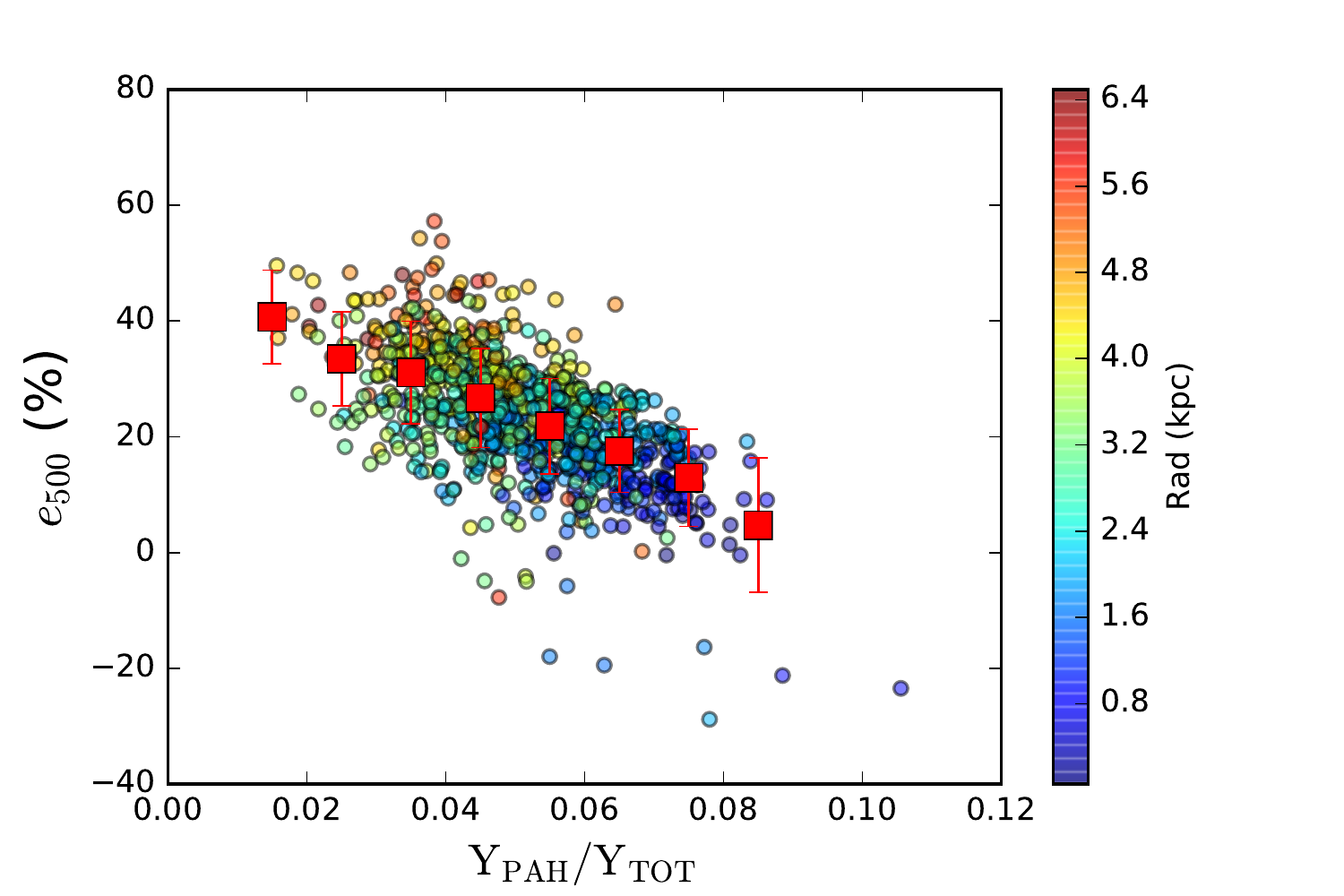}
   \caption{Top: Excess in the SPIRE 500\,\mi\ band versus the molecular gas (left) and dust (right) surface density. Bottom: Excess in the same band versus the SFR (left) and the mass fraction of PAHs (right). Each red dot and its error bars correspond to the mean and the standard deviation of the magnitud represented in $y$ axis. The color bar corresponds to galactocentric distances in kiloparsecs.}
   \label{fig:e_500trends}
\end{figure*}

In this section we try to identify under which ISM conditions the submillimetre excess is prominent. This will shed light onto the mechanisms that might cause the excess.   
In Fig.~\ref{fig:e_500trends} we show correlations between $e_{500}$ and some parameters representative of the physical conditions of the ISM. We find an anti-correlation between $e_{500}$ and the molecular hydrogen surface density mass in agreement with results from \citet{2014A&A...561A..95T}, while no relation at all between $e_{500}$ and the neutral hydrogen surface density mass was found (not shown here). A similar anti-correlation between $e_{500}$ is seen for $\Sigma_{dust}$, $\Sigma_{SFR}$ and the relative mass fraction of PAHs. The anti-correlation between the excess and the dust surface density has also been observed in the LMC by \citet{2011A&A...536A..88G}. The excess is seen in regions with low SFR where the molecular gas and dust surface densities are low as well. These regions correspond to the diffuse ISM outside the main \hii\ regions. Inside the main star-forming regions and molecular clouds the excess is in general close to zero.  

The location of the excess within the galaxy, shown in the present work to be clearly associated with the diffuse ISM,  is consistent with previous results 
given in the literature: \citet{2016A&A...590A..56H}, fitting the global SED of M\,33 with a radiation transfer model, found a higher submillimetre excess in the diffuse SED of M\,33 than in the combined SED of all the \hii\ regions of the galaxy. \citet{2014ApJ...797...85G} performed a pixel-by-pixel SED fitting of the LMC and SMC using different versions of MBBs. For both LMC and SMC, these authors found higher values of $e_{500}$ in the diffuse regions corresponding to low values of SPIRE 250\,\mi\ fluxes \citep[see Fig.\,6 and\,7]{2014ApJ...797...85G}. Finally, \citet{2011A&A...536A..25P} reported a slight difference in the emissivity index between the dense ($\beta\sim$\,2) and the diffuse medium ($\beta\sim$\,1.7) of the Taurus complex, which agrees with the findings presented here.

We thus conclude that there is an excess in the SPIRE 500\,\mi\ band in M\,33 which is significantly higher than the uncertainties (15\%) in the observed  500\,\mi\ flux. The excess is mainly located in the diffuse part of the disc, while the star-forming regions present no excess within the uncertainty in the observed flux in this band. There is also a clear trend of the excess to increase with the galactocentric radius. Although we cannot rule out that the excess might have a dependence with the metallicity within the disc of the galaxy \citep[indeed there is large evidence that the excess is mainly seen in low metallicity galaxies, e.g.][]{2011A&A...532A..56G}, the metallicity gradient in M\,33 is quite shallow \citep[0.045\,dex/$\rm R_{kpc}$,][]{2011ApJ...730..129B} to establish a clear trend between $e_{500}$ and metallicity in this galaxy. The radial trend observed in  Fig.~\ref{fig:excess} could well be a consequence of the fact that at larger radii the fraction of diffuse ISM is higher than in the inner parts of the disc of M\,33. 

\begin{figure*}[h]
\includegraphics[width=0.49\textwidth]{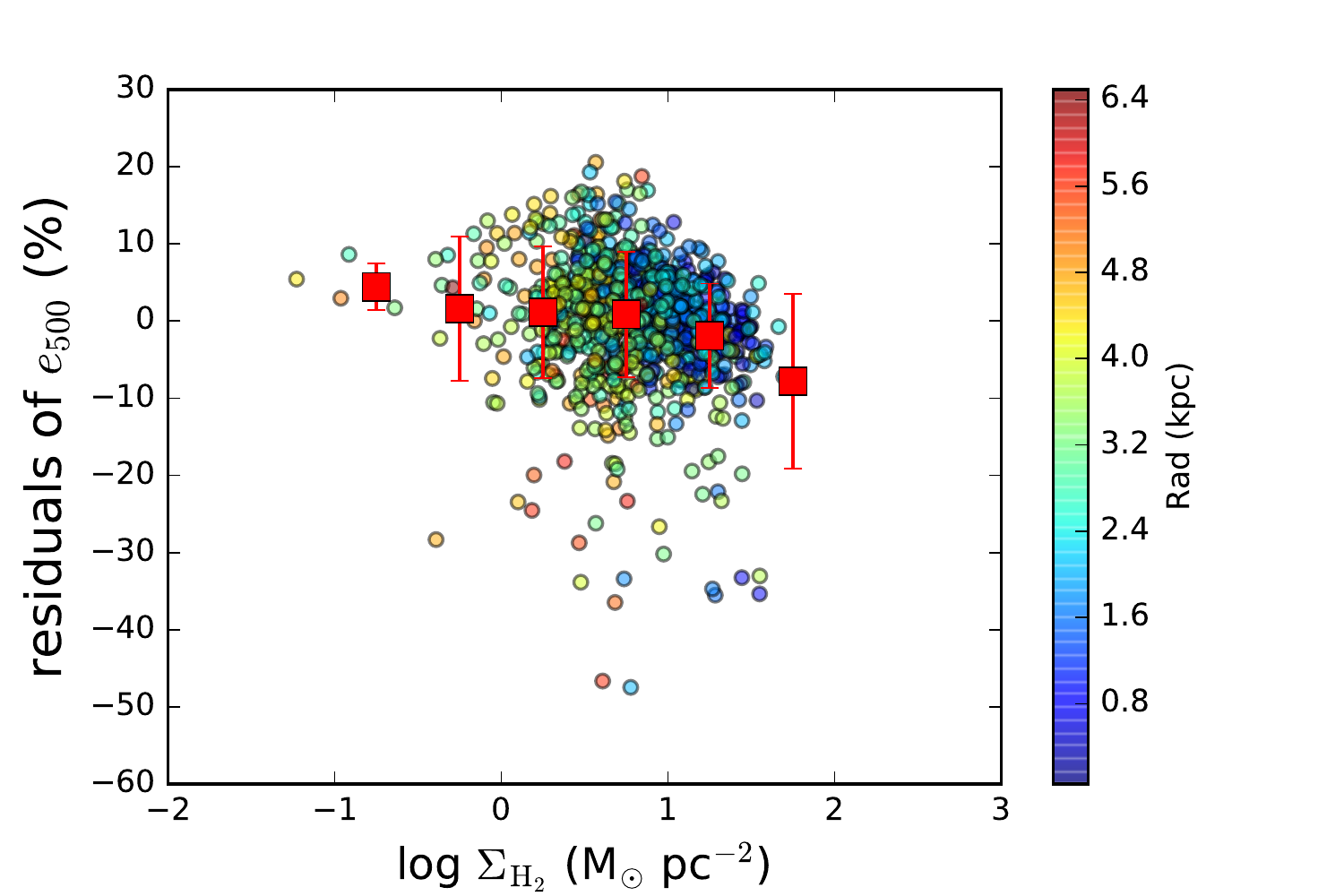}
\includegraphics[width=0.49\textwidth]{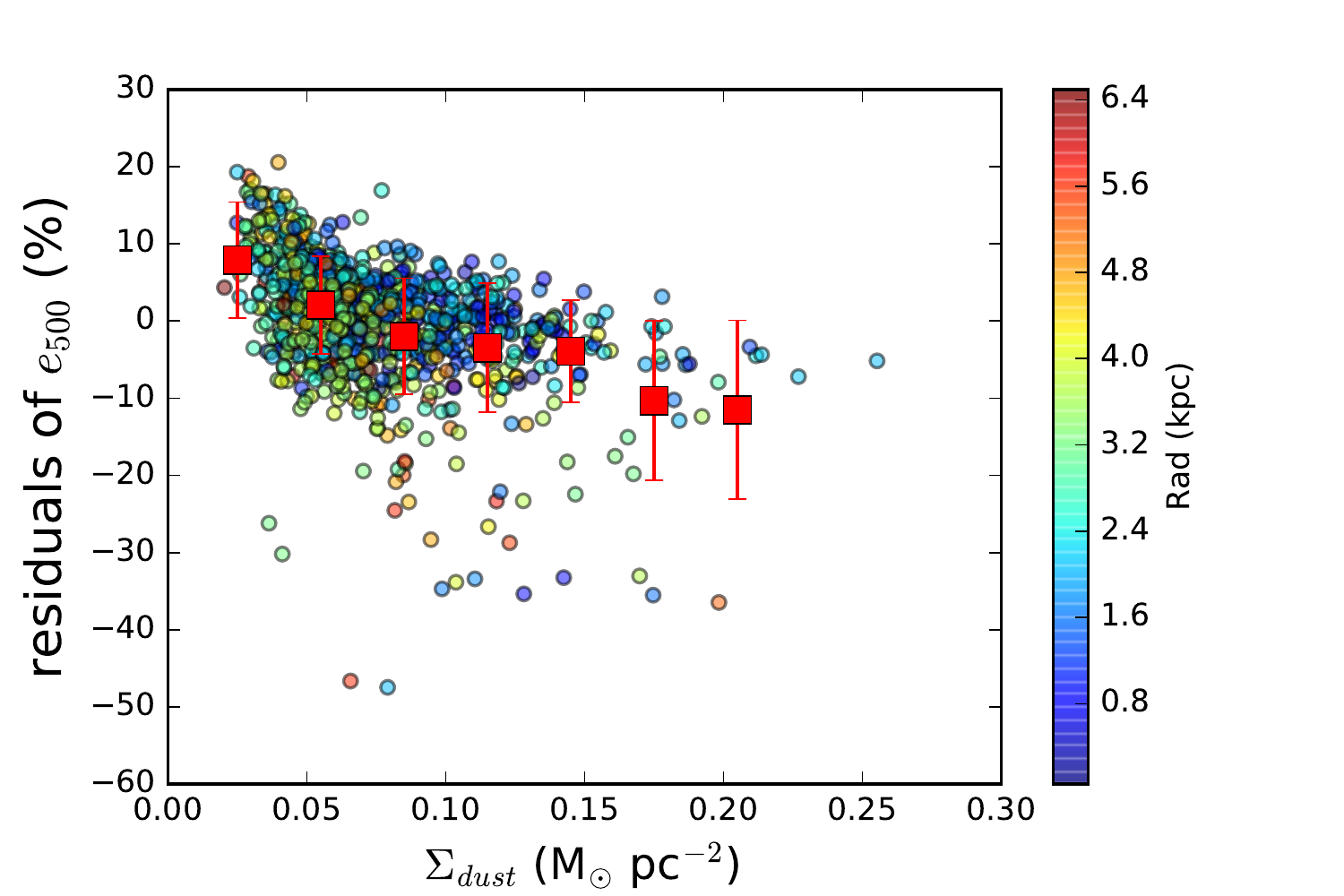}
\includegraphics[width=0.49\textwidth]{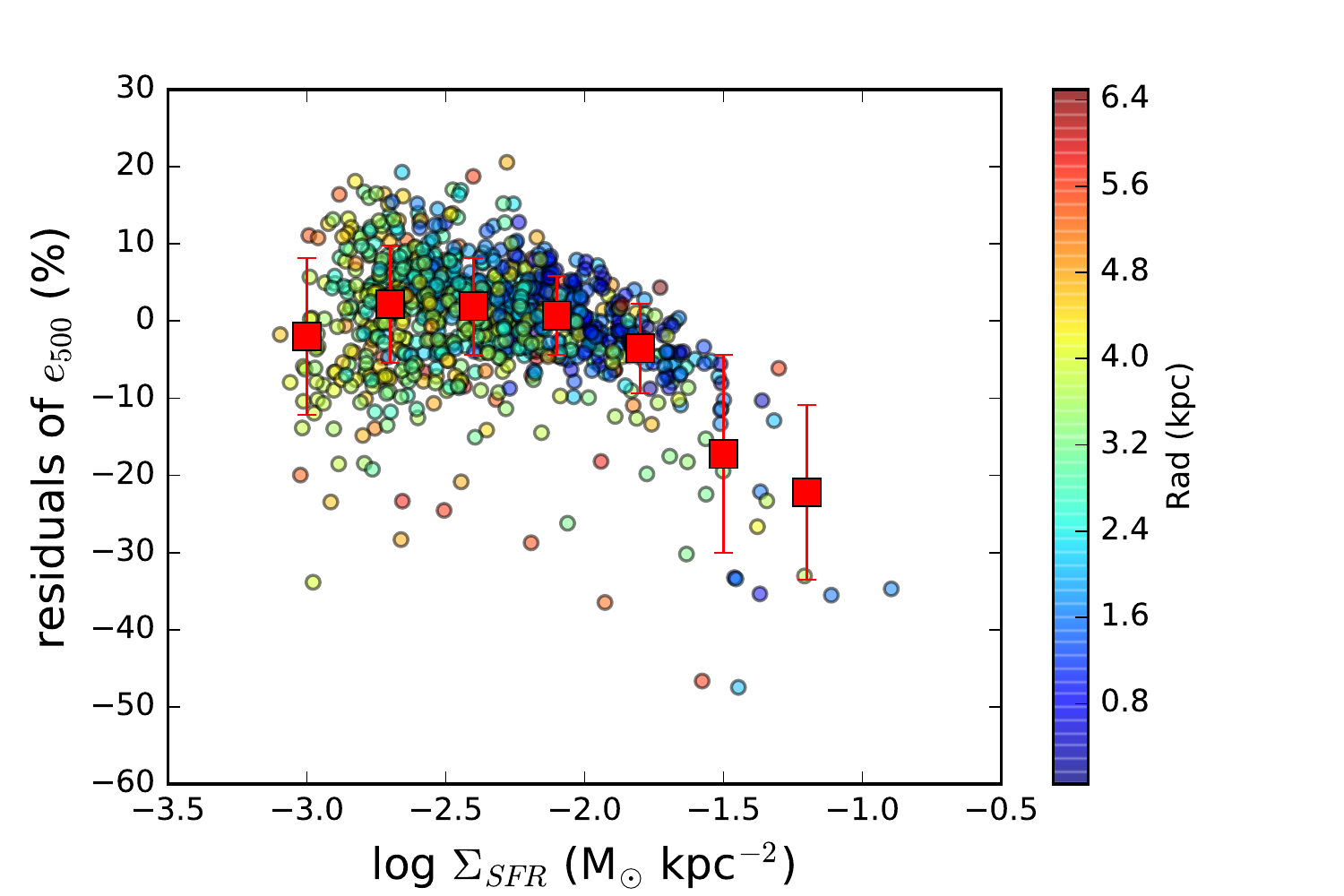}
\includegraphics[width=0.49\textwidth]{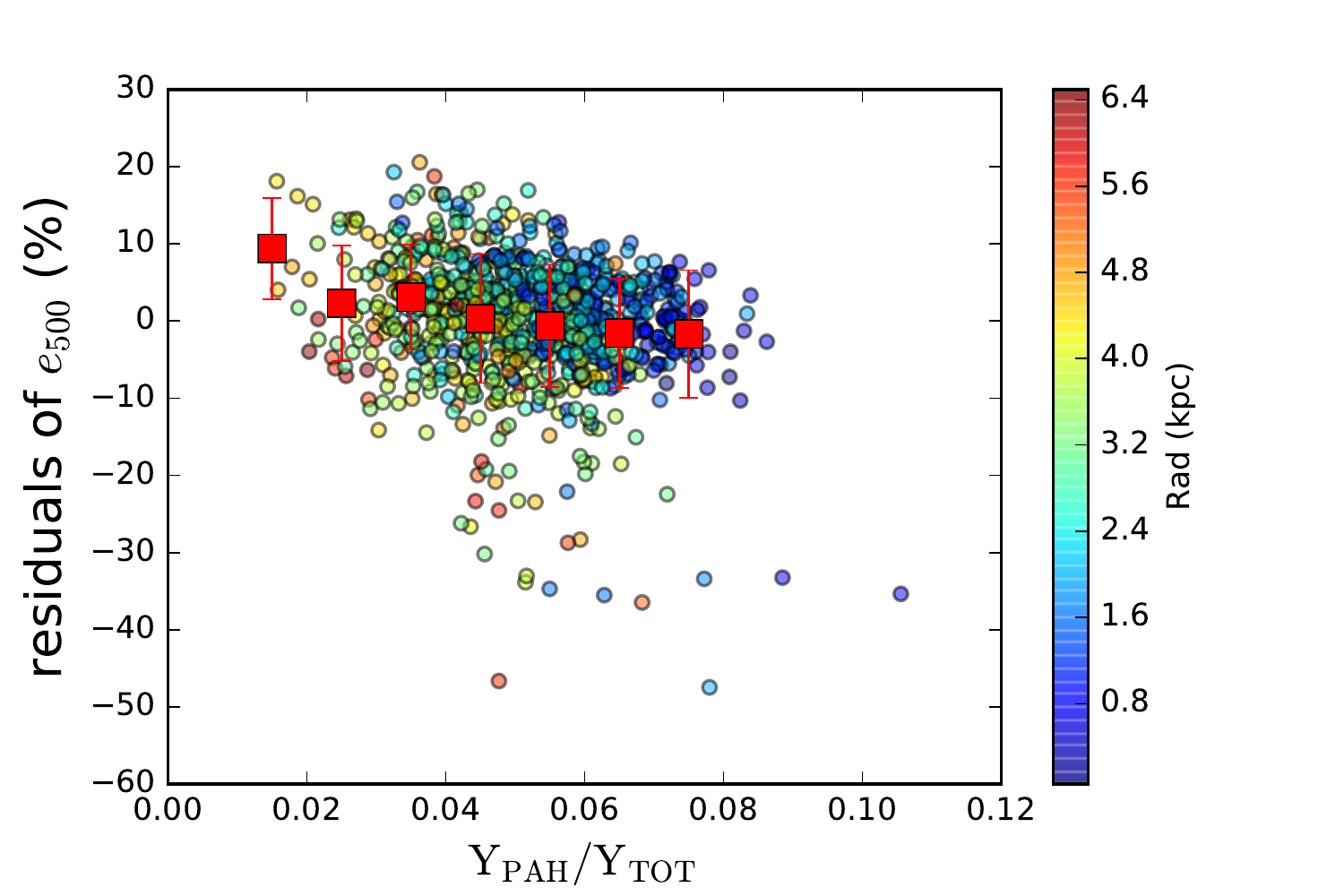}
   \caption{Top: Residual excess in the SPIRE 500\,\mi\ band versus the molecular gas (left) and dust (right) surface density. Bottom: Excess in the same band versus the SFR (left) and the mass fraction of PAHs (right). Each red dot and its error bars correspond to the mean and the standard deviation of the magnitud represented in $y$ axis. The color bar corresponds to galactocentric distances in kpc.}
   \label{fig:rese_500trends}
\end{figure*}

In order to investigate further the location of the excess we analyse the residuals of the excess relative to the radius. For that, we use the linear fit shown in Fig.~\ref{fig:excess} and obtained the deviation of the excess relative to this linear fit. In Fig.~\ref{fig:rese_500trends} we plot the residuals versus the same quantities as in Fig.~\ref{fig:e_500trends}. The correlations between the residuals and the surface density of the molecular gas, dust and the SFR are kept, showing that indeed the excess is more prominent in diffuse regions where these three quantities have low values independent of the galactic radius. There is no correlation between the residuals and \ypah/\ytot, showing that the trend of the excess with \ypah/\ytot\ presented in Fig.~\ref{fig:e_500trends} was mainly driven by the fact that there is a decrease of PAHs with the galactocentric radius in M\,33 as well.  

\subsection{Other dust models}
Several mechanisms have been invoked to explain the submilimeter excess in galaxies: the existence of a large amount of very cold dust, which would require a very high dust mass incompatible with the observations \citep{2002A&A...382..860L,2011A&A...536A..88G,2011A&A...532A..56G,2014ApJ...797...85G}, different dust grain properties producing a lower then 2 emissivity coefficient $\beta$  \citep{2002A&A...382..860L,2011A&A...536A..88G,2014ApJ...797...85G}, emission of magnetic nanograins \citep{2012ApJ...757..103D}, and spinning grains \citep{1998ApJ...508..157D,2010A&A...523A..20B}. We refer the reader to \citet{2016A&A...590A..56H} who analysed in very detail how these mechanisms contribute to the excess observed in the integrated SED of M\,33. These authors showed that fitting the excess with very cold dust requires a high amount of dust mass giving an unreasonable GDR. Besides, the emission of spinning grains and  magnetic nanoparticles are not able to reproduce completely the observed excess in the total SED of M\,33 \citep[see Fig.~9 in][]{2016A&A...590A..56H}.

Here we showed that the excess is present in the diffuse ISM (low gas and dust surface densities) whereas in star-forming regions (high gas and dust surface densities) no excess is found. The physical properties of the  interstellar dust might be completely different in both ISM phases, mainly because the evolutionary stage of the dust might be determined by the conditions of the ISM in both phases. \citet{1994ApJ...433..797J,1996ApJ...469..740J} already proposed a theoretical framework where the dust grains are destroyed due to shocks in the ISM altering the grain size distribution of the dust. Recently, \citet{2013A&A...558A..62J,2017arXiv170300775J} has proposed a new dust model based on three different components: small ($a\simeq$0.4 to $\sim$100\,nm) and large amorphous hydrocarbon grains, a-C(:H), and large amorphous silicate grains. The small grains present a power-law size distribution, while the large grains are assumed to have a log-normal size distribution peaking at radii $\simeq$200\,nm. An extension of this model has been introduced by  \citet{2014A&A...565L...9K} adding a mix of amorphous olivine- and pyroxene-type silicate grains with iron nano-inclusions.  This dust model has been successfully applied to the LMC and SMC by \citet{2017A&A...601A..55C}. 

The \citet{2013A&A...558A..62J,2017arXiv170300775J} dust model, calibrated to reproduce the emission of the diffuse dust in the Milky Way, is mainly valid for dust located in the diffuse ISM where the gas and dust density is low, but it is flexible and can also accommodate the evolution of grains in different ISM phases. The model predicts that large a--C particles have a flat emissivity slope of $\beta\gtrsim$\,1.2-1.3. On the contrary, within the molecular clouds coagulation of particles and a--C:H accretion leads to materials with $\beta\sim$\,1.8-2.5. This model can naturally explain our results where the excess, quantified as $e_{500}$,  is mainly located in diffuse low dense regions of the ISM. \citet{2015A&A...579A..15K} studied the evolution of the dust grains in the model proposed by \citet{2013A&A...558A..62J} and \citet{2014A&A...565L...9K}. They studied the transition from the diffuse ISM to denser regions assuming that the dust in the dense regions is affected by two mechanisms:  i) accretion of C and H from the gas phase onto the surface of the dust grains, and ii) coagulation of grains into aggregates. They modelled the resulting  dust emission and found that both processes make the dust temperature to decrease by 3\,K and the emissivity coefficient  to increase to $\beta$\,=\,2 with respect to the assumed values of the dust model for the diffuse ISM,  $\beta\sim$1.8, \citep[see Fig. 6 in][]{2015A&A...579A..15K}. Our results are consistent with those from  \citet{2015A&A...579A..15K}: the low density ISM medium has a low emissivity coefficient $\beta$, while the more intense star-forming regions surrounded by the molecular clouds and exhibiting higher gas and the dust mass surface density  values, present no signatures of excess, (i.e. emissivity coefficient values close to the one assumed by the dust model from \citet[][]{1990A&A...237..215D}, $\beta$\,=\,2).

\section{Conclusions}~\label{sec:conc}
We performed an analysis of the individual SEDs at scales of $\sim$170\,pc in the whole disc of M\,33. Using a Bayesian statistical approach we fitted the individual SEDs with the classical dust model of \citet{1990A&A...237..215D}. This dust model presents has an emissivity index of $\beta$=2, which allows us to make a direct comparison with MBB fitting reported in the literature. The best fit values representing the observed SEDs are obtained from the PDFs of each input parameter: the best value corresponds to the mean of the PDF, while the uncertainty is the corresponding 16th-84th percentile range for each parameter. Our main conclusions are the following: 
\begin{itemize}
\item The relative fraction of the VSGs,  \yvsg/\ytot, increases at locations of intense star formation, while the relative fraction of BGs, \ybg/\ytot, follows the opposite correlation. These results agree with those presented in \citet{2016A&A...595A..43R} and are consistent with the framework of dust evolution models of  \citet{1994ApJ...433..797J,1996ApJ...469..740J} suggesting dust grain destruction/fragmentation by interstellar shocks in the warm medium. 

\item We derive the GDR over the whole disc of M\,33. The GDR is relatively constant with a value of $\sim$300 up to a radius of 4\,kpc where it starts to increase. The radial variation is consistent with the shallow metallicity gradient of this galaxy. We do not find a strong correlation between the GDR and the strength of the ISRF, as it was suggested by \citet{2011A&A...536A..88G} for the LMC, even when including the dark gas fraction reported by \citet{2017A&A...600A..27G}. 

\item We find a correlation between $\rm G_{0}$, the strength of the ISRF in the solar neighbourhood given by \citet{Mathis:1983p593}, and the SFR at each spatial location in the disc of M\,33. The relation holds when the SFR is derived using a linear combination of \ha\ and 24\,\mi, or FUV and TIR luminosities. This relation shows that $\rm G_{0}$, derived from the models, can account for the amount of the star formation forming at these spatial locations. 

\item We find a submillimetre excess, defined as the fraction of observed luminosity in the SPIRE 500\,\mi\ band above the luminosity derived from the model in the same band, that can be as high as $\sim$50\%, clearly above the uncertainty of the 500\,\mi\ flux. The excess increases at larger radii in agreement with the radial dependence of $\beta$  reported by \citet{2014A&A...561A..95T} for M\,33.

\item The excess is prominent in regions of low surface densities of SFR, dust and gas mass. These regions correspond to the diffuse ISM outside the main star-forming \hii\ regions. 

\item Our findings are consistent with THEMIS dust model presented in \citet{2013A&A...558A..62J,2017arXiv170300775J}. This model is mainly valid for the diffuse ISM and predicts that large a-C(:H) particles exhibit a flat emissivity coefficient $\beta\sim$\,1.2\,-\,1.3. A dust component (composed of carbon or silicate-type dust) with similarly flat dust emissivity coefficients than the large a-C(:H) particles in the THEMIS dust model could be invoked to explain the submillimetre excess in the diffuse disc of M\,33. 

\end{itemize}

\begin{acknowledgements}
We would like to thank the referee for the useful comments that helped to improve the first version of this paper. This work was partially supported by the Junta de Andaluc\'ia Grant FQM108 and Spanish MEC Grants, AYA-2011-24728 and AYA-2014-53506-P. MRP thanks to the Computational service PROTEUS at the Instituto Carlos I. This research made use of APLpy, an open-source plotting package for Python hosted at http://aplpy.github.com of TOPCAT \& STIL: Starlink Table/VOTable Processing Software \citep{2005ASPC..347...29T} of Matplotlib, a suite of open-source python modules that provide a framework for creating scientific plots.

 \end{acknowledgements}

\bibliographystyle{aa} 
\bibliography{references}  
\begin{appendix}

\section{Robustness of the fit}\label{App:robust}
\begin{figure} 
 \includegraphics[width=0.5\textwidth]{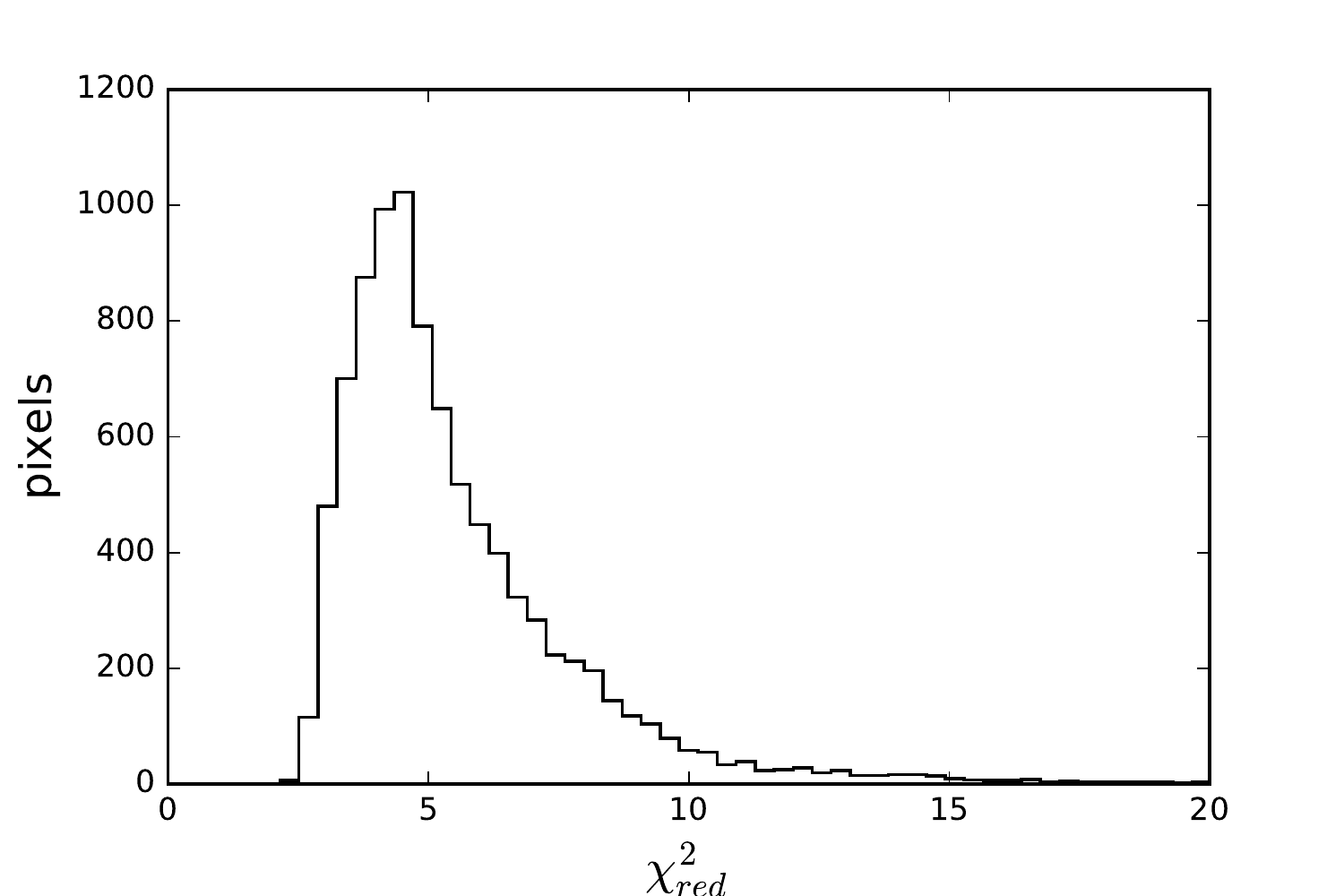}       
   \caption{\chidos\ distribution for all the SEDs in the disc of M\,33.}
   \label{fig:chi2_dist}
\end{figure}

The low values for the \chidos\ in most of the fits (see Fig.~\ref{fig:chi2_dist}) show that in general the fits are reproducing quite well the observed SEDs. As a further step, we also analyse how robust our fitting methodology is. For this purpose we use two procedures: (i) generate {\it mock} SEDs and check if the fitting routine applied to the {\it mock} SEDs gives consistent results as the original ones, and (ii) fit each PDF with a single Gaussian and compare the results to the estimates of the best fit parameter values and corresponding uncertainties. 

\begin{figure*} 
\centering 
\includegraphics[width=0.49\textwidth]{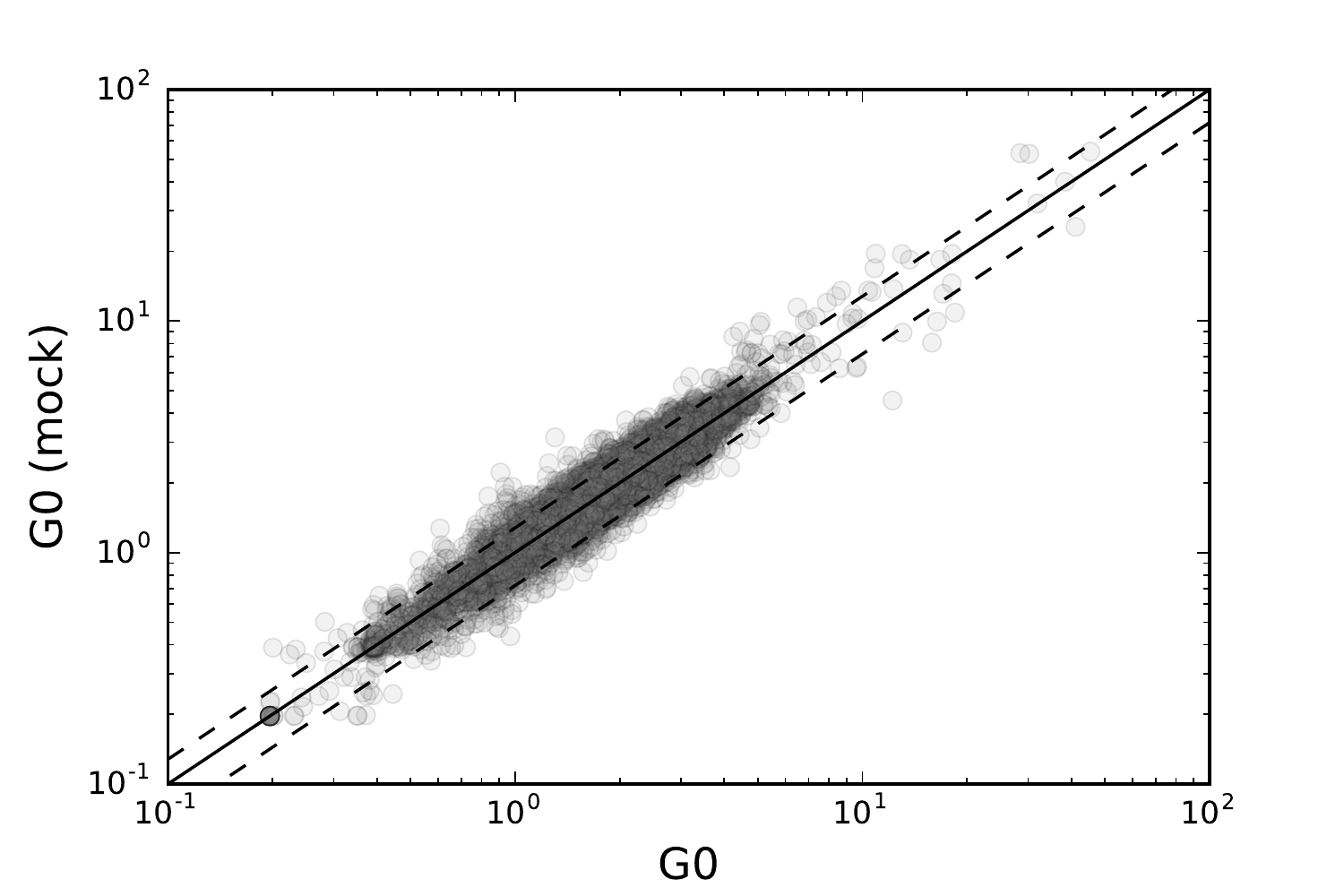}   
\includegraphics[width=0.49\textwidth]{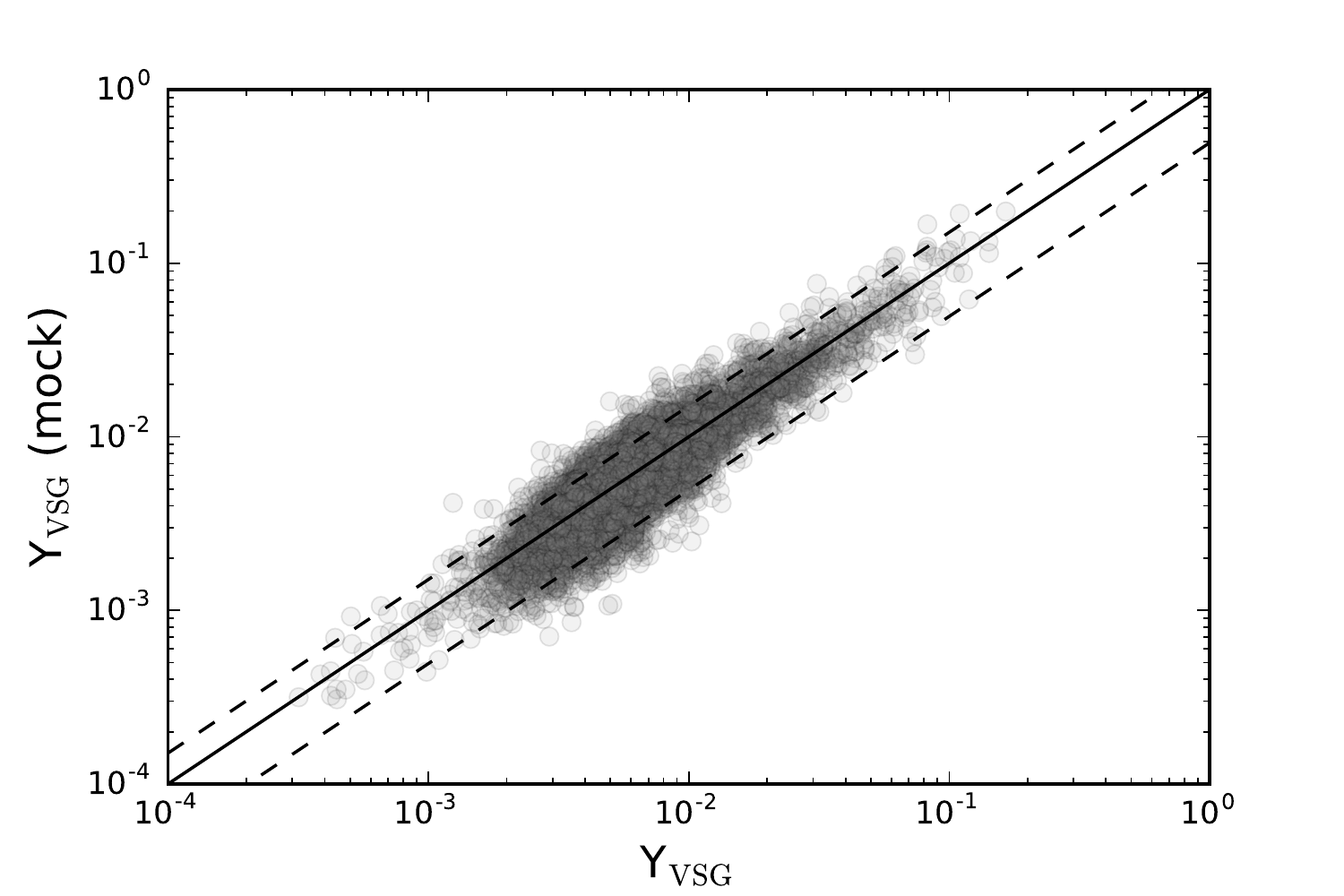}   
\includegraphics[width=0.49\textwidth]{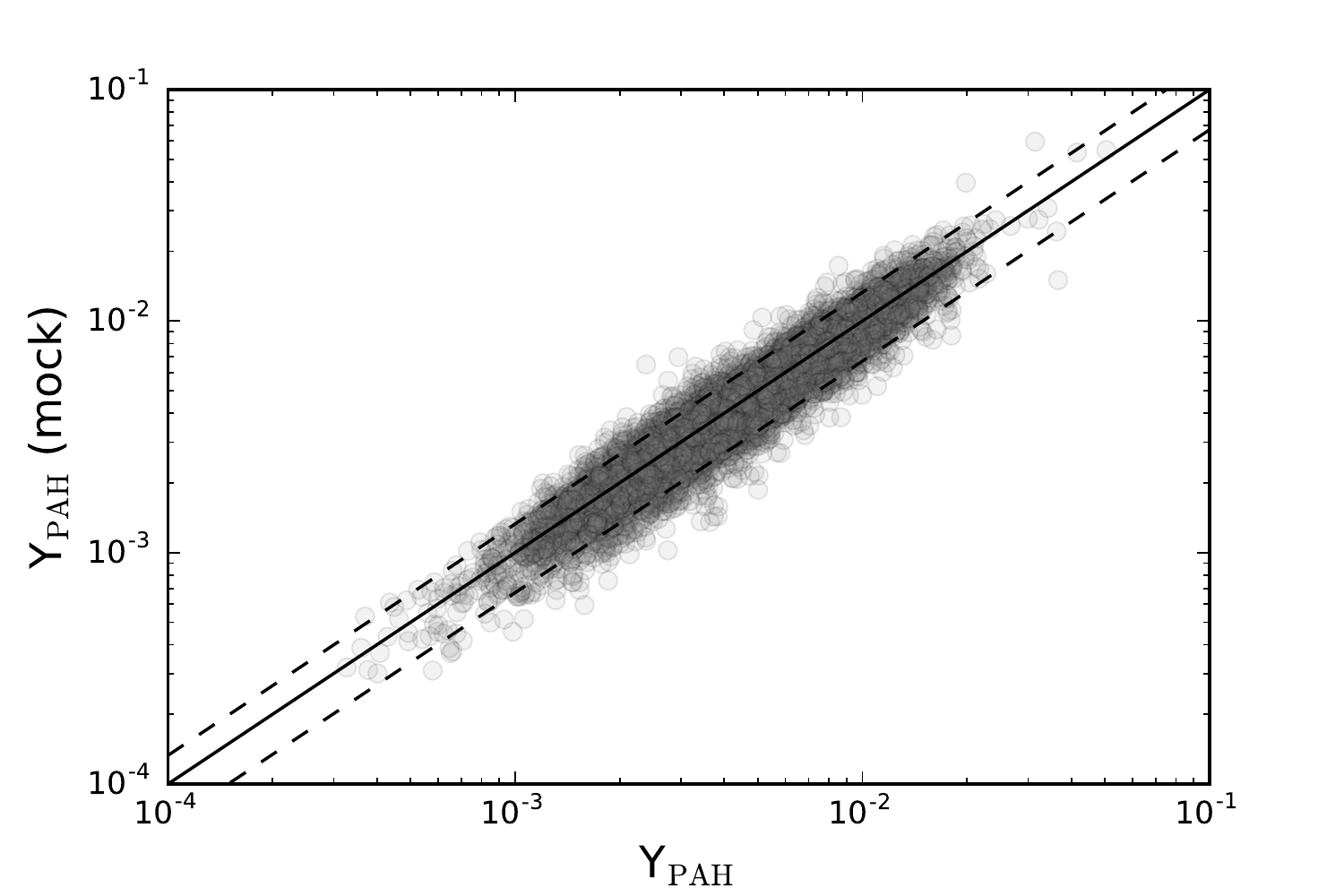}   
\includegraphics[width=0.49\textwidth]{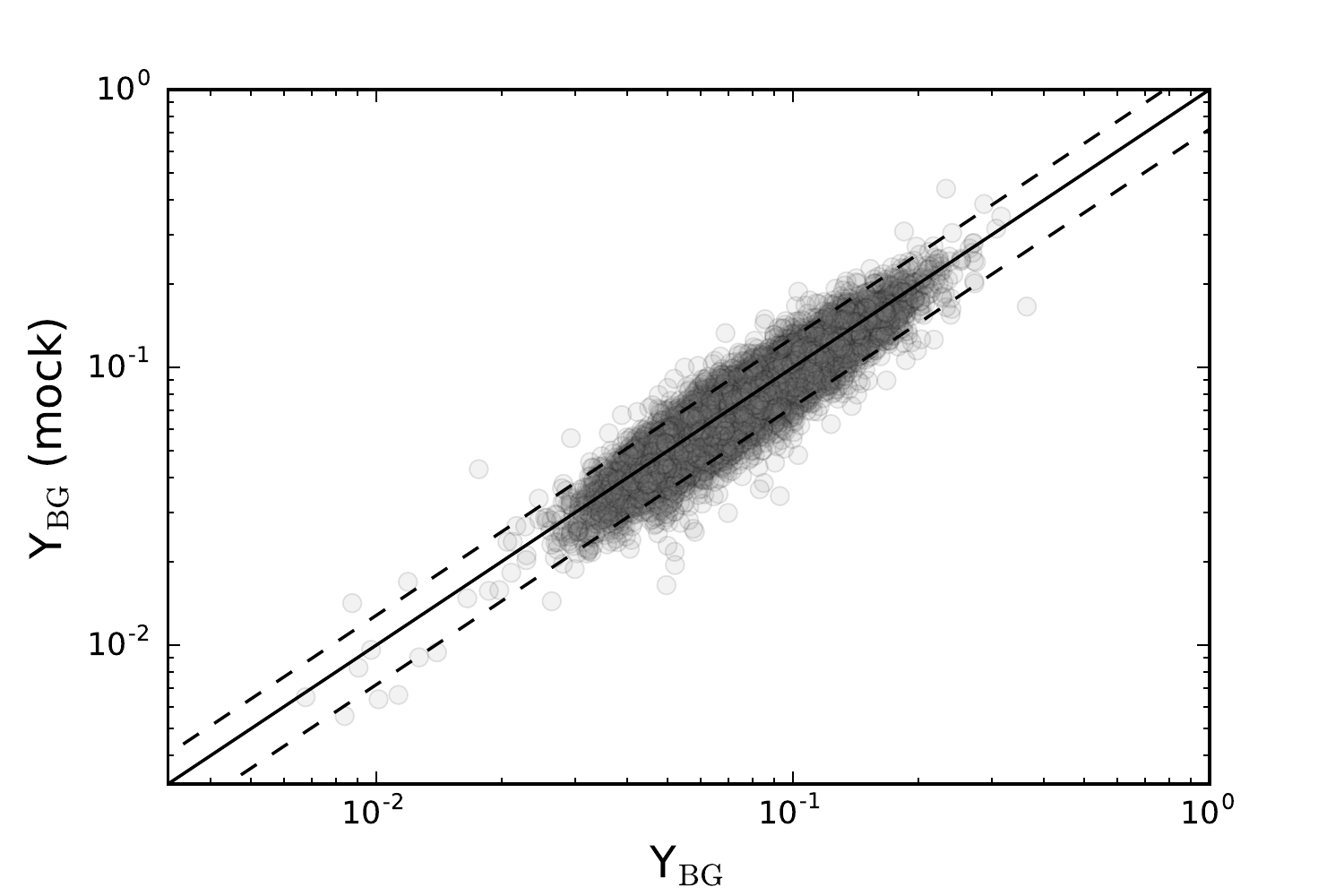}   
\caption{Robustness of the best fit for the $G_{0}$ (top-left), \yvsg\ (top-right), \ypah\ (bottom-left), and \ybg\ (bottom-right). We use the best fit SEDs given by the method described in Section~\ref{sec:sed_fit} and create new SEDs choosing random flux values in each band from a Gaussian distribution having the best fit flux as a mean value and the observational uncertainty as $\sigma$. The best fit parameters for the {\it mock} SEDs are then compared with those initially obtained. The dashed lines represents 1+median$(\sigma)$ of the original best fit parameters.}
\label{robustfit}
\end{figure*}

For the first procedure for each pixel we use the best fit SED and create a new {\it mock} SED. The fluxes of the  {\it mock} SED in each band are chosen from a random Gaussian flux distribution having the best fit flux as a mean value and the observational uncertainty as $\sigma$  \citep[see][]{2012A&A...539A.145B}. We then fit the {\it mock} SED with the same procedure explained in Section~\ref{sec:sed_fit} and obtain the best {\it mock} parameters. The comparison between the best fit and the best {\it mock} parameter values for each pixel is shown in Fig.~\ref{robustfit}. Most of the best {\it mock} parameter values agree with the original ones within 1-$\sigma$, which reinforces that the fitting method is robust.  

\begin{figure*}
\centering 
\includegraphics[width=0.49\textwidth]{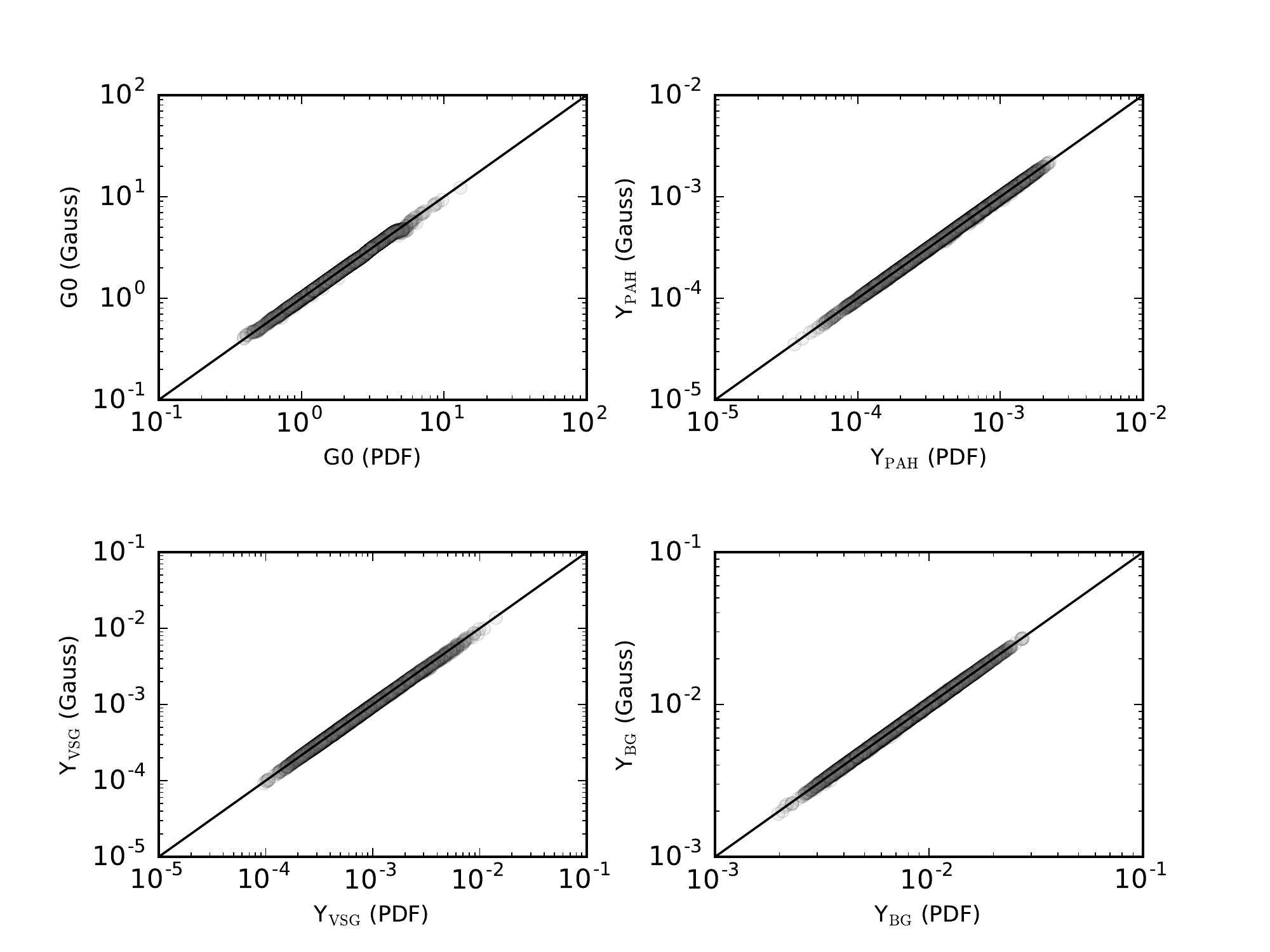}  
\includegraphics[width=0.49\textwidth]{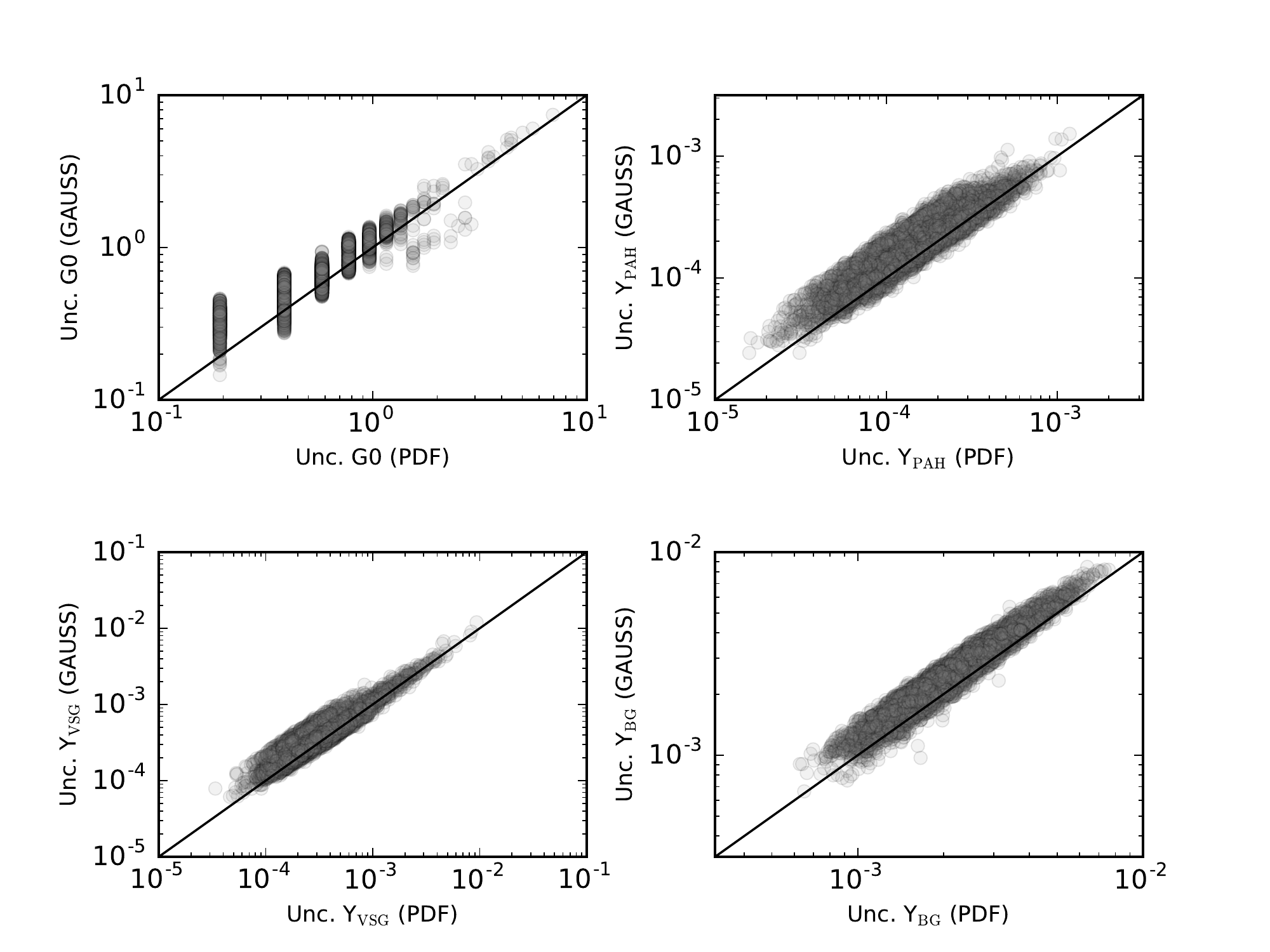}  
\caption{Left: Comparison of the best fit parameter values adopted here and the mean of the Gaussian fitted to each PDF in each pixel and for each parameter. The agreement between each magnitude shows that the PDF are well represented by a single Gaussian. Right: Comparison of the adopted uncertainty -the 16th-84th percentile range- and the FWHM value derived from the Gaussian fit in each pixel. }
\label{fig:Gauss_PDF_compar}
\end{figure*}

For the second procedure we fit the PDF of each parameter with a single Gaussian and compare the mean and the Full Width at Half Maximum (FWHM)  with the best fit parameter value and the corresponding uncertainty. The result is shown in Fig.~\ref{fig:Gauss_PDF_compar}. We can see that the mean of the fitted Gaussian agrees very well with the assumed best fit parameter value for the four input parameters, which shows that in general the PDFs are well fitted with a single Gaussian. Although with higher dispersion, the estimated uncertainties of the PDFs and the FWHM values giving by the Gaussian fit are in agreement as well.

\section{Double-peak PDFs}\label{App:doub}

We make use of the results of the Gaussian fitting to each PDF and consider the existence of double peaks in the PDFs in two ways. First, we check if there are secondary peaks in the PDF outside the main one. For that, we look for peaks with an intensity higher than 0.3 the maximum of the fitted Gaussian and located at a distance larger than 7$\sigma$ from the main peak. We do this for each parameter and pixel. We confirm there are no secondary peaks outside the main one of the PDF distribution.

\begin{figure*}
\centering 
\includegraphics[width=\textwidth]{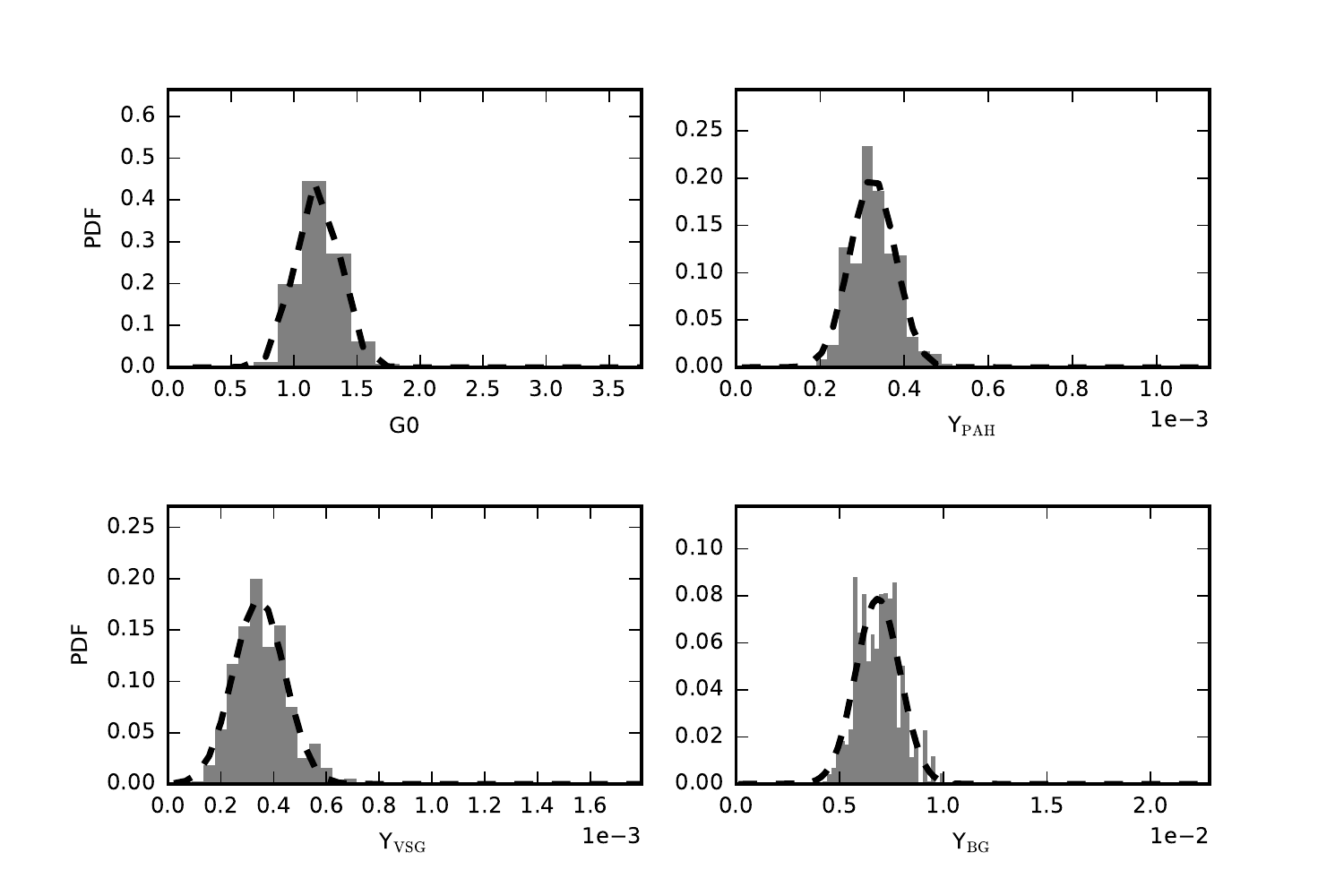} 
\caption{PDF for pixel (70, 189) where a double peak in the BG PDF has been detected. Our program looks for secondary peaks with 0.5 times the intensity of the maximum of the PDF and located at a distance higher than 0.5 times the mean FWHM, where the mean has been taken over all the pixels in each parameter. The adopted uncertainty of \ybg\ for this pixel -the 16th-84th percentile range- accounts for the double peak of the PDF for this particular pixel.}
\label{fig:dbp}
\end{figure*}

Second, it might also happen that the PDF distribution could be described with two peaks close to each other, as it is shown in the lower-right panel of Fig.~\ref{fig:dbp}. To detect these cases, we look for peaks having an intensity higher than 0.5 the maximum of the PDF distribution and separated a distance higher than one $<\sigma>$, where $<\sigma>$ is the mean $\sigma$ over all pixels in each parameter. We detect double peaks in 28(0.3\%), 108(1.3\%), 240(2.9\%) and 136(1.6\%) PDFs for $\rm G_{0}$, \ypah, \yvsg, and \ybg, respectively. The low fraction of PDFs with double peaks confirms the assumption of single-peaked PDFs and reinforces the results presented here. 

\begin{figure*} 
 \includegraphics[width=\textwidth]{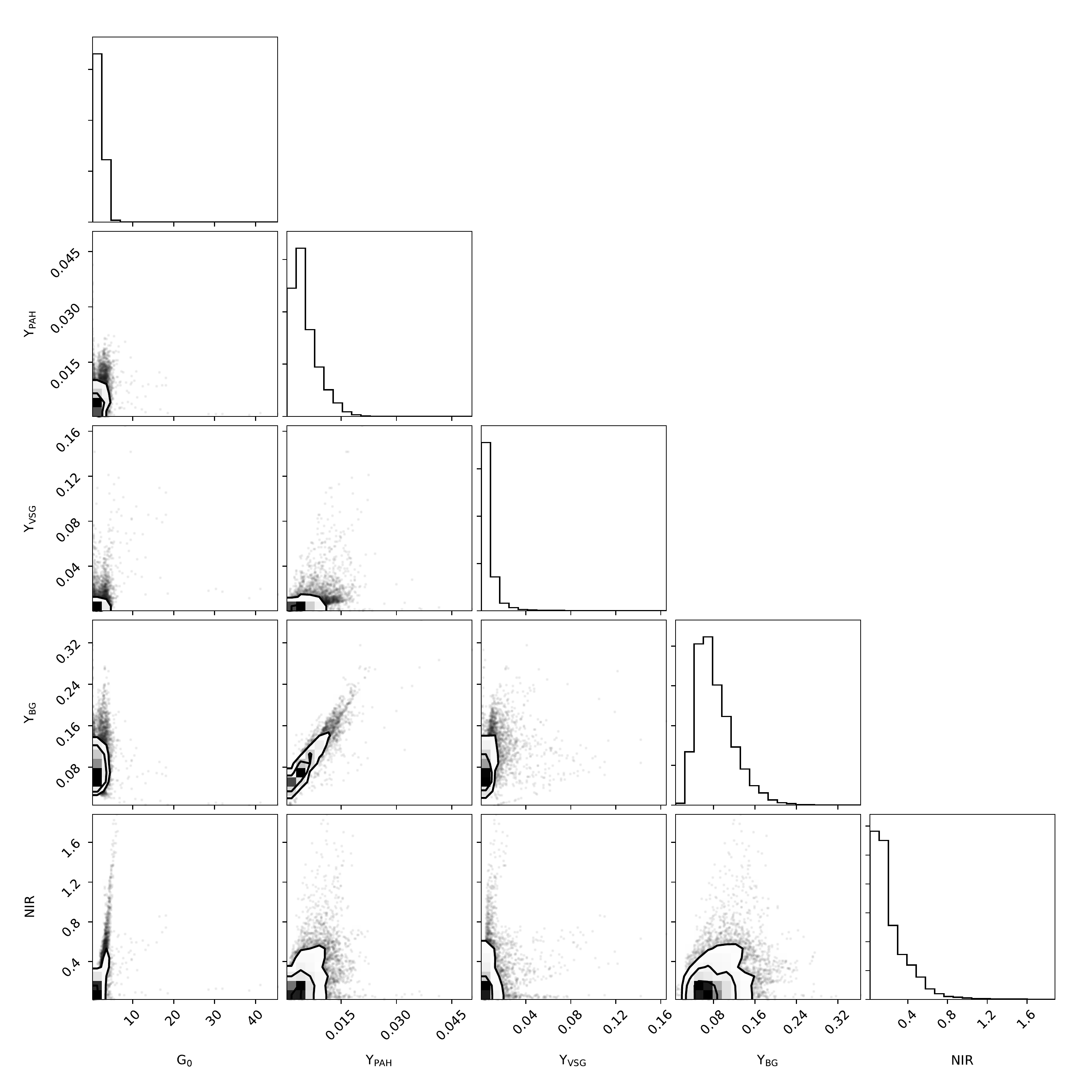}       
   \caption{Relations between the five parameters fitted with our technique described in Section~\ref{sec:sed_fit}.}
   \label{fig:allparamplot}
\end{figure*}

\end{appendix}
\end{document}